%% file: main.tex
\documentclass[a4paper,11pt]{article}
\pdfoutput=1 
\usepackage{jheppub} 
\usepackage[numbers]{natbib}
\usepackage[T1]{fontenc} 
\usepackage{subfig}
\usepackage{hyperref}
\usepackage{float}
\usepackage{feynman}
\usepackage{tikz}
\usepackage{tikz-feynman}
\usepackage{xcolor}
\usepackage{adjustbox}
\usepackage{blindtext}
\usepackage{multirow}
\usepackage{verbatim}
\usepackage{bm}
\usepackage[normalem]{ulem}

\newcommand{\OO}{\ensuremath{\mathcal{O}}}
\newcommand{\QQ}{\ensuremath{\mathcal{Q}}}

\renewcommand{\phi}{\ensuremath{\varphi}}
\newcommand{\sss}{\scriptscriptstyle}
\newcommand{\sst}{\scriptstyle}
\newcommand{\Op}[1]{\OO_{\sss #1}}
\newcommand{\Opp}[2]{\OO_{\sss #1}^{\sss #2}}
\newcommand{\Qp}[1]{\QQ_{\sss #1}}
\newcommand{\Qpp}[2]{\QQ_{\sss #1}^{\sss #2}}

\newcommand{\cp}[1]{c_{\sss #1}}
\newcommand{\Cp}[1]{C_{\sss #1}}
\newcommand{\cpp}[2]{c_{\sss #1}^{\sss #2}}
\newcommand{\Cpp}[2]{C_{\sss #1}^{\sss #2}}
\def\lra#1{\overset{\text{\scriptsize$\leftrightarrow$}}{#1}}
\newcommand{\sw}{s_{\sss W}}
\newcommand{\cw}{c_{\sss W}}
\newcommand{\mw}{m_{\sss W}}
\newcommand{\mz}{m_{\sss Z}}
\newcommand{\mt}{m_{t}}
\newcommand{\ttbar}{t\bar{t}}
\newcommand{\ttZ}{t\bar{t}Z}

\newcommand{\twll}{tW\ell\ell}

\newcommand{\mwz}{m_{\sss WZ}}
\newcommand{\gztr}{g^{\sss Z}_{t_{\sss R}}}
\newcommand{\gztl}{g^{\sss Z}_{t_{\sss L}}}

\newcommand{\gzbl}{g^{\sss Z}_{b_{\sss L}}}

\newcommand{\gwz}{g_{\sss WZ}}
\newcommand{\bWtZ}{b\,W\to t\,Z}

\title{\boldmath Single top production in association with a $WZ$ pair at the LHC in the SMEFT}

\author[a,b]{Hesham El Faham,}
\author[a,c]{Fabio Maltoni,}
\author[d]{Ken Mimasu,}
\author[e]{Marco Zaro}

\affiliation[a]{Centre for Cosmology, Particle Physics and Phenomenology (CP3),\\ Universit\'e Catholique de Louvain,\\ Chemin du Cyclotron, B-1348 Louvain la Neuve, Belgium}
\affiliation[b]{Inter-University Institute for High Energies (IIHE), Vrije Universiteit Brussel, \\Pleinlaan 2, 1050 Brussels, Belgium}
\affiliation[c]{Dipartimento di Fisica e Astronomia, Universit\`a di Bologna and INFN, Sezione di Bologna,\\ Via Irnerio 46, 40126 Bologna, Italy}
\affiliation[d]{Theoretical Particle Physics and Cosmology Group, Department of Physics, King’s College London, London WC2R 2LS, U.K.}
\affiliation[e]{TIFLab, Universit\`a degli Studi di Milano and INFN, Sezione di Milano,\\ Via Celoria 16, 20133 Milano, Italy}

\emailAdd{hesham.el.faham@vub.be}
\emailAdd{fabio.maltoni@uclouvain.be}
\emailAdd{ken.mimasu@kcl.ac.uk}
\emailAdd{marco.zaro@mi.infn.it}

\preprint{
\begin{flushright}
{CP3-21-62}\\
{KCL-PH-TH/2021-86}\\
{TIF-UNIMI-2021-19}
\end{flushright}
}

\abstract{We study single top quark production in association with a $WZ$ pair at the LHC in the context of the Standard Model (SM) and the Standard Model Effective Field Theory (SMEFT). A significant advantage of $tWZ$ compared to other EW top production processes is its sensitivity to unitarity-violating behaviour induced in its $2\to2$ sub-amplitudes through modified EW interactions. At NLO in QCD, $tWZ$ interferes with $t\bar{t}Z$ and $t\bar{t}$ and a method to meaningfully separate it from these overlapping processes needs to be employed. In order to define $tWZ$ production for total rates and differential distributions, we consider the approaches proposed in the literature for similar cases and find that diagram-removal procedures  provide reliable results both for the SM and the SMEFT in a suitably defined phase-space region. We provide robust results for total and differential cross sections for $tWZ$ at 13 TeV, including the six relevant dimension-6 operators ($\Op{tW}$,$\Op{tZ}$,$\Op{tG}$,$\Opp{\phi Q}{(-)}$,$\Opp{\phi Q}{(3)}$,$\Op{\phi t}$), also matching short-distance events to parton shower.}


\begin{document}
\maketitle
\flushbottom

\section{Introduction}
\label{sec:intro}
In spite of the astounding precision by which the Standard Model (SM) has been verified against a wide range of experimental measurements, many questions and challenges remain open. Some of these, such as the origin of the matter/anti-matter asymmetry of the Universe and the existence of dark matter, can be associated to the electroweak symmetry breaking scale, i.e., where the weak bosons, the Higgs boson and the top quark reside. The planned runs of the LHC and the future High-luminosity (HL-LHC) phase provide, for the first time, the possibility of systematically studying the properties of these particles and their interactions.  

Precise measurements at the LHC have the potential to  unveil deviations from the expectations of the SM and therefore find indications of the existence of new physics phenomena. To maximally exploit the wealth of measurements expected in the coming years, two ingredients are found to be essential. 
The first is to have predictions for SM processes to a precision that is better than or, at least, comparable to the expected experimental uncertainties. In addition, such predictions should be in a form that can be directly employed by the experimental collaborations, {\it i.e.}, to be directly used in their event simulations. A vigorous research program in this direction is being pursued~\cite{EuropeanStrategyforParticlePhysicsPreparatoryGroup:2019qin}. The second key element is an interpretation framework which can fully exploit information coming from different measurements, observables or experiments operating at different energy scales. UV complete models can be used to this end, yet due the plethora of possibilities that are still viable, both in terms of models and parameter space, a systematic search is an extremely challenging endeavour. 
An alternative, very powerful approach is that of assuming new physics to reside  at a scale $\Lambda$, higher than those directly probed (which is justified a posteriori by the very absence of any sign of new physics at colliders) and employ an Effective Field Theory (EFT) approach~\cite{Weinberg:1978kz,Buchmuller:1985jz,Leung:1984ni}. A particularly simple choice is that of the Standard Model EFT (SMEFT), which assumes that the gauge symmetries of the SM are linearly realised, and adds towers of (gauge-invariant) operators of higher dimensions to the SM Lagrangian. The SMEFT provides a consistent and calculable framework where deviations from the SM can be predicted in type and pattern (yet not in size) and systematically analysed. In addition, SMEFT predictions can be improved in terms of strong and weak corrections, in the same way as those of the SM. Together with the expansion in operator dimension, this provides a fully predictive, consistent and improvable framework to interpret precision measurements. 

Such an approach is being developed and applied to the electroweak, Higgs and top sectors by performing global fits to publicly available data~\cite{Brivio:2019ius,Hartland:2019bjb,Ethier:2021bye,Ellis:2020unq,Buckley:2015lku,Miralles:2021dyw}. These first studies, on the one hand, prove the feasibility and usefulness of using the SMEFT as an interpretation framework, and on the other, provide indications on which directions (operators) in the fit can be constrained and whether complementary information and/or new strategies would be helpful. Two features of gauge-invariant, higher dimensional operators, as encapsulated in the SMEFT, are particularly relevant for this work. First, contributions from said operators can lead to energy-growth in scattering amplitudes that can be visible in suitably constructed observables~\cite{Farina:2016rws}. Second, multiple processes can be used to search for signs of operators that link different fields, including those that do not explicitly feature the supposedly relevant particles in the initial or final state. This occurs most frequently in operators involving the Higgs field, which, though the Goldstone degrees of freedom, connects to amplitudes involving the longitudinally polarised $W$'s and $Z$'s~\cite{Henning:2018kys}. 

The sensitivity to new physics of the SMEFT interpretation obviously also depends on the accuracy of the SMEFT predictions themselves. This is very easy to understand if one keeps in mind that the power of the SMEFT relies in establishing correlations among possible deviations in the interactions. The more precise the predictions for such correlations are, the higher the sensitivity is to new physics effects. As a consequence, an intense effort has been invested in obtaining next-to-leading order (NLO) accurate predictions, and in particular the (normally) dominant QCD corrections, for SMEFT predictions at the LHC in the top~\cite{Zhang:2013xya,Zhang:2014rja,Degrande:2014tta,Franzosi:2015osa,Zhang:2016omx,Bylund:2016phk,Maltoni:2016yxb,Rontsch:2014cca,Rontsch:2015una,Degrande:2018fog} and Higgs~\cite{Hartmann:2015oia,Ghezzi:2015vva,Hartmann:2015aia,Gauld:2015lmb,Mimasu:2015nqa,Degrande:2016dqg} sectors. Recently, the full automation of NLO corrections in QCD has been achieved~\cite{Degrande:2020evl}, allowing this level of accuracy for virtually any process of interest at the LHC.  

In this work, we focus on a very specific process, i.e. the production of single top quark in association with a pair of weak bosons, a $W$ and a $Z$. At the tree level, this process is very similar to single top associated production with a $W$ boson, except with the added emission of a $Z$ boson from any of the external (or internal) lines, i.e., $gb\to tWZ$. While characterised by a relatively small cross section, of the order of 100 femtobarns, this process features unique unitarity-violating behaviour induced in its sub-amplitudes when top quark EW interactions are modified~\cite{Dror:2015nkp,Maltoni:2019aot} and therefore provides an enhanced sensitivity to new physics deviations. The final goal of our work is to explore whether such a feature could be exploited to constrain/discover new physics once enough luminosity will be collected at HL-LHC, and an initial, LO study in this direction has confirmed the promising potential~\cite{Keaveney:2021dfa}. 
To this end, in analogy to what happens for $tW$ production, the first step is to provide an operational definition of $tWZ$ that allows to meaningfully include NLO QCD  corrections. 

At leading order (LO) accuracy and in the five-flavor scheme (5FS), $tWZ$ can be easily identified  through the tree-level partonic process $gb \to tWZ$. At NLO, real corrections of the type $gg \to tWbZ$ arise that can feature a resonant (anti-)top quark in the intermediate state and therefore overlap with $gg \to t\bar{t}Z$, $t \to \bar bW$, i.e. $t\bar{t}Z$ production at LO, as well as with $gg \to \bar{t}t$, $\bar t\to \bar bWZ$, i.e., $\bar{t}t$ production at LO. It is clear that this poses a two-fold problem. First, the cross section of $t\bar{t}Z$ is roughly five times larger than $tWZ$ while $\bar{t}t$ cross section is larger by few orders of magnitude, at LO. A na\"ive computation would therefore lead to a poorly-behaved perturbative expansion on the one hand, and on the other, to a loss of the overall precision due to the "LO overlap" with other final states whose predictions are currently known to much better accuracy (NLO in QCD/EW interactions or even at NNLO in QCD). Such resonant contributions need therefore to be subtracted. Different methods to solve this issue have been proposed that achieve a subtraction of resonances that is local in phase-space~\cite{Frixione:2008yi,White:2009yt,Weydert:2009vr,Re:2010bp,Binoth:2011xi,GoncalvesNetto:2012yt,Gavin:2013kga,Gavin:2014yga,Demartin:2016axk,Frixione:2019fxg}. Here we follow that of Ref.~\cite{Demartin:2016axk,Frixione:2019fxg}. In short, one subtracts the resonant contributions, yet, due the quantum mechanical nature of process, ambiguities are introduced  whose impact on the analyses need to be carefully assessed, both for the SM as well as for the SMEFT amplitudes. In this respect, as we will see, achieving a meaningful definition valid at NLO accuracy for $tWZ$ production is more complicated than $tW$ or even $tWH$ and needs an \emph{ad hoc} treatment.  

The plan of the paper is as follows. In Sec.~\ref{sec:def} we briefly discuss and propose an operational definition of $tWZ$ production in the SM that can be used to calculate predictions up to NLO in QCD. To this aim we make use of the subtraction techniques introduced and employed for $tW$~\cite{Frixione:2008yi}, showing that, apart from a necessary addition, they also work for $tWZ$.  In Sec.~\ref{sec:SMEFT} we discuss in detail the features of $tWZ$ in the context of the SMEFT, In Sec.~\ref{sec:results} we present the results of our study, with particular emphasis on understanding the sensitivity to the various dimension-6 operators entering in the process also as a function of the partonic energy. We conclude with a summary in Sec.~\ref{sec:conclusions}. 

\section{$tWZ$ production in the Standard Model}
\label{sec:def}
\begin{figure}[h!]
    \centering
    \includegraphics[width=\textwidth]{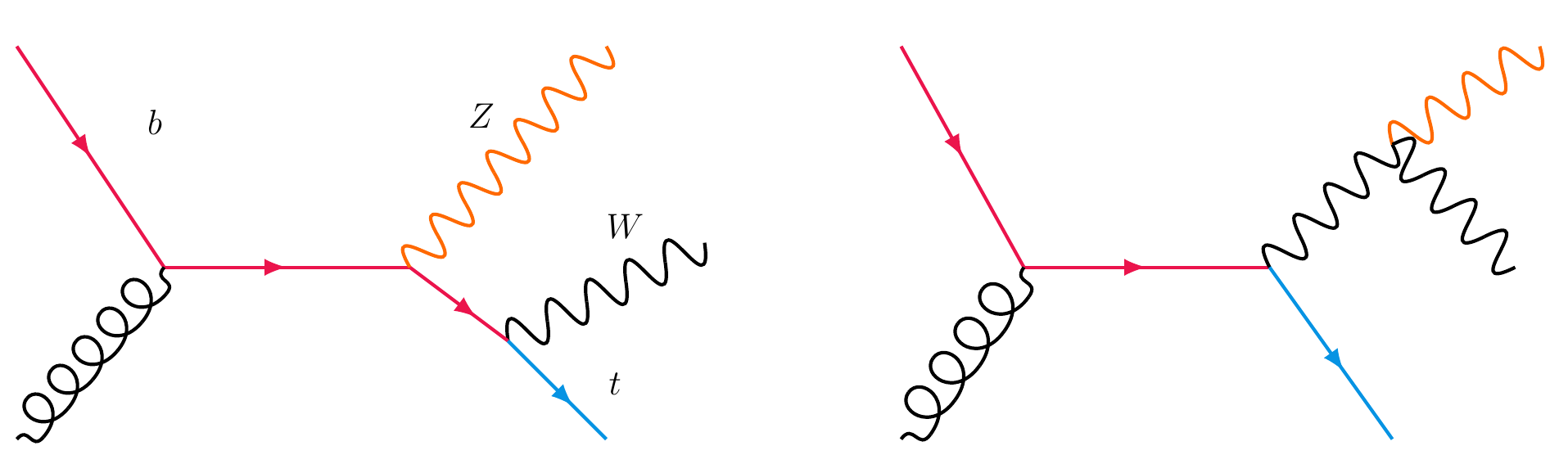}
    \caption{\label{LO}
    {Representative diagrams for $tWZ$ production in the 5FS at leading order in the gauge couplings. One can think of this process as $tW$ production (featuring two diagrams) with the subsequent Bremsstrahlung of the $Z$ boson from each of the lines (internal or external, except for the gluon) giving rise to a total of eight diagrams.}}
\end{figure}

\begin{figure}[h!]
    \centering
    \includegraphics[width=\textwidth]{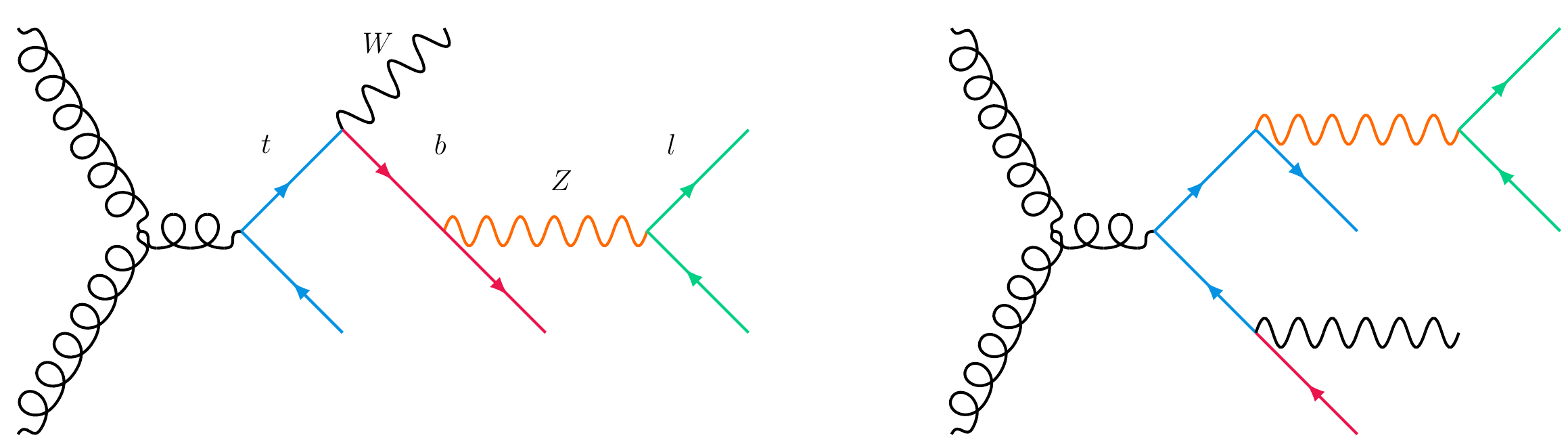}
    \caption{\label{feyn_intro_diagrams}
    Representative diagrams of the $gg$-initiated production of $t\bar t$ (\emph{left}) and $t\bar{t}Z$ (\emph{right}) processes with a potentially resonant (anti-)top quark leading to the $t\bar \ell \ell Wb$ final state.}
\end{figure}

At the leading order, $tWZ$ production can be easily defined if the 5FS is employed. In this case, the process is simply $gb\to tWZ$, shown in Fig.\ref{LO}.
The situation, however, changes at NLO where complications arise from the real radiation processes, and in particular, from  $gg (b \bar b) \to tWZ b$. These processes feature contributions where
an intermediate (anti-)top quark can go on-shell and become resonant. These contributions actually correspond to a $t \bar t$ final state, with $\bar t\to WZ \bar b$, or to $ t\bar t Z$, with $\bar t\to W \bar b$, as displayed in Fig.~\ref{feyn_intro_diagrams}.
These amplitudes share the same $tWZb$ final state, yet they are naturally associated to different underlying partonic processes, i.e., $t\bar t$ and $t\bar t Z$. Even though superficially appearing at the same order in the perturbative expansion in $\alpha_S$ and $\alpha_W$, they are in fact enhanced in the perturbative expansion when the top is close to being on-shell by a factor of $1/\alpha_W$ coming from the top width in the top propagator.  Due to the 
narrowness  of the top quark, including these diagrams potentially leads to very large corrections, which spoil the perturbative convergence of the $tWZ$ cross section and obscures its physical meaning. 
Technically, a problem of consistency also arises: top quarks appear in the same diagram on the one hand 
as stable asymptotic states, and on the other hand as intermediate ones. In the former case, their width should be set to zero, in order to fulfill local cancellation of infra-red divergences, whereas in the latter a finite top quark width must be kept, in order to regularise the resonant propagator. In general, this violates QCD gauge invariance. 

A possible solution to the problems illustrated above consists of considering a leading order process in terms of (pseudo)-stable final states, {\it i.e.}, $p p \to b \bar b W^+ W^- Z$ by employing a four-flavor scheme (4FS) and treating the top quark width in the complex mass scheme~\cite{hep-ph/9904472,hep-ph/0505042}. This approach could have advantages, yet it also poses non-trivial challenges. Ignoring the decays of the vector bosons and considering only the leading term $\alpha_S^2 \alpha_W^3$, it can be easily computed at tree level. Albeit at sizeable computational cost, modern techniques could also allow the automatic computation of NLO QCD corrections. In so doing, all processes mentioned above, i.e.,  doubly resonant $t\bar t$ and $t \bar t Z$ processes, singly resonant $tW^-Z$ and $\bar t W^+ Z$ and non-resonant contributions would be accounted for simultaneously, together with their interference\footnote{The computation of $p p \to b \bar b W^+ W^- Z$ including the decays of the vector bosons is available at NLO QCD accuracy with massless $b$ quarks~\cite{Bevilacqua:2019cvp}. In this case, the $b$ quarks must be resolved in order to obtain a finite prediction, hence the phase-space for $tWZ$ cannot be probed.}.
Promoting such an NLO calculation to a full event generator could also be envisaged, following, for instance, the approach proposed for $W^+ b W^- \bar b$ in \cite{Jezo:2016ujg}. While possible in principle, such an approach is not currently practical for two main reasons. First, such NLO computations are not available and implementing them  would entail considerable work. In any case, once available they will be very computationally expensive, not least in our use-case, in which we investigate a wide parameter space of BSM effects in this process. Second, such an approach would encounter complications in the experimental workflow, as some of the processes that would be included in the computation, and considered as backgrounds, {i.e.}, $t \bar t$, are known to a higher accuracy and are currently simulated in an independent way. Finally, at the theoretical level, one would face the question of the reliability of a 4FS for such a high-$Q^2$ process, i.e., how to re-sum potentially large logarithms $\log m_b^2/Q^2$. 
It is therefore useful to consider an alternative approach that is sufficiently accurate to define the signal at NLO accuracy and that can be directly employed by the experimental searches/measurements. To this aim, we develop a procedure to remove the additional resonant contributions appearing at NLO and yield an operative definition of $tWZ$. 

As mentioned in the Introduction, methods to achieve the removal of resonances appearing in higher-order corrections have been established for quite some time, and successfully applied to the related processes, $tW$ and $tWH$.
Resonant overlap removal methods that can be applied
at the differential level fall into two broad categories, dubbed diagram removal (DR) and
diagram subtraction (DS). In the latter, a suitable gauge-invariant subtraction term is constructed in order to locally suppress the contribution of the resonant process (it should be noted that there exist infinite possibilities to build such subtraction terms, which result in many alternative definitions of DS).  In the former, resonant diagrams are simply set to zero,
thus requiring a much simpler implementation, but possibly inducing some 
gauge dependence of the final result. A variant of the DR scheme, which retains the interference between resonant and non-resonant diagrams, is also possible, often dubbed as DR with interference (DR-I) or DR2 (in this case, the standard DR is referred to as DR1). In this work, we will employ the DR schemes for their ease of use. We also note that, in the cases of $tW$ and $tWH$, the DS results were found to lie in between DR1 and DR2~\cite{Demartin:2016axk}, such that the difference between the latter can serve as a measure of the uncertainty associated to these schemes.

The use of DR and DS techniques has, so far, been limited to the case where resonant diagrams correspond to $1\to 2$ decays, and (mostly) to cases where a single resonant process overlaps with each real-emission matrix element. In fact, these cases can now be treated in a fully-automatic manner~\cite{Frixione:2019fxg}. Instead, $tWZ$ production, or more specifically
its version where the decay products of the $Z$ boson are considered, $\twll$, has two underlying resonant processes, one of which proceeds through a $1\to 3$ decay (or $1\to 4$ if the $\twll$ final state is considered) see the left-hand diagram of Fig.~\ref{feyn_intro_diagrams}. We therefore consider it worthwhile to spend some time to describe the  structure of our process, in order to establish the notation. 

The general amplitude, $\mathcal{M}$, can be decomposed into a non-resonant and a resonant part,
\begin{equation}
    \mathcal{M} = \mathcal{M}_{\rm non-res} + \sum_i \mathcal{M}_{{\rm res}, i},    
\end{equation}
where the index $i$ labels different resonant processes (in the case at hand, $t \bar t$ and $t \bar t Z$). Once the amplitude is squared, one has
\begin{equation}
    |\mathcal{M}|^2 = \left|\mathcal{M}_{\rm non-res}\right|^2 + 2\Re \left(\mathcal{M}_{\rm non-res} \sum_i \mathcal{M}_{{\rm res}, i}\right) +   \left|\sum_i \mathcal{M}_{{\rm res}, i}\right|^2 .
\end{equation}
In the case of standard DR procedure (DR1), one discards the resonant terms before squaring the amplitude, so that
\begin{equation}\label{eq:DR1}
    |\mathcal{M}|^2_{\rm DR1} \equiv \left|\mathcal{M}_{\rm non-res}\right|^2 .
\end{equation}
In the case of diagram-removal with interference (DR2), one only discards the squared resonant matrix elements in the squared amplitude, but keeps their interference with the non-resonant part:
\begin{equation}\label{eq:DR2}
    |\mathcal{M}|^2_{\rm DR2} \equiv \left|\mathcal{M}_{\rm non-res}\right|^2 + 2\Re \left(\mathcal{M}_{\rm non-res} \sum_i \mathcal{M}_{{\rm res}, i}\right) .
\end{equation}
It has to be noted that, in the case of DR2, one could in principle keep the interference terms between different resonant amplitudes. However, in the following we will not consider these terms. 

Whether to include the interference between the resonant and the non-resonant contributions is a matter of how the predictions are going to be employed: the aim of this paper is to use the $tWZ$ process to constrain new physics effects via an EFT parameterisation. Thus, an operative definition of $tWZ$ as an independent process must be provided, and interferences with other processes, including the resonant ones, must be reduced as much as possible. Such a reduction can be achieved by designing suitable cuts which suppress the resonant contributions. In the case of top resonances, a suitable cut is to veto the bottom-tagged jet ($b$-jet) that would emerge from the top decay~\cite{Frixione:2008yi}. As a general rule, these tend to be harder, since the $b$-quarks originating from real emission diagrams are instead enhanced in the soft and collinear limits. Indeed, we find that imposing such a cut completely kills 
the difference between DR1 and DR2, implying a negligible overlap with the resonant processes, and a satisfactory level of control on the theoretical uncertainty associated to the subtraction method.

\begin{figure}[t!]
    \centering
    \includegraphics[trim=0.0cm 5.0cm 0.0cm 0.0cm, clip,width=.48\textwidth]{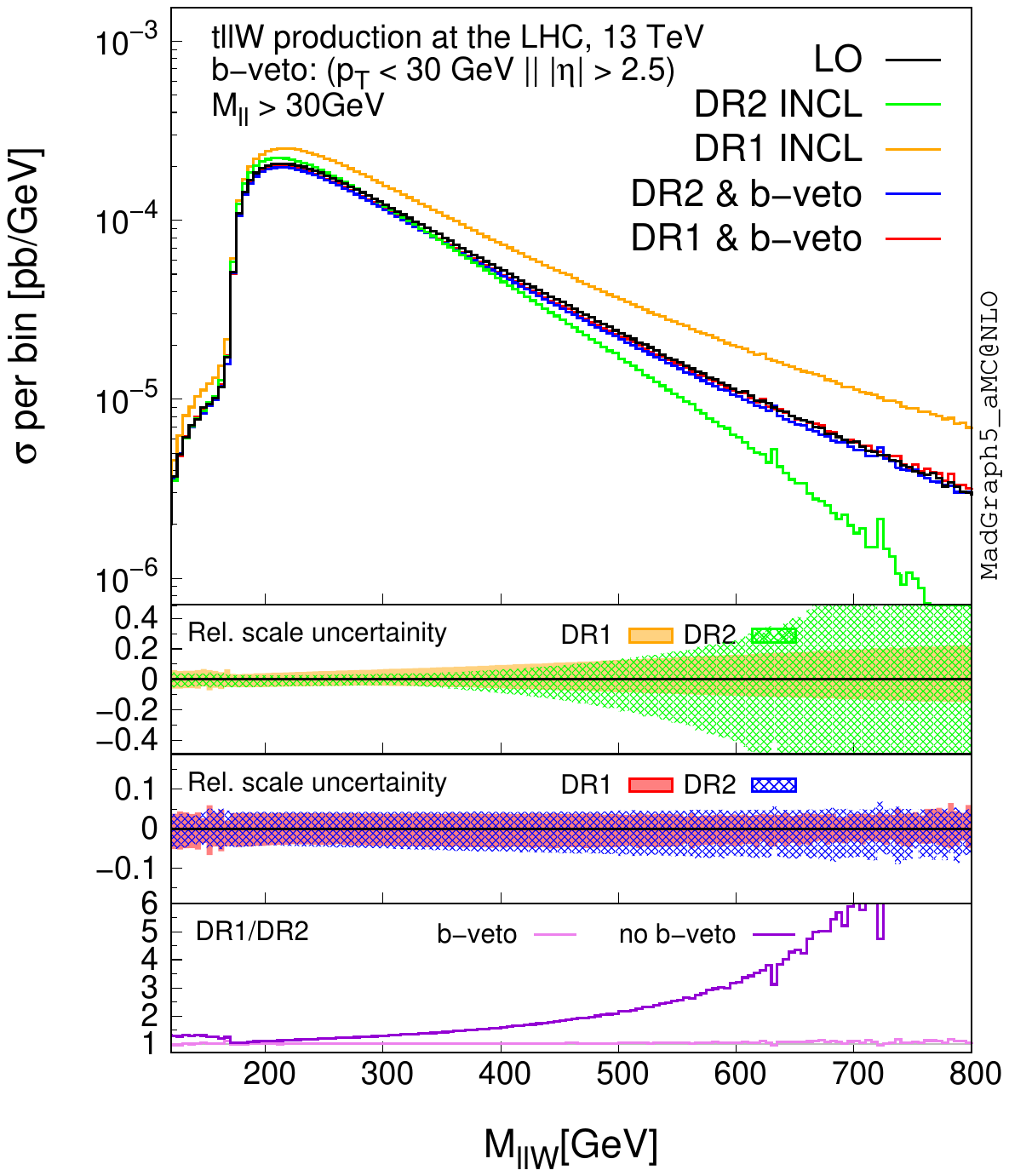}
    \includegraphics[trim=0.0cm 5.0cm 0.0cm 0.0cm, clip,width=.48\textwidth]{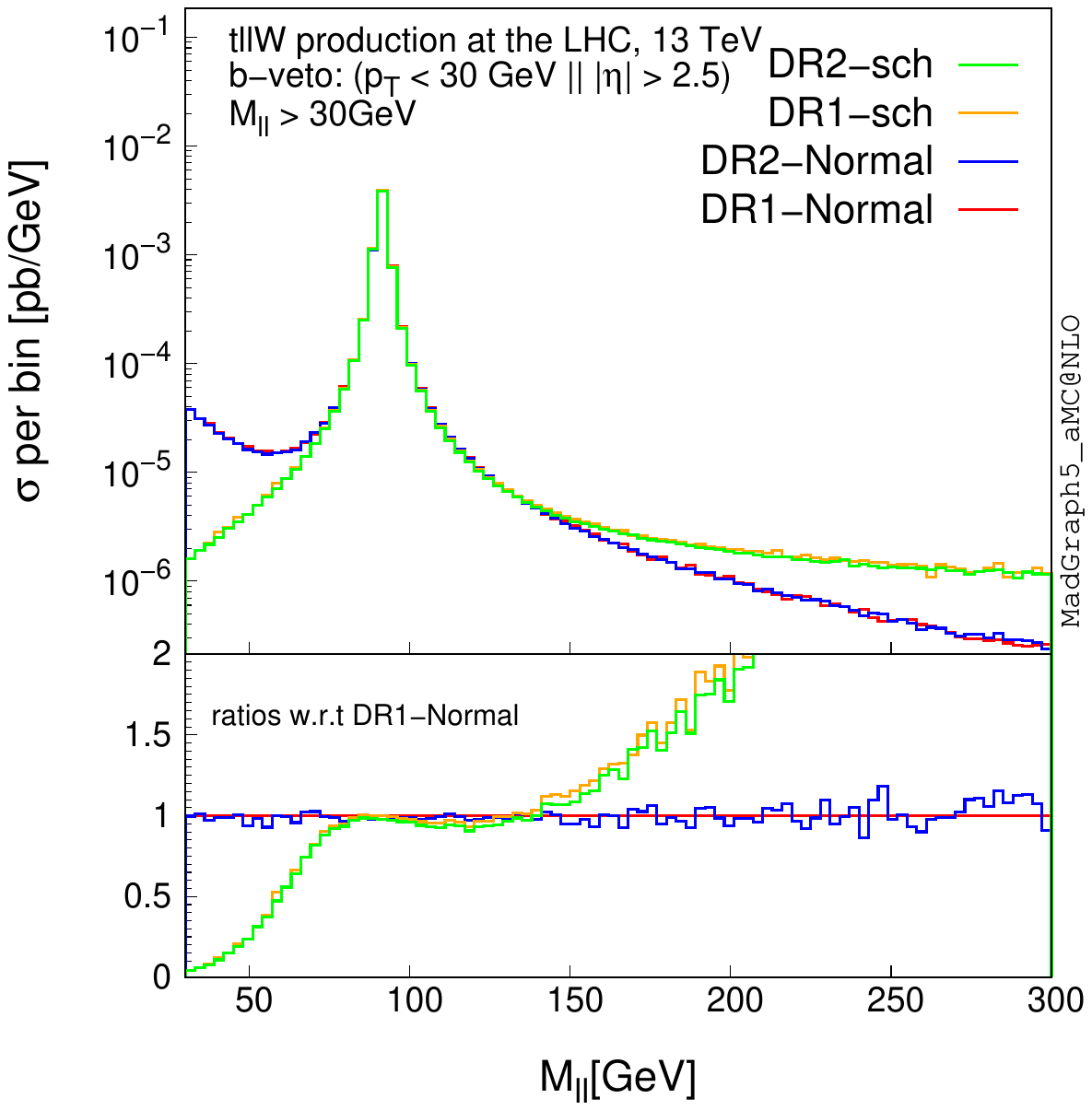}
    \caption{\label{fig:sm_main_text}  (\emph{Left}) The differential cross section in the invariant mass bins of the $W\ell^+\ell^-$ system for the $\twll$ process at NLO in QCD. Predictions are given for the DR1 and DR2 overlap removal schemes
    before and after including a veto on central ($|\eta|<2.5$) or hard ($p_T>30$ GeV) $b$-quarks. The first two insets depict relative scale uncertainties and the third  shows the DR1/DR2 ratio.  (\emph{Right}) DR1 and DR2 predictions of the invariant mass of the lepton pair, $M_{l^{+}l^{-}}$. Full predictions  labelled `Normal' case, are shown alongside the `sch' case, which only includes diagrams with a $Z$ boson in the $s$-channel. The inset shows the ratios of the distributions w.r.t to the `Normal' DR1 prediction.}
\end{figure}

The left panel of Fig.~\ref{fig:sm_main_text} shows the differential cross section predictions in the $W\ell^+\ell^-$ invariant mass bins for the $tW\ell\ell$ process at NLO in QCD in the DR1 and DR2 schemes, both at inclusive-level and after imposing a veto on central ($|\eta|<2.5$) or hard ($p_T>30$ GeV) $b$-quarks. We also apply a lower cut on the dilepton invariant mass of $m_{\ell\ell}<30$ GeV to suppress the photon-mediated contributions. The first two insets show the relative scale uncertainties of the DR1 and DR2 predictions for the inclusive and $b$-veto cases, respectively. The third inset shows the ratio of the DR1 and DR2 differential cross sections for the two cases. The clear discrepancy between DR1 and DR2 at the inclusive-level indicates that a significant overlap exists in the full phase-space, while in the $b$-veto case, the disagreement essentially disappears. This is in line with the observed behaviour in previous studies of the $tW$~\cite{Frixione:2008yi} and $tWH$~\cite{Demartin:2016axk} processes. We also note that the  discrepancy between DR1 and DR2 as well as the scale uncertainty of the latter grows significantly as function of energy, indicating a pathology in the predictions. This is likely due to the fact that one is probing regions of phase-space where the potentially resonant diagrams are no longer so, causing gauge-invariance issues. Away from the resonant region, the diagrams lose their special status and should therefore be retained to ensure a well behaved amplitude.

\begin{figure}[h!]
    \centering
    \includegraphics[trim=0.0cm 6.5cm 0.0cm 0.0cm, clip,width=.6\textwidth]{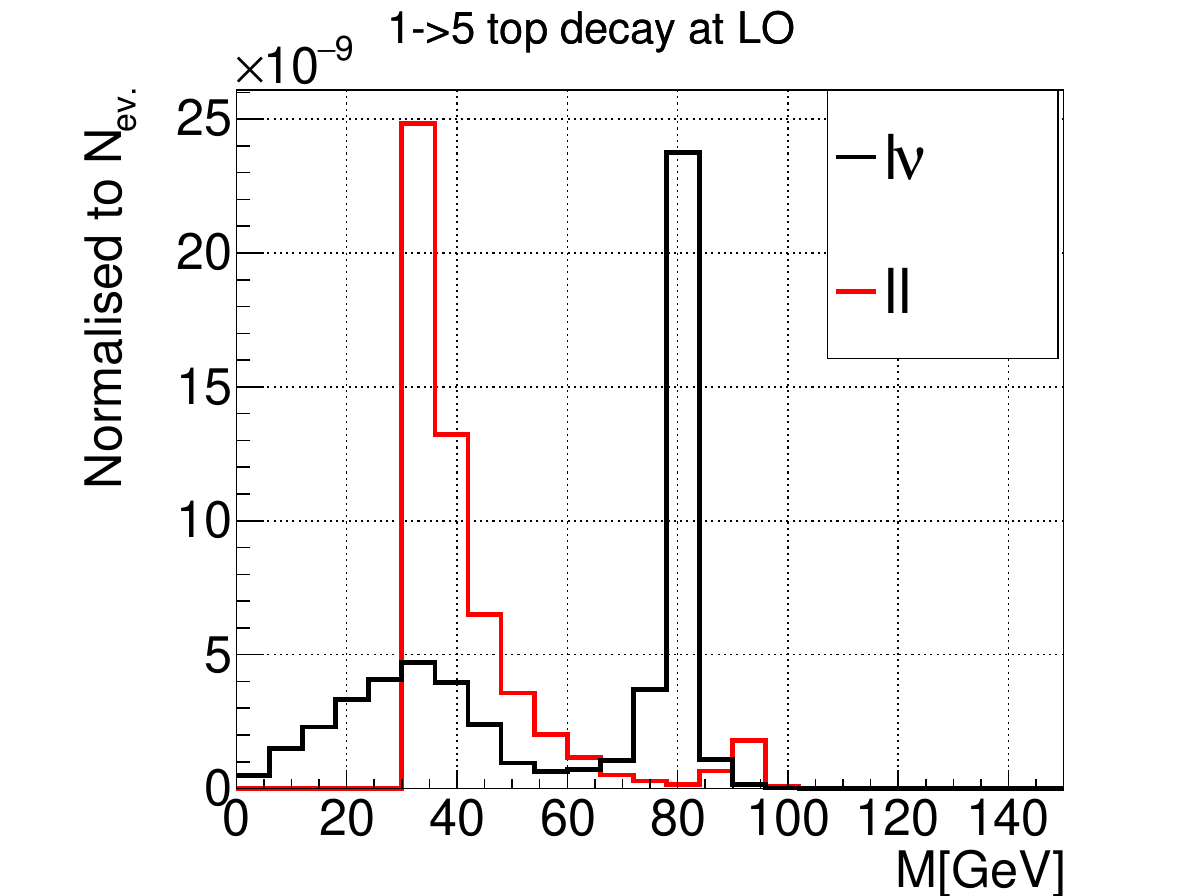}
    \caption{\label{fig:top_decay} The normalised invariant mass distributions of the $\ell\nu$ and $\ell\ell$ pairs in the 5-body, $t\to b e^{+}e^{-} \mu^+\nu_{\mu}$ decay, where different lepton flavours were simulated to distinguish those predominantly coming from intermediate neutral and charged gauge bosons. The top decay is performed in the 5FS and using the complex mass scheme at LO. A minimum invariant-mass cut of 30 GeV is imposed on the same-flavour, opposite-charge lepton pair. The neutral final state invariant mass is dominantly away from the $Z$ pole. }
\end{figure}

Another simplification of the process definition is that, in the EFT study that follows, a stable $Z$ boson will be considered \emph{in lieu} of the dilepton pair. This simplification is justified by the fact that, if a $b$-veto is enforced, effects due to diagrams not involving a $Z \to \ell \ell$ splitting only matter far away from the $Z$ boson peak, where the cross section is suppressed. If we select only diagrams with a resonant $Z$ boson and compare the cross section with the case where all diagrams are retained, the dilepton invariant mass distributions depart from each other for less than 10\% in the range [70, 150] GeV, as displayed in the right panel of Fig.~\ref{fig:sm_main_text}. While the cross section contribution continues to grow with lower invariant masses, in practice, it would be suppressed by cuts and lepton trigger requirements. Moreover, the $tWZ$ component can easily be isolated by requiring the dilepton invariant mass to lie around the $Z$-peak (see, \emph{e.g.}, the $tZq$ analysis of Ref.~\cite{ATLAS:2020bhu}).
With this simplification, we can safely neglect the $t \bar t $, $t \to W b \ell \ell$ resonant process. Indeed, inspecting the five-body decay
$t\to b \ell^+ \ell^- {\ell^+}' \nu$ shown in Fig.~\ref{fig:top_decay}, the opposite-charge, same-flavour lepton-pair invariant mass
peaks at low values, and shows a suppression at the $Z$ mass. 
Besides, $t\to W Z b$ occurs extremely close to the top threshold, and in practice never contributes to the cross section.

\section{ $tWZ$ in SMEFT}
\label{sec:SMEFT}
\subsection{Probing new physics via high-energy top quark scattering}
\begin{figure}[h!]
    \centering
    \includegraphics[width=\textwidth]{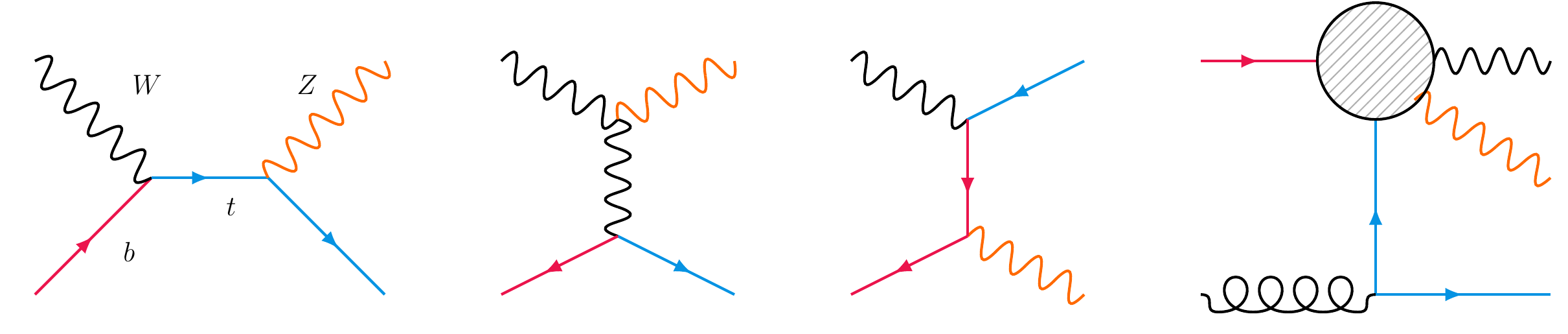}
    \caption{\label{fig:feyn_eft} SM diagrams for the $\bWtZ$ sub-process. The far right diagram shows the embedding of the $\bWtZ$ sub-amplitude, represented by the blob, into $gb\to tWZ$ production at a hadron collider}
\end{figure}

Being a rare EW top production process, $tWZ$ is a potential probe of the weak couplings of the top quark that are, to date, relatively poorly measured. Although they cannot compete in terms of total rate with more traditional probes such as $t\bar{t}Z$, these EW-driven processes have the significant advantage of being sensitive to unitarity-violating behaviour induced by modified EW interactions~\cite{Dror:2015nkp,Maltoni:2019aot}. Searching for the characteristic energy growth in EW top quark scattering amplitudes is therefore a complementary avenue for searching for new physics via new interactions. The relevant sub-amplitude probed by $tWZ$ is $\bWtZ$, whose tree-level, SM Feynman diagrams are given in Fig.~\ref{fig:feyn_eft}. Its embedding into the $tWZ$ process at a hadron collider is also shown.
Clearly, this scattering will be affected by modified weak interactions of the top and bottom quark, as well as the $SU(2)$ gauge bosons self-interactions. What is not immediately obvious is that such modifications lead to an unacceptable growth with energy, beyond the maximal $E^0$ dependence admitted by perturbative unitarity for a $2\to2$ scattering amplitude. This indicates that the theory has a finite range of validity in energy, beyond which new, unitarising physics must appear, and clearly motivates the use of SMEFT to classify the possible deviations from SM interactions. 

The general application of this concept to top EW scatterings has been studied in detail for a wide range of amplitudes and associated collider processes in Refs.~\cite{Dror:2015nkp,Maltoni:2019aot}, elaborating on the connection between modified SM interactions and SMEFT operators, as well as the emergence of energy growth via contact interactions between fermion and gauge/Goldstone boson currents. Going from amplitudes to the observable cross sections is key, in order to determine the interference structure of the new physics contributions with the corresponding SM amplitudes. The interference term is the leading, dimension-6 contribution in the $1/\Lambda$ SMEFT power counting.
In most cases, when the scattering states involve longitudinal polarisations of gauge bosons, the energy-growing SMEFT helicity configurations coincide with the leading high-energy SM ones, such that their interference also grows with energy. This provides an enhanced, leading-order sensitivity to the coefficients, motivating searches that exploit these channels.

\subsection{Conventions, relevant degrees of freedom \&  existing bounds}
In this study, we consider SMEFT contributions to $tWZ$ via the $\bWtZ$ sub-amplitude with a particular flavor symmetry assumption that singles out top quark interactions denoted by
\begin{align}
    U(3)_l\times U(3)_e \times U(2)_q\times U(2)_u\times U(3)_d\equiv U(2)^2\times U(3)^3.
\end{align}
The subscripts refer to the five fermion representations of the SM. This minimal relaxation of the universal, $U(3)^5$, scenario allows chirality-flipping interactions for the top quark such as dipole interactions and those modifying the Yukawa coupling. Furthermore, it also allows the 3$^{\text{rd}}$ generation quark doublet and right-handed top quark fields to receive independent modifications to their couplings to gauge bosons via the current operators. Our notation and operator conventions follow those of Refs.~\cite{Aguilar-Saavedra:2018ksv,Degrande:2020evl}, where more detailed discussions of the flavor symmetry baseline can be found. There are eight relevant operators consistent with the flavor symmetry assumption, summarised in Tab.~\ref{tab:relevant_operators} alongside recent limits on their Wilson coefficients from global fits\footnote{Note that, in keeping with the assumed flavor symmetry, the right-handed charged current operator, 
$\Op{\phi tb}=
i\big(\tilde{\phi}^\dagger\,{D}_\mu\,\phi\big)
\big(\bar{t}\,\gamma^\mu\,b\big)$
, and the operator modifying the right-handed $Zb\bar{b}$ coupling, $\Op{\phi b}=
i\big(\phi^\dagger\,\lra{D}_\mu\,\phi\big)
\big(\bar{b}\,\gamma^\mu\,b\big)$ 
are omitted, although they do contribute to the $\bWtZ$ amplitude. The interference term of $\Op{\phi tb}$ and the entire energy-growing contribution from $\Op{\phi b}$ are suppressed by the $b$-quark mass.}.

Although it does not enter in $\bWtZ$ scattering, the top quark chromomagnetic moment, $\Op{tG}$, does affect the $tWZ$ process, through the coupling of the initial state gluon to the top quark. 
\begin{table}[ht!]
    \centering
    \input{tab_operators}
    \caption{\label{tab:relevant_operators}
    SMEFT operators affecting $\bWtZ$ and/or $tWZ$ in the $U(2)^2\times U(3)^3$ flavor assumption. Operators denoted by $\OO$ align with the \texttt{SMEFTatNLO}~\cite{Degrande:2020evl} conventions used in this work, while those denoted by $\QQ$ are the original Warsaw basis operators to which they are related by a simple linear rotation. Individual and marginalised limits from recent global fits are given both at linear ($\mathcal{O}(\Lambda^{-2})$) and quadratic ($\mathcal{O}(\Lambda^{-4})$) level, for $\Lambda=1$ TeV.
    }
\end{table}
In keeping with the conventions of Ref.~\cite{Aguilar-Saavedra:2018ksv}, two subsets of operators are rotated to align the $t\bar{t}Z$ coupling directions, with Wilson coefficients, $c_i$, defined in terms of the Warsaw basis coefficients, $C_i$, as follows:
\begin{align}
    \begin{split}
        \cp{tW}&=\Cp{tW},\,\cp{tZ} = -\sin{\theta_{\sss W}}\Cp{tB}+\cos{\theta_{\sss W}}\Cp{tW}\\
        \cpp{\phi Q}{(3)}&=\Cpp{\phi Q}{(3)},\,\cpp{\phi Q}{(-)} = \Cpp{\phi Q}{(1)}-\Cpp{\phi Q}{(3)},
    \end{split}
\end{align}
with the corresponding operator relations given in Tab.~\ref{tab:relevant_operators}, where the original Warsaw and our top-specific basis operators are denoted by $\QQ$ and $\OO$, respectively. Since different, global fits have worked in terms of one these two conventions, we quote limits for both cases separately. Although the two bases are trivially related, the different analyses include different datasets and therefore obtain different confidence intervals.
The first three entries of Tab.~\ref{tab:relevant_operators} are of the bosonic type, in that they do not contain any fermion fields. They are primarily constrained by non-top data, such as EW precision data and diboson production. Interestingly, the typical assumption that these are much better constrained than the remaining top quark operators does not necessarily hold any longer, particularly at marginalised-level. Nevertheless, we expect improvements in constraints in this sector to continue to be driven by non-top data and therefore restrict our analysis to the other operators:
\begin{align}
\label{eq:top_ops}
    \Opp{\phi Q}{(3)},\,\Opp{\phi Q}{(-)},\,
    \Op{\phi t},\,\Op{tW},\,\Op{tZ},\,\Op{tG}.
\end{align}

\subsection{The $\bWtZ$ amplitude \label{sec:bwtz}}
In this section, we summarise the discussion of Ref.~\cite{Maltoni:2019aot} relating new interactions in the top/EW sectors and energy growth in $\bWtZ$ scattering. Dimension-6 operator contributions to $2\to2$ scatterings can have, maximally, a quadratic dependence on the energy, $E$.
Inspection of the interaction vertices of Fig.~\ref{fig:feyn_eft} 
indicates the types of modified interactions that will affect the amplitude. Considering modifications of the existing SM interactions first, we see that all possible neutral and charged top and bottom gauge interactions appear ($t\bar{t}Z$, $b\bar{b}Z$ and $tbW$), as well as the $W^+W^-Z$ triple-gauge coupling. Note that in the high-energy (massless) limit, it is the gauge couplings of the left handed top and bottom quarks that are relevant. 

Energy growth induced by modifications of these types of vertices can be understood to originate from the violation of delicate unitarity cancellations in the SM that relate particle masses and their EW interactions. By this logic, one can rule out a shifted $tbW$ vertex as a possible source of energy growth, since it appears in every diagram, meaning that it can only lead to a global re-scaling of the amplitude. The remaining interactions, on the other hand, do participate in said unitarity cancellations. The leading energy growth is found to be in the left-handed, fully longitudinal helicity configuration, $(\lambda_{\sss b},\lambda_{\sss W},\lambda_{\sss t},\lambda_{\sss Z})=(-,0,-,0)$. It grows as $E^2$, and is proportional to 
\begin{equation}
    \label{eq:bwtz_leading}
    \mathcal{A}(-,0,-,0)\propto\sqrt{s(s+t)}\,(\gzbl - \gztl + \gwz) \,,
\end{equation} 
where $\gzbl$, $\gztl$ and $\gwz$ denote generic couplings for the left-handed $b\bar{b}Z$, $t\bar{t}Z$ and the triple-gauge interactions, respectively, and $s,t$ are the usual Mandelstam variables. Assigning SM values to each coupling results in an exact cancellation, underlining the fact that all three are related by non-Abelian gauge-invariance of the theory. Several sub-leading sources of energy growth are also present, including:
\begin{align}
        \label{eq:bwtz_subleading}
        \begin{split}
        \mathcal{A}(-,-,-,0)= \mathcal{A}(-,0,-,-)\propto\sqrt{-t}\,(\gzbl - \gztl + \gwz)\,,\\
        \mathcal{A}(-,0,+,0)\propto\sqrt{-t}\,(2m_W^2(\gzbl - \gztr + \gwz) - \gwz m_Z^2)\,.
        \end{split}
\end{align}
The first corresponds to a change from longitudinal to transverse polarisation of the $W$ or $Z$ boson with respect to Eqn.~\ref{eq:bwtz_leading} and carries the same cancellation structure, differing only by a factor of $\sim m_{\sss W,Z}/\sqrt{s}$ associated to the helicity flip. The second has a rather non-trivial cancellation structure between gauge charges of two different fermion representations, gauge boson self-interactions and the gauge boson masses. The specific couplings involved along with the appearance of gauge boson masses strongly suggest a connection to the EW symmetry breaking mechanism, even though the Higgs does not explicitly participate in the scattering.

Tab.~\ref{tab:bwtz_hel} summarises the contributions to this scattering, focusing on the top-specific operators of Eqn.~\eqref{eq:top_ops}. The effective theory generalises beyond the `anomalous couplings' description above and one can identify the SMEFT origin of the three example amplitudes as coming from various combinations of the three current operators, $\Opp{\phi Q}{(3)}$, $\Opp{\phi Q}{(-)}$ and $\Op{\phi t}$. The maximal growth from $\Opp{\phi Q}{(3)}$ can be identified with a dimension-6 contact-term, in Feynman gauge,  between the charged fermion and Goldstone boson  currents,
\begin{align}
    \Opp{\phi Q}{(3)} \supset (\bar{t}_{\sss L}\gamma^\mu b_{\sss L})( G^{\sss 0}\lra{\partial}_\mu G^+) + \mathrm{h.c.}
\end{align}
In the high-energy limit, the Goldstone equivalence theorem~\cite{Cornwall:1974km} identifies amplitudes involving the Goldstone bosons with those of the longitudinal gauge degrees of freedom, which explains the $E^2$ dependence induced by this operator in the fully longitudinal configuration.
The SMEFT also admits new Lorentz structures not present in the SM, including the weak dipole interactions relevant for this process. They predict maximal energy growth in the mixed transverse-longitudinal configurations which, like Yukawa couplings, flip fermion helicity. $\Op{tW}$ and $\Op{tZ}$ specifically couple the right-handed top and the left-handed bottom. Several sub-leading contributions arise from helicity-flips of the external states, suppressed by a corresponding power of $\sim m/\sqrt{s}$ (except for the $b$-quark, which we consider to be massless).

The schematic high-energy behaviour of the corresponding SM helicity amplitudes are also given in Tab.~\ref{tab:bwtz_hel}, from which we can infer the likely behaviour of the associated SMEFT interference term. The SM has leading ($\propto$ constant) energy growth in three configurations. The fully-longitudinal, left-handed one coincides with the leading energy growth from $\Opp{\phi Q}{(3)}$, such that we expect an energy-growing interference term. All other sources of energy growth in the SMEFT have a SM counterpart with an opposite energy dependence, such that the interference term is not expected to display significant growth. This is expected from the so-called ``non-interference'' theorems from helicity selection rules in $2\to2$ scattering at dimension-6, involving at least one transverse gauge boson~\cite{Azatov:2016sqh}. The cross-section contributions of all operators that display growth should be dominated by the $|\mathcal{A}_{\sss EFT}|^2$ component in the very high-energy limit. At intermediate energies, there will be an interplay between the various contributions that we will try to uncover in the Sec.~\ref{sec:hel_xs}.

\subsection{Advantages of $tWZ$ \label{sec:twz_advantages}}
The $tWZ$ process has several interesting advantages when considered as an indirect probe for new physics in top quark interactions. Firstly, as shown in the previous sections, it is sensitive to unitarity-violating behaviour in $\bWtZ$. This underlines its importance with respect to $t\bar{t}Z$, which proceeds dominantly via the QCD-induced channel and mainly receives re-scalings of its total rate from these operators. EW-induced $t\bar{t}Z/W$ production do embed interesting EW top scatterings. However, they are not only dwarfed by their QCD-induced counterparts but also, at LO, feature an off-shell EW state in the $s$-channel that suppresses any interesting energy growth\footnote{ We note that part of the NLO EW corrections to $\ttbar X$ involve the so-called $tX$ scattering diagrams, which do not suffer the aforementioned $s$-channel suppression~\cite{Frederix:2017wme,Frederix:2020jzp}. These also serve as good candidates to study top EW interactions~\cite{Dror:2015nkp,Maltoni:2019aot}.}. Moreover, these processes can also be modified by the large number of four fermion operators that mediate top pair production, further diluting sensitivity to EW top interactions. 

$tWZ$ also compares favourably to related EW processes that also embed $\bWtZ$, namely $tWj$ and $tZj$. While $tWj$ embeds the $\bWtZ$ amplitude, this final state corresponds to a real-radiation contribution to $tW$ production at NLO. This QCD-induced process does not embed $\bWtZ$ and therefore would likely reduce the sensitivity to several EW top couplings. Furthermore, even at LO, $tWj$ only differs from $t\bar{t}$ production and decay by a single $b$-tag, which may pose additional challenges in reconstructing the final state. This is in contrast to $tWZ$, where only NLO corrections lead to an overlap with $t\bar{t}(Z)$ and also, weakly, $\ttbar$. It was found in Ref.~\cite{Maltoni:2019aot}, that the $tZj$ process showed significant promise in probing EW top couplings, particularly given that it has recently been measured for the first time at the LHC. However, some of the expected energy growth at interference-level from $\Opp{\phi Q}{(3)}$ was not observed, due to phase-space cancellations. An initial study of the $tWZ$ process found that the expected energy growth was recovered in this channel. Finally, $tWZ$ is not affected by any top quark four fermion operators at tree-level, contrary to all of the other processes discussed so far. We therefore believe that it constitutes a worthy and complementary probe of new top quark interactions, and therefore strongly motivates the dedicated study presented here.

\subsection{Embedding of $\bWtZ$ amplitudes into $tWZ$ \label{sec:hel_xs}}
As a precursor to our main study, we begin by exploring the embedding of our $\bWtZ$ helicity amplitudes into $tWZ$ in the high-energy limit. By comparing the impacts on the full process, with those expected from our helicity amplitude analysis, we can quantify the extent to which the expected high-energy behaviour is retained in realistic collider observables and can therefore genuinely help us improve our understanding of top EW interactions. We simulate the contribution from each operator in Eqn.~\eqref{eq:top_ops} with \texttt{MadGraph5\_aMC@NLO}, using the \texttt{SMEFTatNLO} model~\cite{Degrande:2020evl}. Exploiting the \texttt{MadGraph5\_aMC@NLO} option for handling the polarisation of initial and final states~\cite{BuarqueFranzosi:2019boy}, we split the cross section by $W$ and $Z$ boson helicity, where `T' and `0' denote transverse ($\lambda_i=\pm$) and longitudinal ($\lambda_i=0$) polarisations, respectively,  and appropriately summing/averaging over all others. The polarisations are defined from projections of the spin component along the particle momentum in the centre-of-mass frame of the partonic collision, not the lab-frame. 

\begin{figure}[h!]
    \centering
    \includegraphics[width=.45\textwidth]{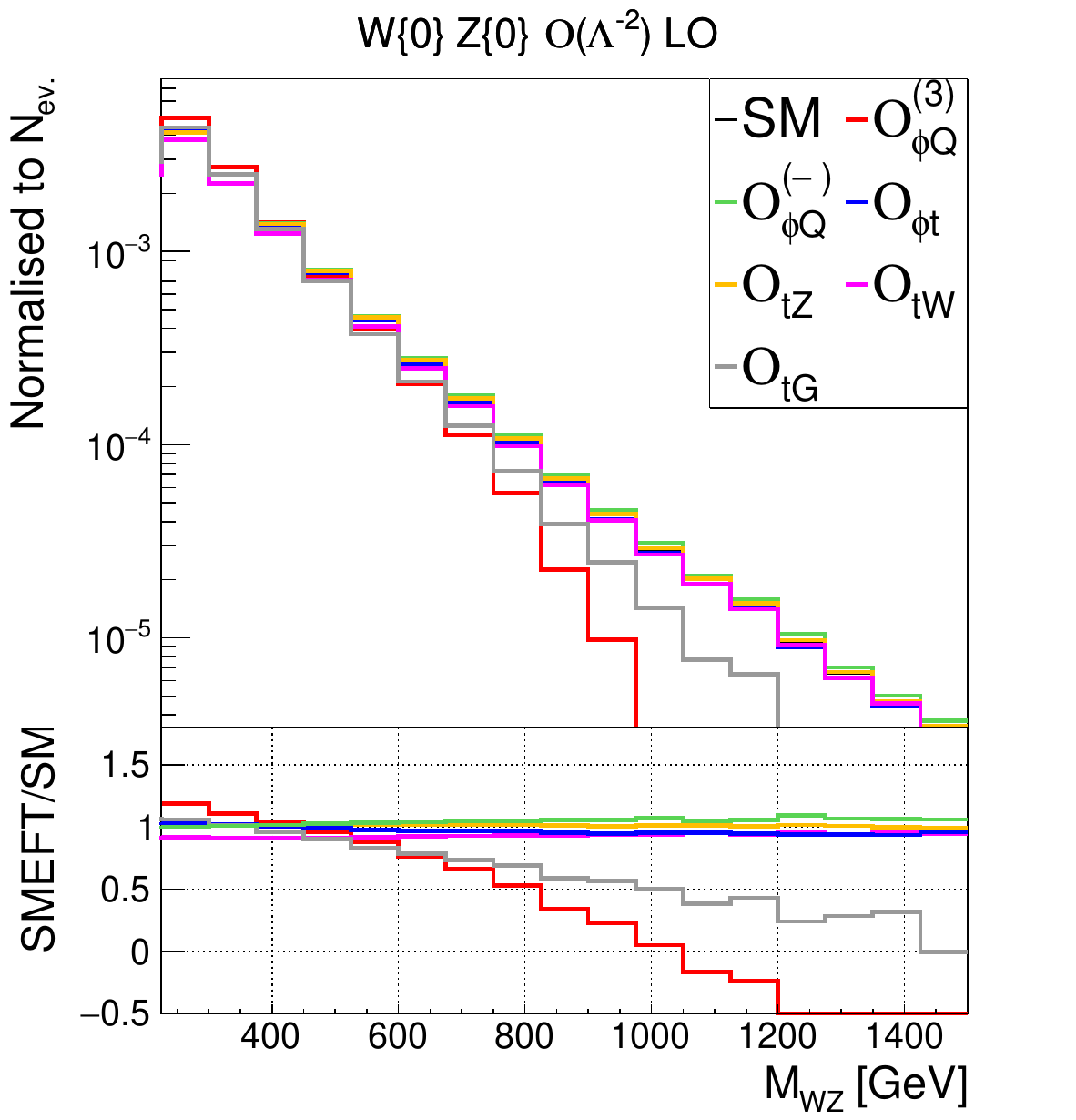}
    \includegraphics[width=.45\textwidth]{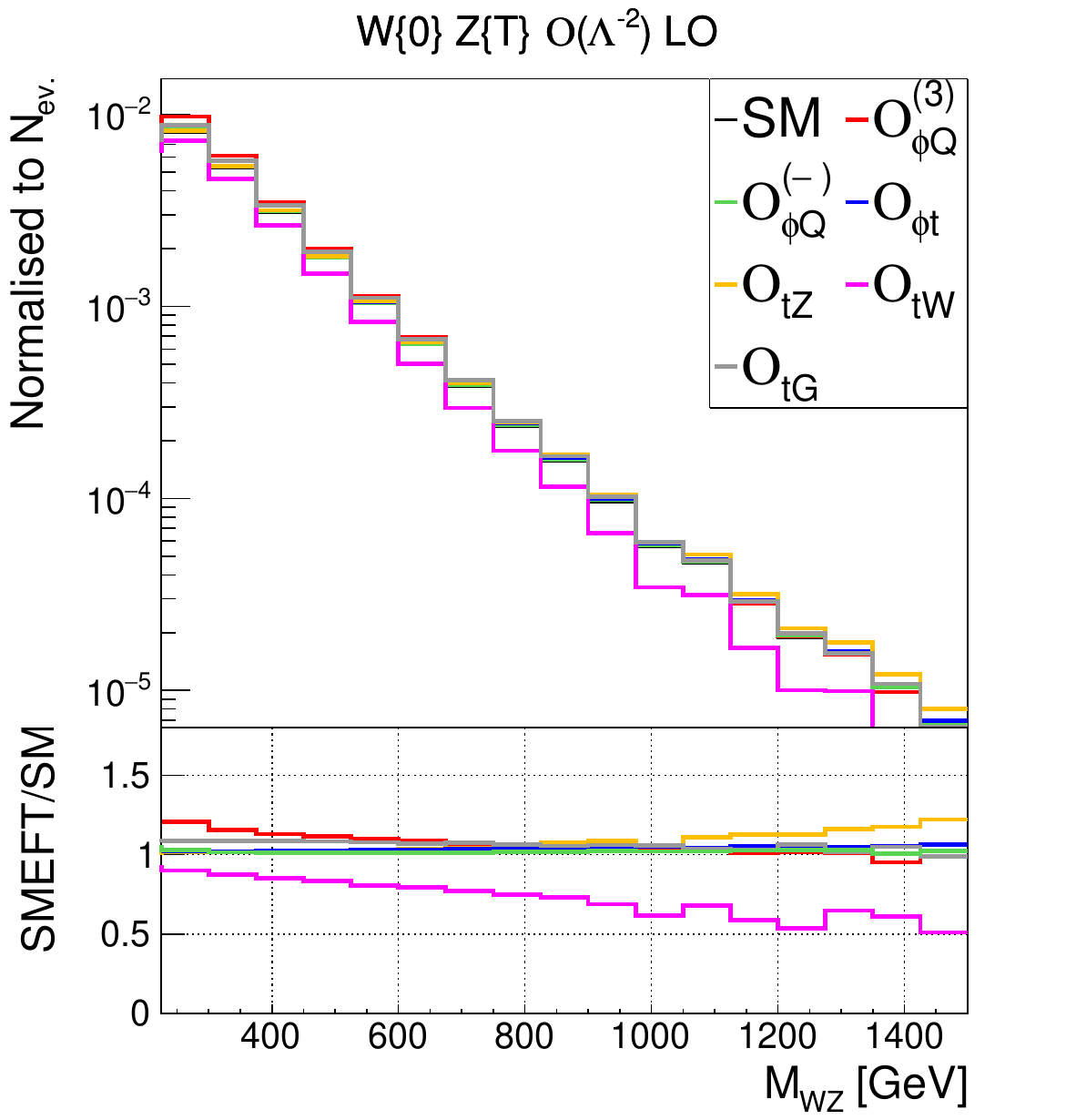}\\
    \includegraphics[width=.45\textwidth]{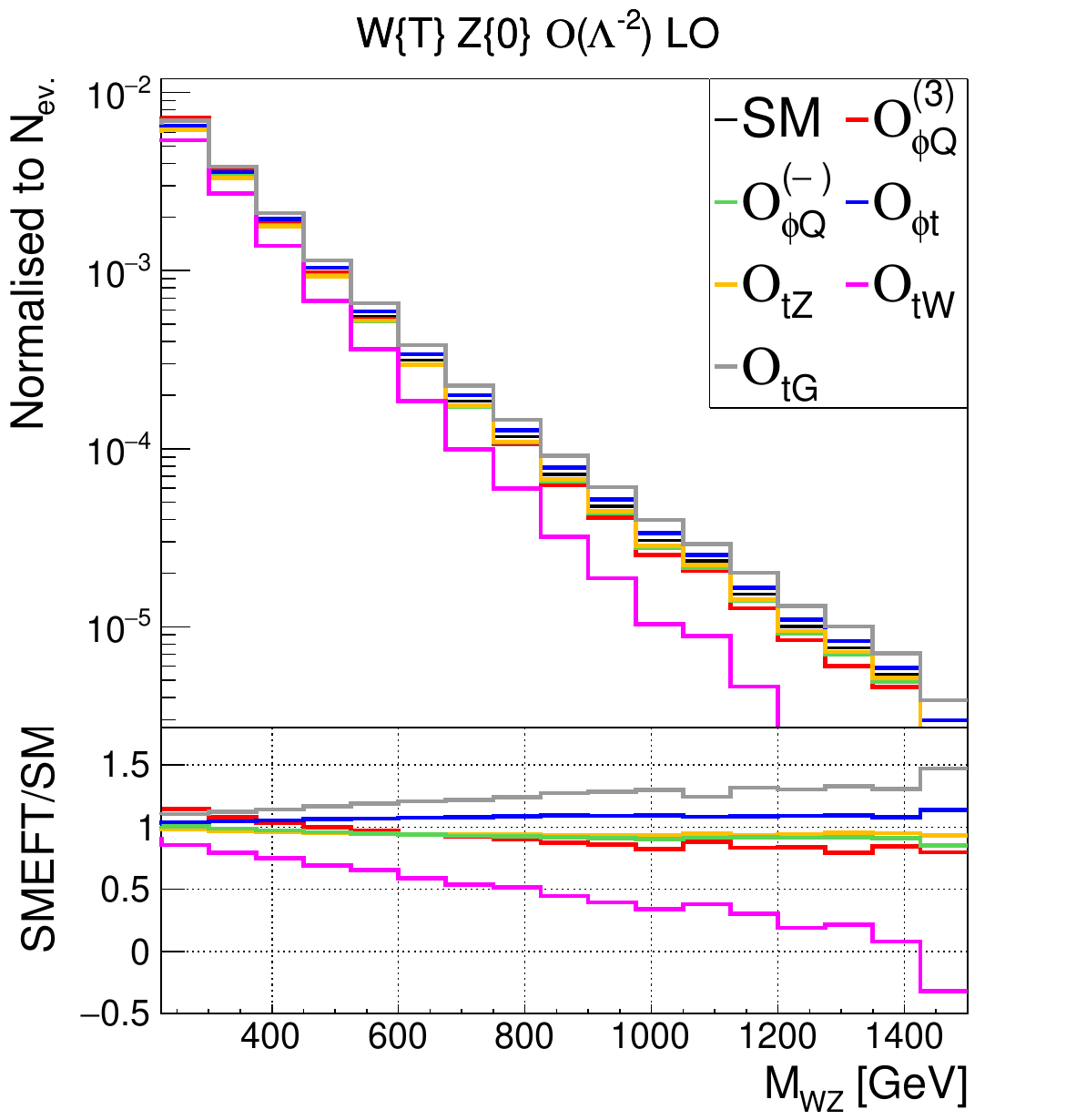}
    \includegraphics[width=.45\textwidth]{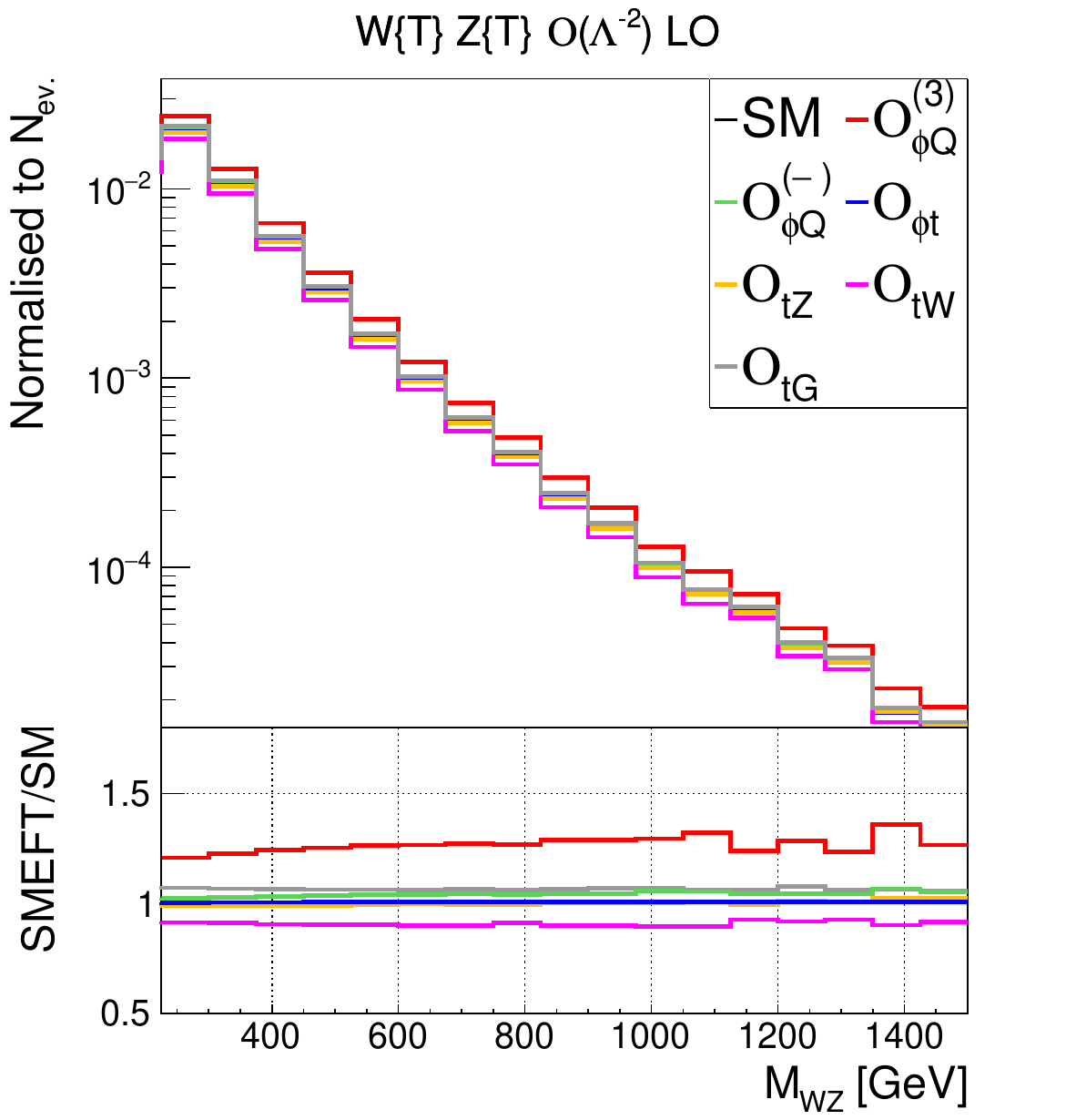}
    \caption{\label{fig:hel_int} Leading order SM and SMEFT contributions to the differential cross-section of the $tWZ$ process in the $WZ$ invariant mass for different $W$ and $Z$ helicity configurations. SMEFT predictions are given for $c_i/\Lambda^2=$ 1 TeV$^{-2}$, and truncated at interference-level ($\mathcal{O}(\Lambda^{-2})$). The curly brackets indicate the helicity eigenstates following the \texttt{MadGraph5\_aMC@NLO} notation where \texttt{\{0\}} and \texttt{\{T\}} refer to longitudinal and transverse polarisations, respectively. The inset shows the ratio of the SM+SMEFT contribution to the SM alone.}
\end{figure}

Fig.~\ref{fig:hel_int} shows the differential cross sections in the invariant mass of the $WZ$ system, $\mwz$, used here as a proxy for the $\bWtZ$ sub-amplitude scattering energy, $\sqrt{s}$. SMEFT contributions are truncated at interference-level ($\mathcal{O}(\Lambda^{-2})$), with $c_i/\Lambda^2=1$ TeV$^{-2}$.
Starting with the fully longitudinal configuration (upper left panel), we can clearly see the expected high-energy enhancement of the $\Opp{\phi Q}{(3)}$ contribution, especially in the inset that shows the ratio of each operator interference term to the SM. Interestingly, at low energies the main effect of this operator results in a net positive shift with respect to the SM, while the energy growing term has the opposite sign, resulting in a sign change of the new-physics contribution around $\mwz=500$ GeV. The relative impact reaches 100\% at $\mwz=1$ TeV for this value of the coefficient, beyond which point the $\mathcal{O}(\Lambda^{-2})$ truncated prediction is clearly not physical, since it predicts a negative cross section. There is also an important, energy-enhanced effect from $\Op{tG}$, which can be understood from the fact that, at LO, large $\mwz$ involves a recoil against a high-$p_T$ top quark, which will be enhanced by the derivative interaction of the chromomagnetic operator. $\Op{tW}$ appears to induce a relatively mild effect that is enhanced at low energies. Following the expectations from Tab.~\ref{tab:bwtz_hel}, all other operator contributions do not induce energy-growing interference effects in this configuration.

In fact, taken at face-value, Tab.~\ref{tab:bwtz_hel} implies that we should not observe any further energy-growing interference effects in any other configuration. Instead, we observe that some of the ``non-interfering'' operators are still able to yield something, suggesting that finite-mass effects remain important in the energy range we have studied. In both mixed longitudinal-transverse configurations (upper right and lower left panels), energy-growth is evident for $\Op{tW}$ and also mildly for $\Op{tZ}$. $\Op{tG}$ exhibits growth for $\lambda_W,\lambda_Z=\pm,0$ but, strangely, not in the other mixed configuration. Similarly, $\Opp{\phi Q}{(3)}$ effects persist, albeit at a milder level and we see a small effect from $\Op{\phi t}$. The fully transverse configurations (lower right panel) do not show any evidence of enhanced interference effects but do receive energy-constant contributions from $\Opp{\phi Q}{(3)}$, $\Op{tW}$ and $\Op{tG}$. 

\begin{figure}[h!]
    \centering
    \includegraphics[width=.45\textwidth]{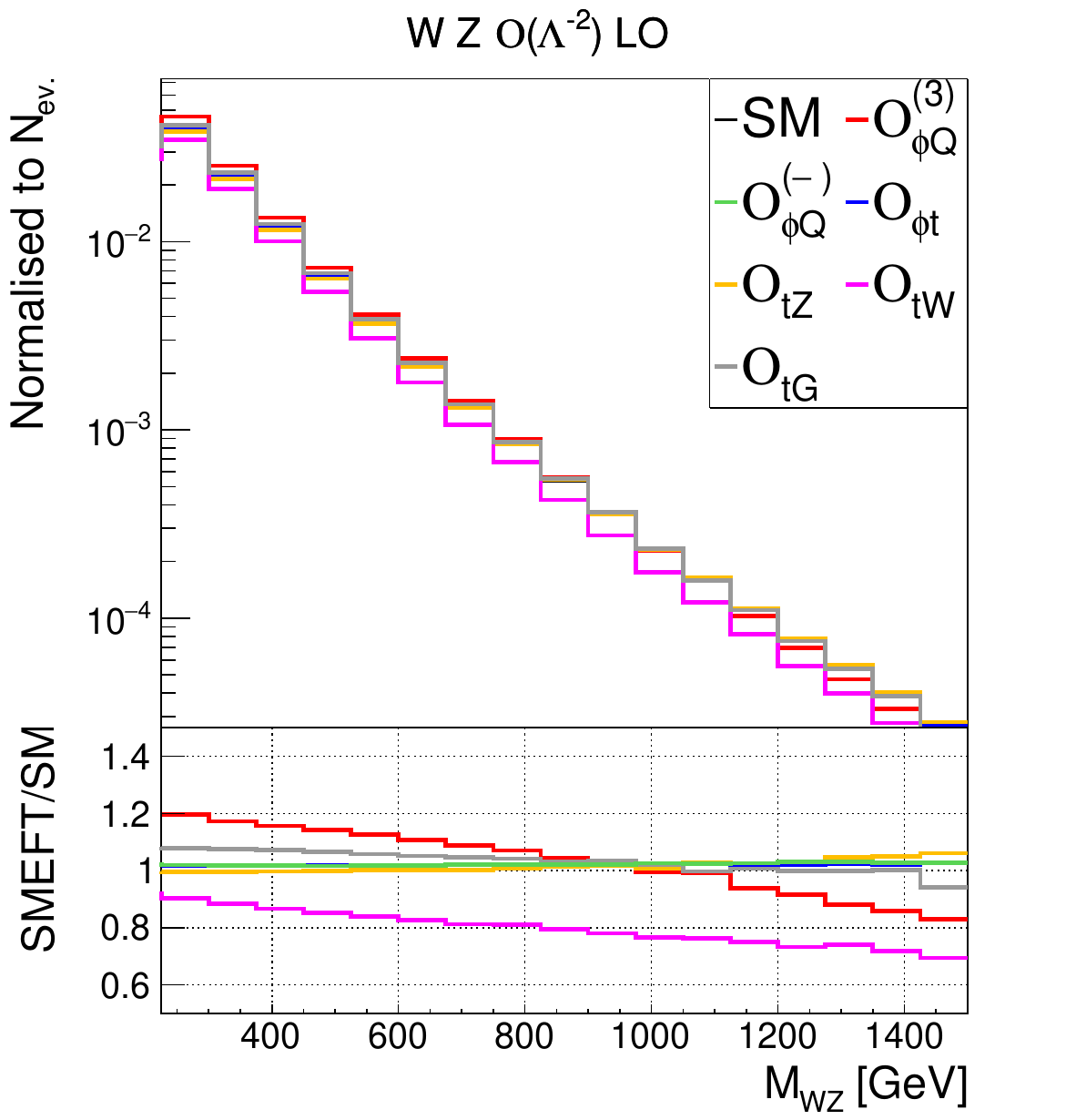}
    \includegraphics[width=.45\textwidth]{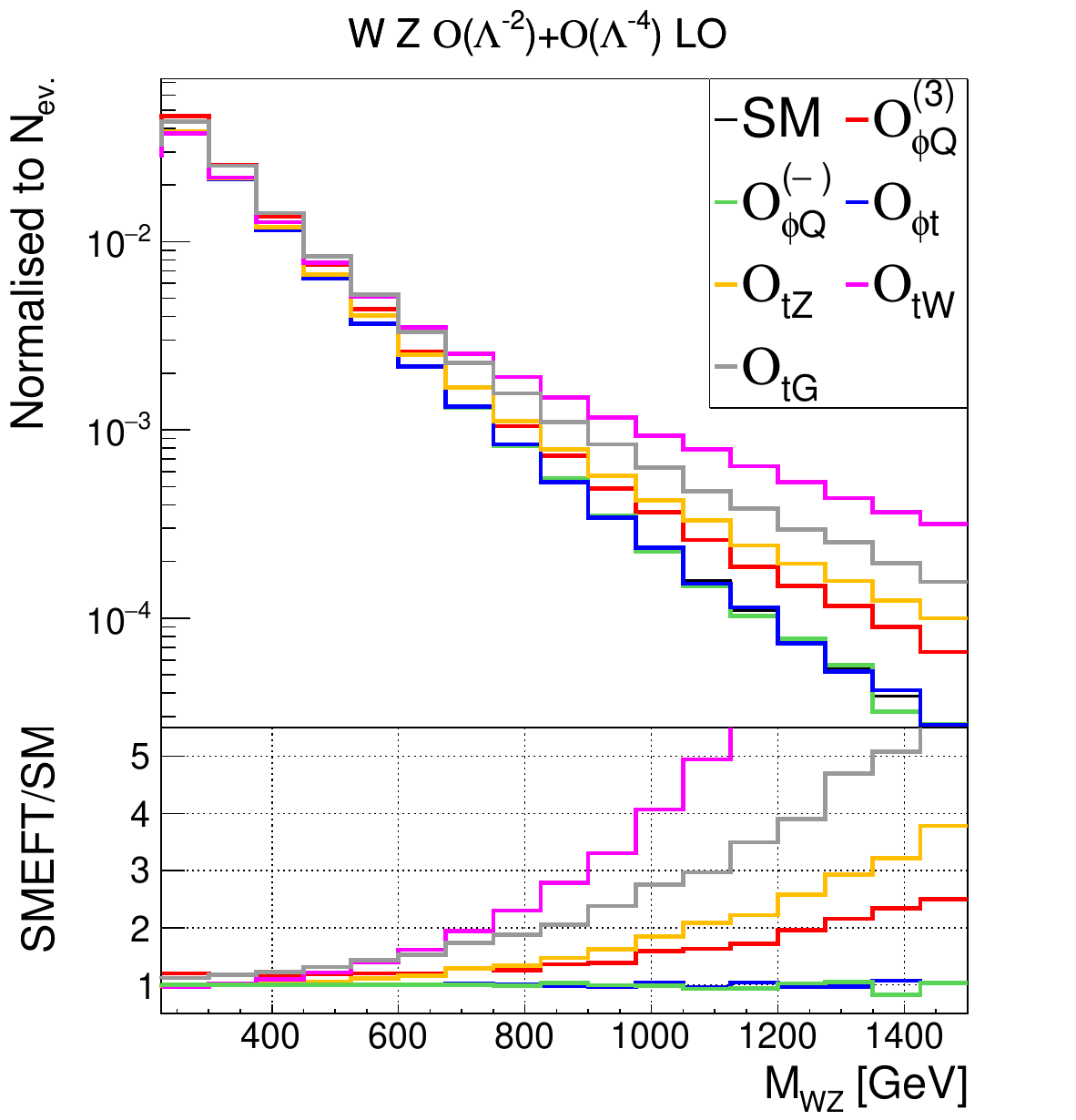}
    \caption{\label{fig:hel_sum} Same as Figs.~\ref{fig:hel_int} (\emph{left}) and ~\ref{fig:hel_full} (\emph{right}), but showing contributions to the total, unpolarised cross-section.}
\end{figure}

It is interesting to note the hierarchy in SM rates among the different configurations, the fully-transverse being the largest (56.20 $fb$), followed by the mixed transverse-longitudinal (47.64 $fb$) and finally the fully-longitudinal (12.57 $fb$). The sensitivity of measurements that are blind to the gauge boson polarisations will be weighted by the relative size of each configuration, diluting the impact of the rarer modes. This is evident in the left panel of Fig.~\ref{fig:hel_sum}, which shows the interference effects of each operator on the total, unpolarised differential cross section. $\Opp{\phi Q}{(3)}$ and $\Op{tW}$ have a smaller relative impact and the crossing point of the former, from a net positive to negative correction, is shifted to higher $\mwz$ by the positive and constant effect in the fully-transverse configuration. The contributions from all other operators are quite suppressed for this point in parameter space, being diluted by their lack of effect on the fully-transverse configuration. The right panel of Fig.~\ref{fig:hel_sum}, includes the quadratic, $\mathcal{O}(\Lambda^{-4})$ terms on the unpolarised cross section. For this point in parameter space ($c_i/\Lambda^2=1$ TeV$^{-2}$), the quadratic terms start to dominate around $\mwz=300-400$ GeV, yielding significant energy growth for all operators except for $\Opp{\phi Q}{(-)}$ and $\Op{\phi t}$. One can therefore anticipate that the latter two operators are not likely to be significantly constrained by high-energy measurements in $tWZ$. The polarised breakdown of the full SMEFT contributions are shown in Fig.~\ref{fig:hel_full} in Appendix~\ref{app:bwtz}. We note that, ultimately, the relative importance of quadratic terms inherently depends on both the energy scales probed and the constraining power of the data. It should not be used as a measure of the EFT validity until a global fit is performed.

\section{SMEFT Technical setup and results}
\label{sec:results}
Having established a technical procedure and physical phase-space region in which consistent, precise $tWZ$ predictions can be generated in the SM, and followed with a detailed study of the anatomy of this process in the SMEFT, we are ready to combine the two to produce the main results of this work. In this section, we study the $tWZ$ process in the SMEFT using high-energy observables, e.g. $\mwz$, that maximise sensitivity to the SMEFT by enhancing unitarity violating effects. As discussed in Sec.~\ref{sec:def}, in the high-energy phase-space regions, the contribution of the $\ttbar$ overlap becomes non-resonant, and thus the contributions should not be removed. 

Therefore, once we enter the SMEFT sensitive phase-space, only the $\ttZ$ overlap is of concern. 
In presenting results for the SMEFT, we also make the simplifying assumption of keeping the $Z$ boson stable. On one side, this maps better to the discussion on unitarity violating behaviour in $\bWtZ$, and on the other, the approximation was found to be quite good in our study of the SM process, especially when the $\ttbar$ overlap is not relevant. Overall we consider it an unnecessary practical complication.

As with our SM results, the $tWZ$ process is simulated via \texttt{MadGraph5\_aMC@NLO}~\cite{Alwall:2014hca, Frederix:2018nkq} at NLO-QCD accuracy in the SMEFT using \texttt{SMEFTatNLO}. In contrast to the SM results, we used the recommended fixed factorisation, renormalization, and EFT renormalization scales $\mu_F$, $\mu_R$, and \verb|mueft| of 172 GeV$\sim(\mt+\mw+\mz)/2$, since \texttt{MadGraph5\_aMC@NLO} does not evolve operator coefficients. The proton PDFs and their uncertainties are evaluated employing reference sets and error replicas from the \texttt{NNPDF3.1 NLO}  global fit~\cite{NNPDF:2017mvq} in the 5FS, in which the bottom quark is taken to be massless. We do not make any parton-level cuts. The $b$-veto as a method of suppressing the overlap of resonant contributions in this process has proved effective for the SM and we also confirm that it continues to work for the SMEFT. We will now mainly show DR2 results in presenting our predictions and discussing the impact of SMEFT operators. All of the operators also contribute in principle to the overlapping processes, and we take this into account in our simulation, although in practice the $b$-veto should suppress these effects. One should nevertheless take this into account when considering the overlapping processes as backgrounds in a SMEFT interpretation. Finally, some of the operators can also modify the top quark width. Our predictions, however, do not model this effect, since we assume the top to be stable, except in the resonant diagrams where, when DR/DS is applied, the width only acts as a regulator. In these cases its precise value does not matter as long as it is negligible with respect to all other relevant scales since the resonant squared amplitude is never evaluated.

\subsection{Fixed Order (FO) predictions}
\begin{figure}[h!]
    \centering 
    \includegraphics[width=.80\textwidth]{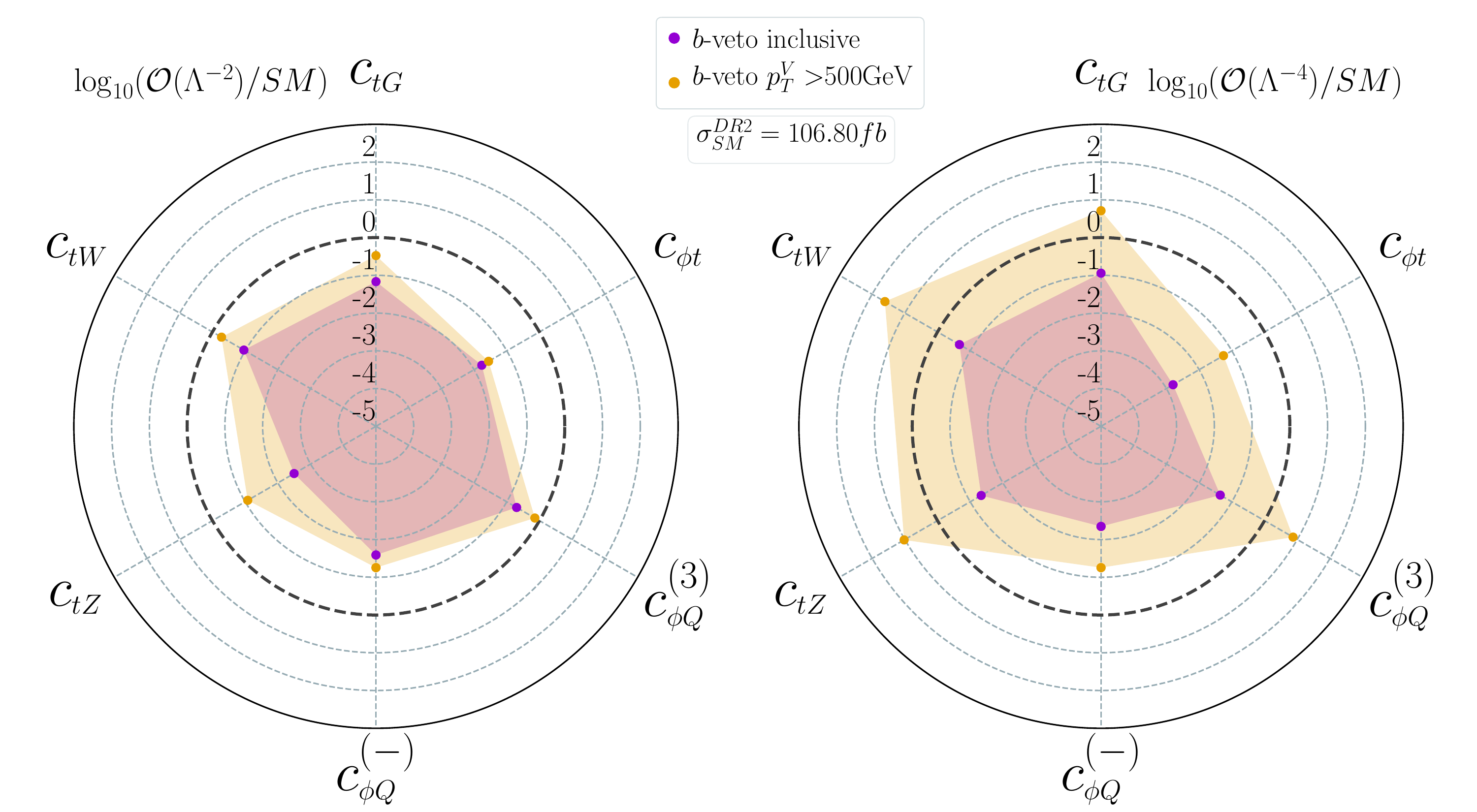}
    \caption{\label{fig:radar_NLO} 
    Relative impact of the linear (\emph{left}) and quadratic (\emph{right}) SMEFT contributions to the $tWZ$ process at NLO accuracy in QCD, applying our DR2 overlap removal and a veto on central ($|\eta|<2.5$) or hard ($p_T>30$ GeV) $b$-jets. They are obtained by dividing the cross section contributions over the inclusive (magenta) and high-energy (orange) phase-space regions by the corresponding SM prediction. The high-energy phase-space region is defined by requiring both the $W$ and $Z$ boson $p_T$ to be greater than $500$ GeV.}
\end{figure}

Before showing fully differential results, in Tab.~\ref{tab:sm_results_tab} we present the SM inclusive cross section predictions as well as cross section predictions in a high-energy region of the phase-space at LO and NLO in QCD. The high-energy region is defined by requiring the $p_T$ of the $W$ and $Z$ bosons to be above 500 GeV. For SMEFT contributions, Tab.~\ref{tab:merged_smeft_table} summarises the inclusive cross sections of the linear ($\mathcal{O}(\Lambda^{-2})$) and quadratic contributions ($\mathcal{O}(\Lambda^{-4})$) for each of the 6 relevant top quark operators at LO and NLO. The numbers assume $c_i/\Lambda^2=1$ TeV$^{-2}$. The impacts of the SMEFT operators on the inclusive cross section and in the high-energy region, following the presentation of Ref.~\cite{Maltoni:2019aot}, are presented in Tab.~\ref{tab:smeft_impact}. 
\begin{table}[ht!]
\input{sm_results_tab}
\caption{\label{tab:sm_results_tab}The SM contributions [fb] to the inclusive and the high-energy $tWZ$ production at LO and NLO, including QCD scale uncertainties, for the LHC $\sqrt{s}=$ 13 TeV.
These results are for DR1 and DR2 predictions when applying a veto on $b$-quarks with $|\eta|<2.5$ or $p_T > 30$ GeV. The stability of the SM DR1 and DR2 cross sections signifies the efficiency of the $b$-veto.}
\end{table}
\begin{table}[h!]
\input{merged_tab_smeft}
\caption{\label{tab:merged_smeft_table}The SMEFT contributions [fb] to inclusive $tWZ$ production, at linear and quadratic levels, LO and NLO, including QCD scale uncertainties, for the LHC $\sqrt{s}=$ 13 TeV
and $\frac{c_i}{\Lambda^{2}}$ = 1 TeV$^{-2}$. These results are for DR1 and DR2 predictions when applying a veto on $b$-quarks with $|\eta|<2.5$ or $p_T > 30$ GeV. 
}
\end{table}
The radar charts shown in Fig.~\ref{fig:radar_NLO} depict the relative impact of each operator with respect to the corresponding SM prediction at NLO. Left and right panels show the effect of the interference term and squared term, respectively, for $c_i/\Lambda^2=1$ TeV$^{-2}$. Purple (orange) points correspond to the impact on the inclusive (high-energy) phase-space.
Numerical versions of the radar chart results are tabulated in Tab.~\ref{tab:smeft_impact}, which include the LO predictions as well as a ``$K$-factor'', defined as the ratio of the NLO and LO impacts. Since the NLO process is defined up to the diagram removal scheme and the $b$-veto implementation, we do not present any traditional $K$-factor values, since they depend on the former. 
\begin{table}[h!]
    \input{SMEFT_impacts}

    \caption{\label{tab:smeft_impact} 
    Relative SMEFT operator contributions (DR2 predictions) to inclusive and high-energy $tWZ$ production. The cross sections are normalised by the corresponding LO(NLO) SM predictions in the inclusive and high-energy regions shown in Tab.~\ref{tab:sm_results_tab}. Individual operator impacts are split into contributions from interference with the SM, $\mathcal{O}(\Lambda^{-2})$, and pure SMEFT amplitude squared, $\mathcal{O}(\Lambda^{-4})$. The NLO numbers are represented in Fig~\ref{fig:radar_NLO}, and $K$-factors are defined as the ratio between NLO and LO impacts.}
\end{table}
The LO results agree with those presented in Ref.~\cite{Maltoni:2019aot}, noting that, as discussed in Sec.~\ref{sec:SMEFT}, a slightly different operator convention is used. One can see the enhancement at high energies of the $\Opp{\phi Q}{(3)}$ interference term, as expected from Tab.~\ref{tab:bwtz_hel}. Further, energy growing interference behaviour is observed for the dipole operators, with the very large enhancement of $\Op{tZ}$ indicative of a cancellation at inclusive level being lifted by the cuts. Although it is naively unexpected from the helicity amplitude computation, interfering growth was observed for $\Op{tW}$ in the LO differential distribution shown in Fig.~\ref{fig:hel_sum}, and is confirmed here. Finally, the neutral current operators $\Opp{\phi Q}{(-)}$ and $\Op{\phi t}$ do not display significant energy-growing interference, in line with expectations. 

\begin{figure}[h!]
    \centering
    \includegraphics[trim=0.0cm 5.0cm 0.0cm 0.0cm, clip,width=.45\textwidth]{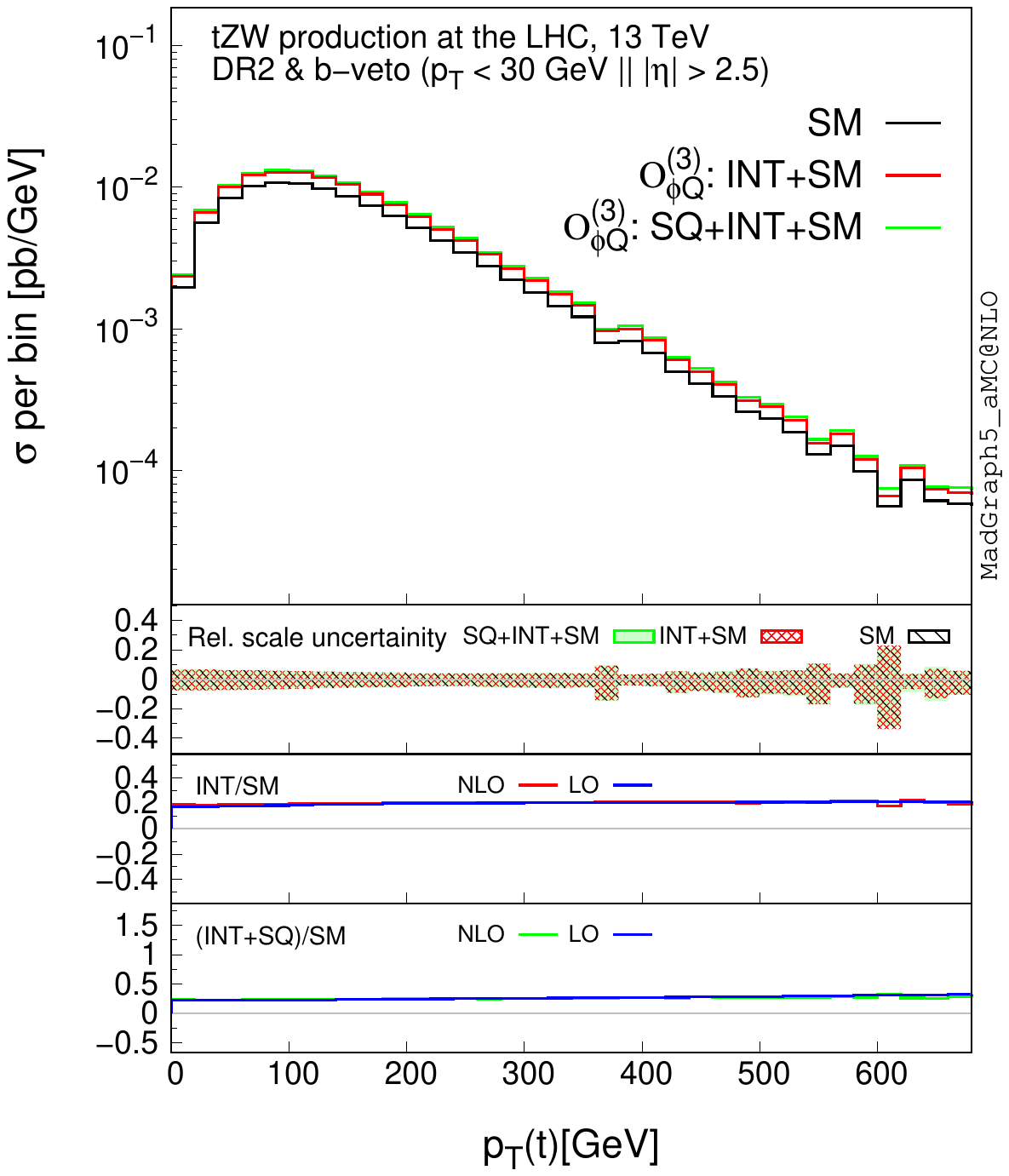}
    \includegraphics[trim=0.0cm 5.0cm 0.0cm 0.0cm, clip,width=.45\textwidth]{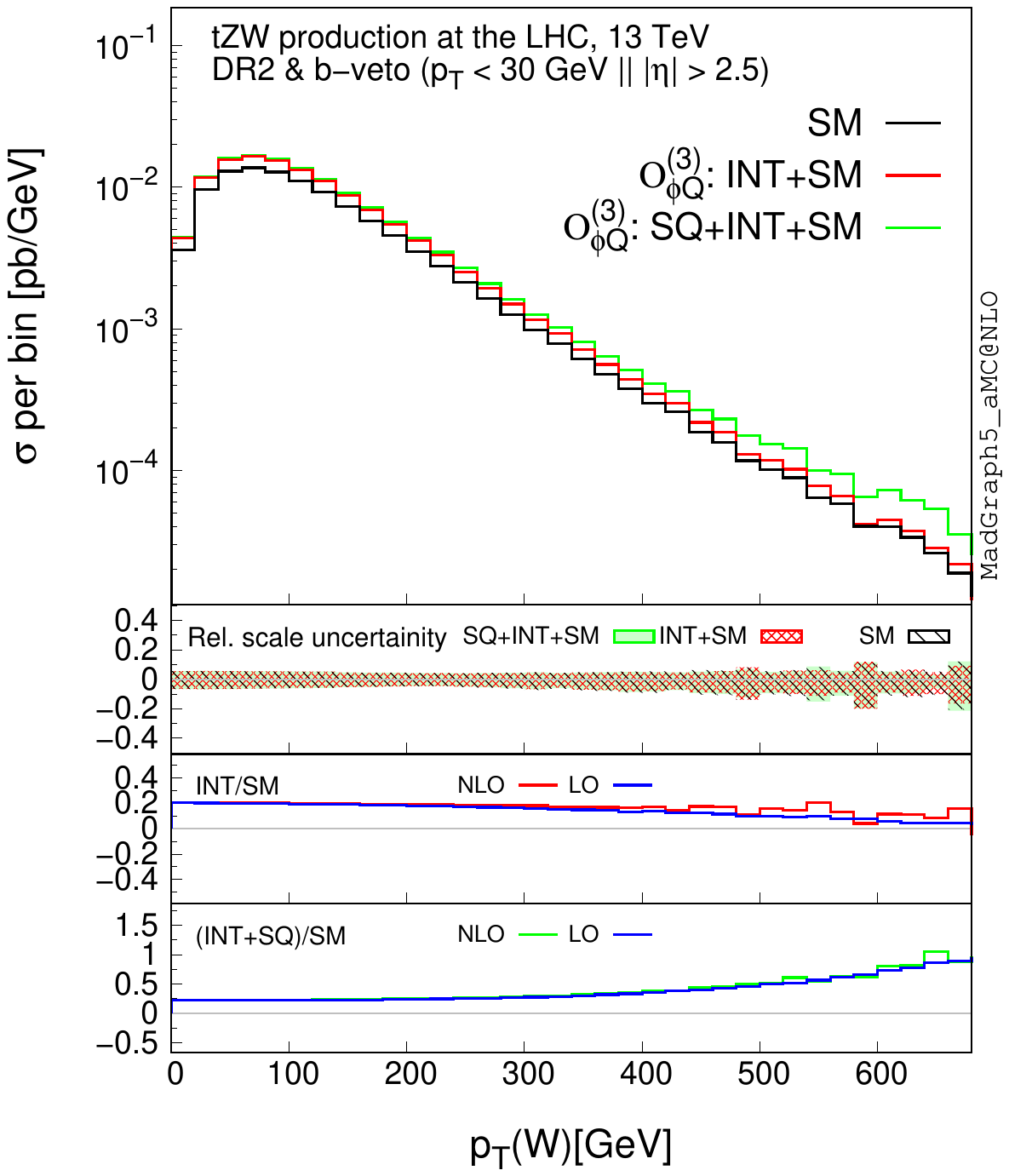}\\
    \includegraphics[trim=0.0cm 5.0cm 0.0cm 0.0cm, clip,width=.45\textwidth]{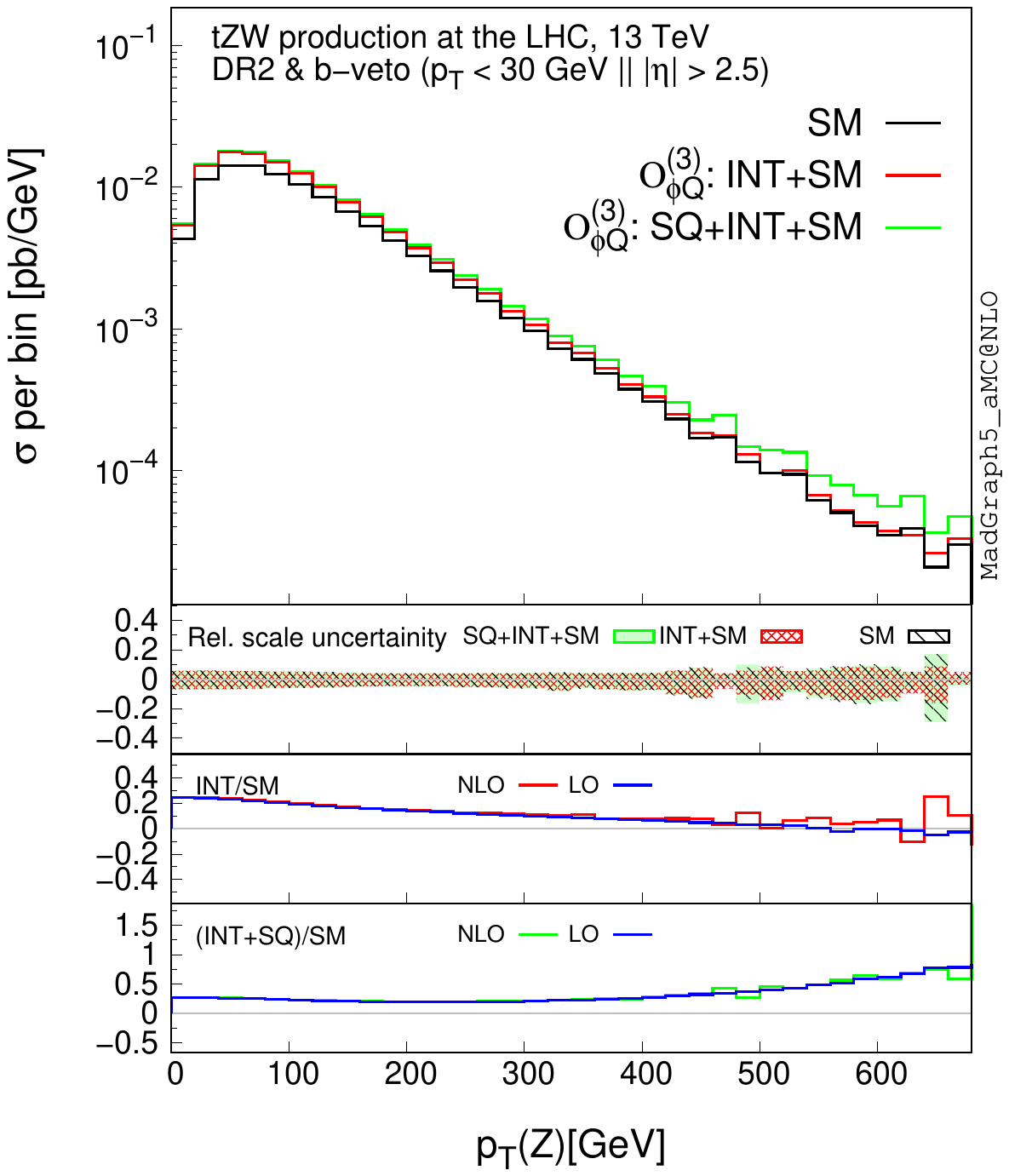}
    \includegraphics[trim=0.0cm 5.0cm 0.0cm 0.0cm, clip,width=.45\textwidth]{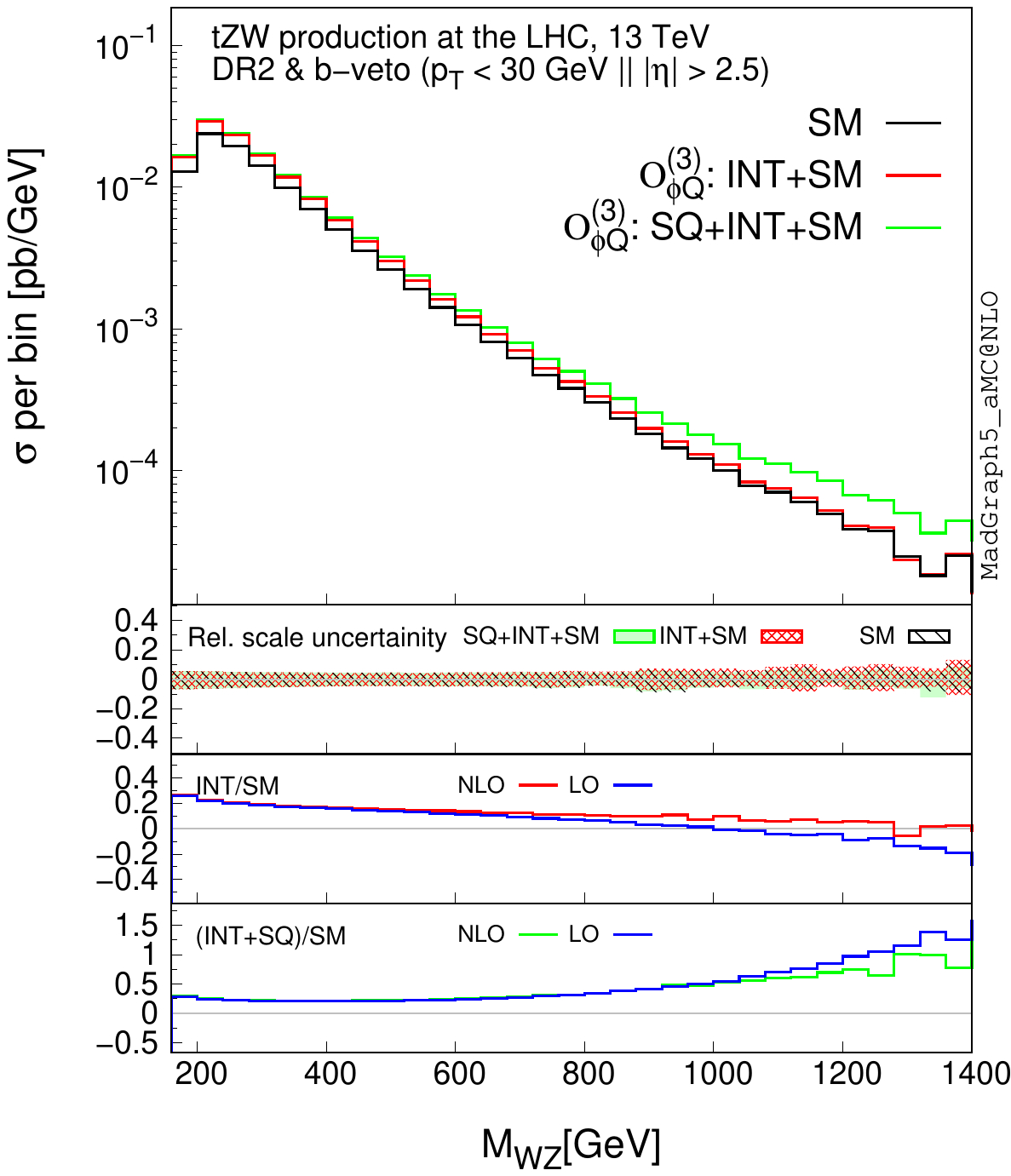}
    \caption{\label{fig:cpq3} Clockwise from top left: the transverse momentum of the (anti-)top quark, $p_T(t)$, of the $W^{\pm}$ boson, $p_T(W)$, the invariant mass of the $WZ$ pair, $M_{WZ}$, and the transverse momentum of the $Z$ boson, $p_T(Z)$ at DR2 for the $\Opp{\phi Q}{(3)}$ SMEFT operator when $t\bar{t}Z$ is removed and $b$-veto is imposed, at fixed order NLO accuracy. The legends correspond to different computations at different SMEFT contributions. The first inset shows the scale variations in the process, while the middle and the last insets show (at LO and at NLO) the interference and the full SMEFT contribution relative to the SM, respectively. The relative scale uncertainties are computed on the summed cross sections and not separately on the interference and the quadratic SMEFT contributions.}
\end{figure}

\begin{figure}[h!]
    \centering
    \includegraphics[trim=0.0cm 5.0cm 0.0cm 0.0cm, clip,width=.45\textwidth]{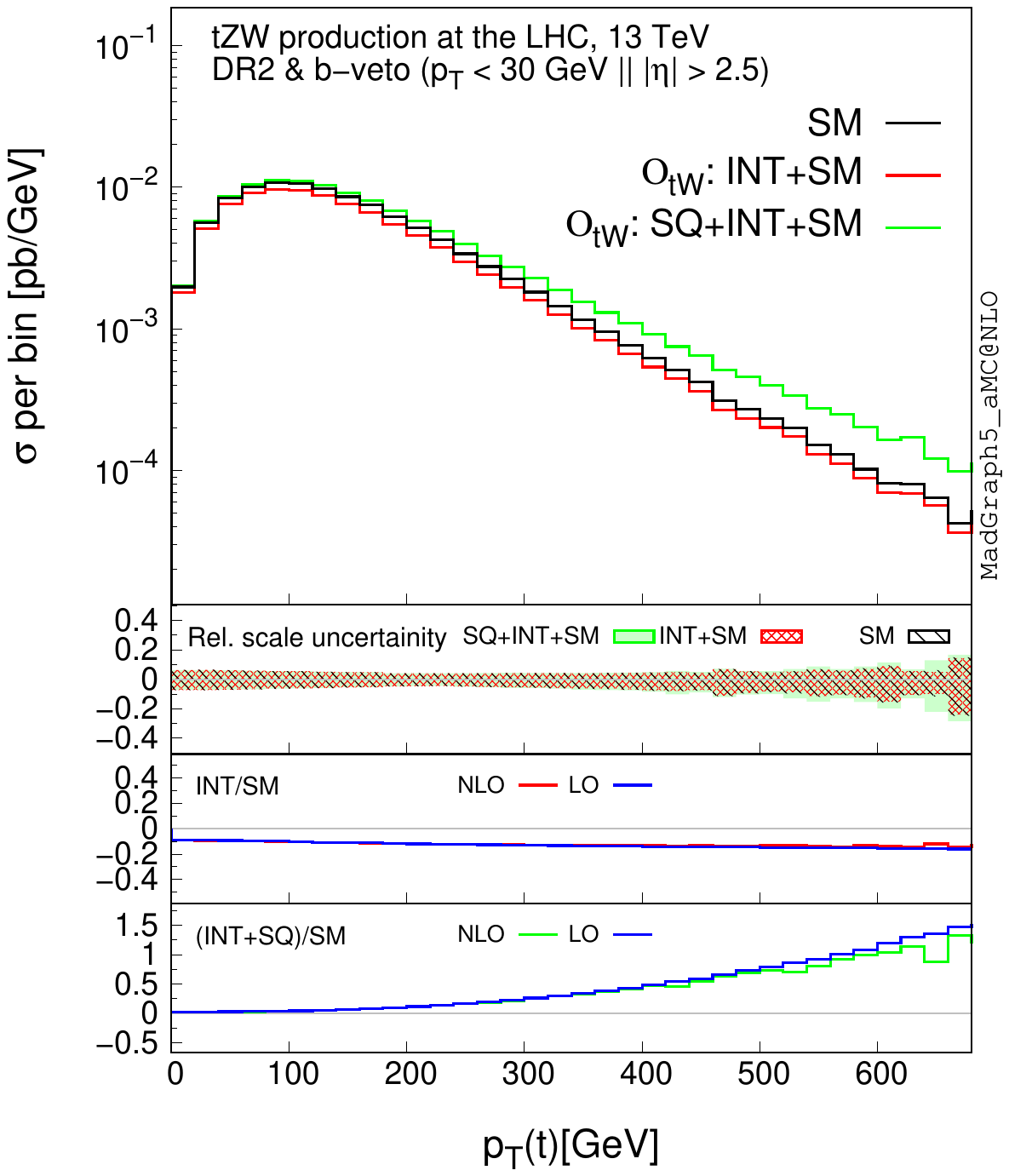}
    \includegraphics[trim=0.0cm 5.0cm 0.0cm 0.0cm, clip,width=.45\textwidth]{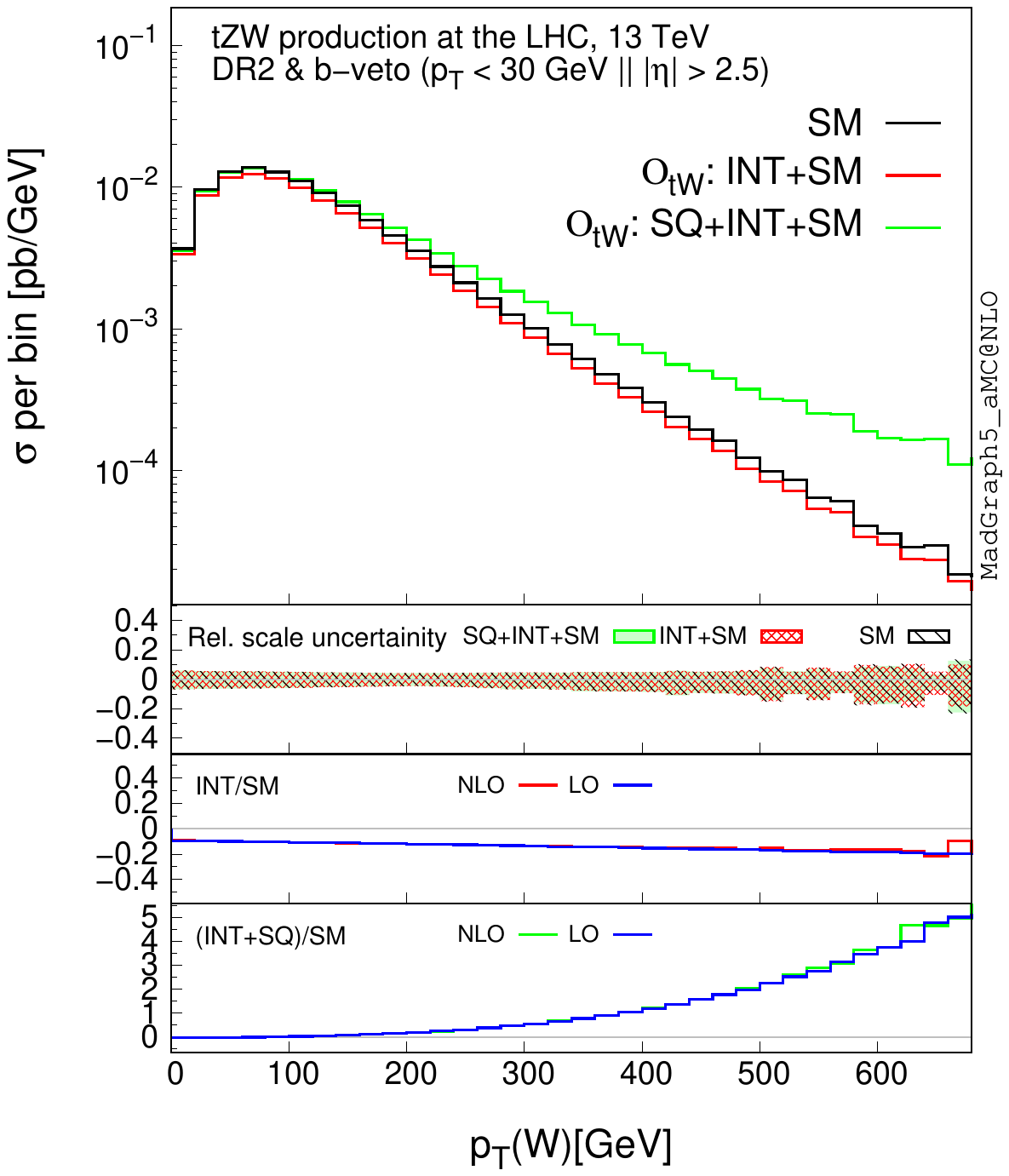}\\
    \includegraphics[trim=0.0cm 5.0cm 0.0cm 0.0cm, clip,width=.45\textwidth]{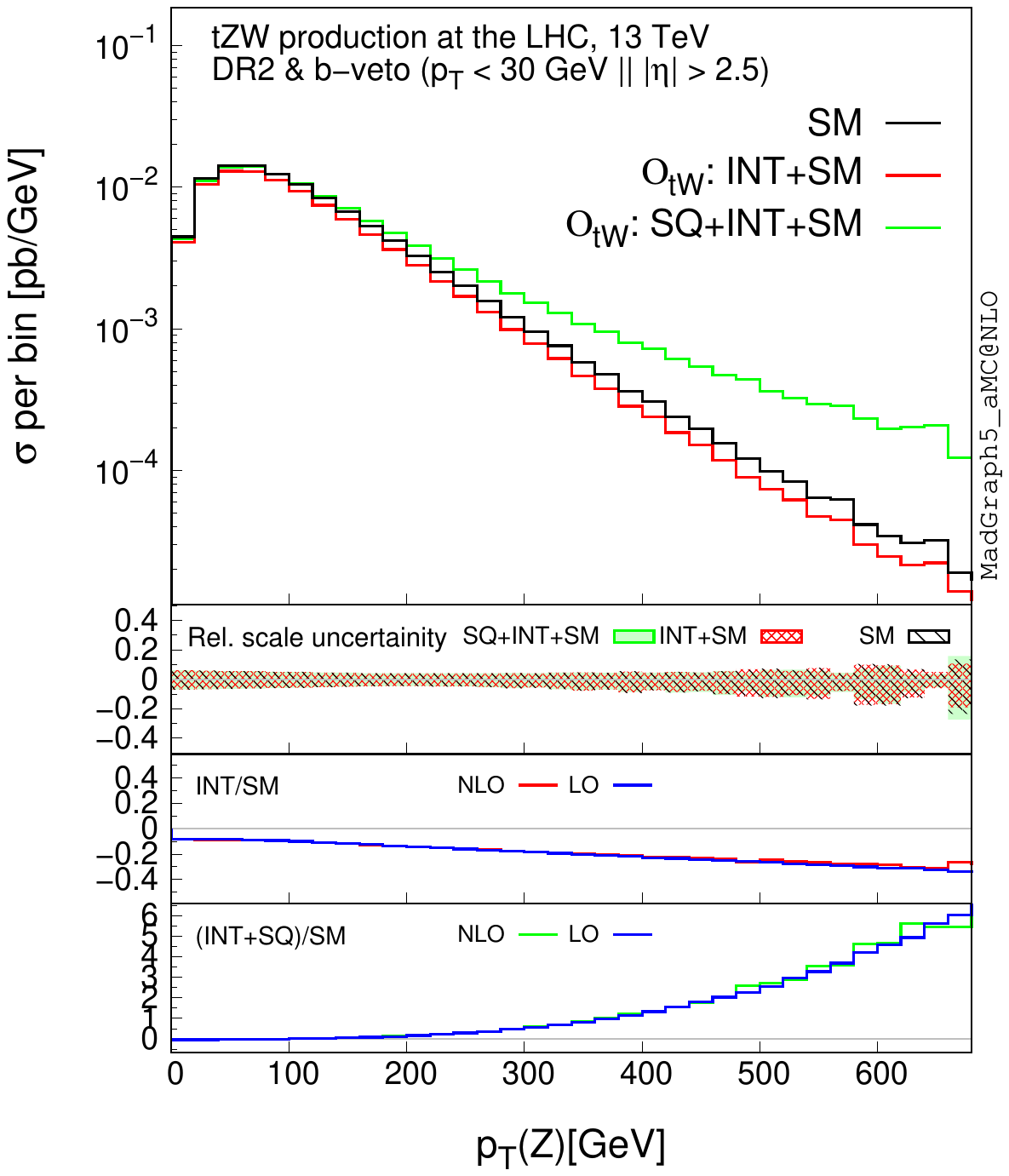}
    \includegraphics[trim=0.0cm 5.0cm 0.0cm 0.0cm, clip,width=.45\textwidth]{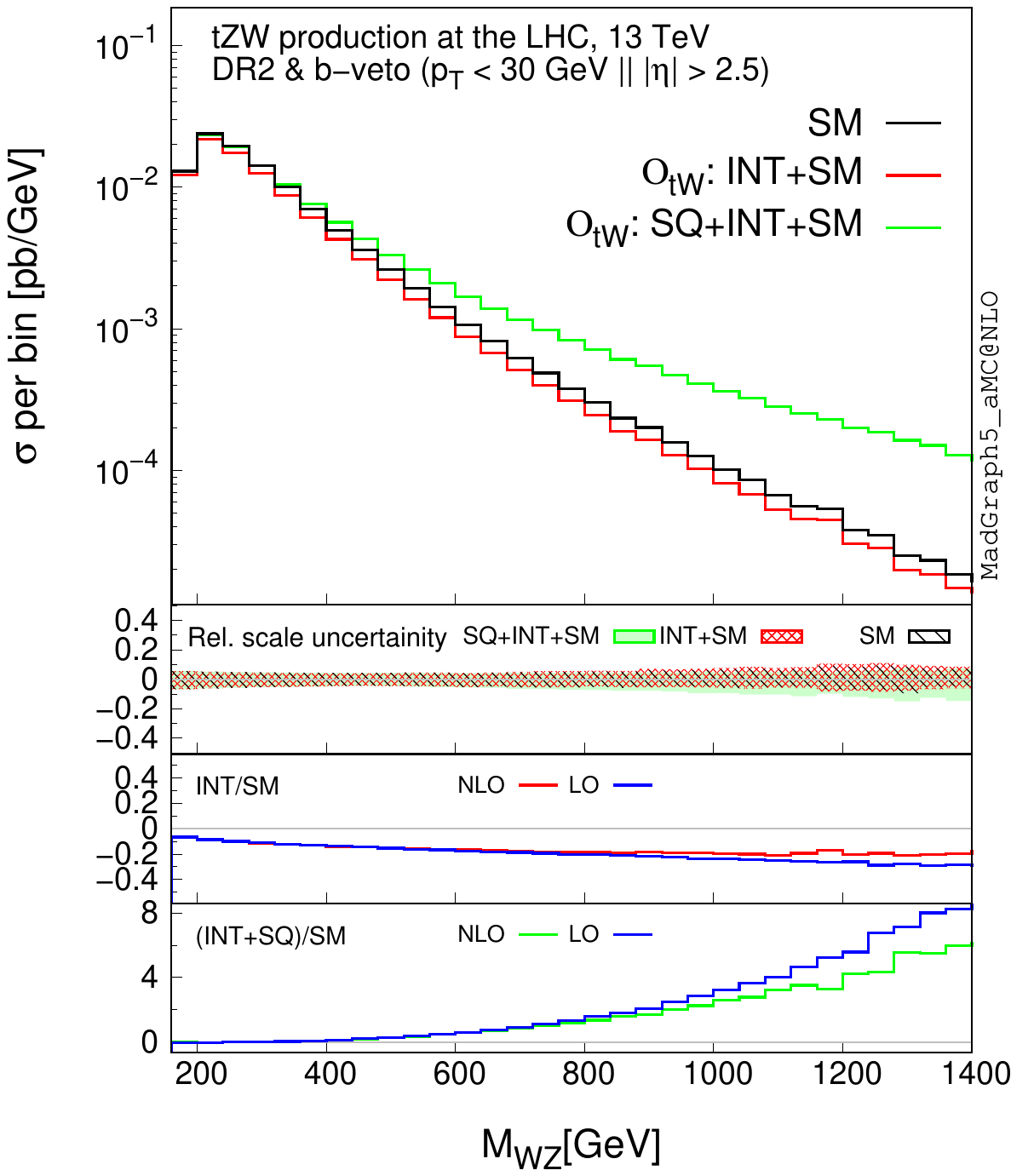}
    \caption{\label{fig:ctw} Same as Fig.\ref{fig:cpq3} but for the $\Op{tW}$ operator}
\end{figure}

\begin{figure}[h!]
    \centering
    \includegraphics[trim=0.0cm 5.0cm 0.0cm 0.0cm, clip,width=.45\textwidth]{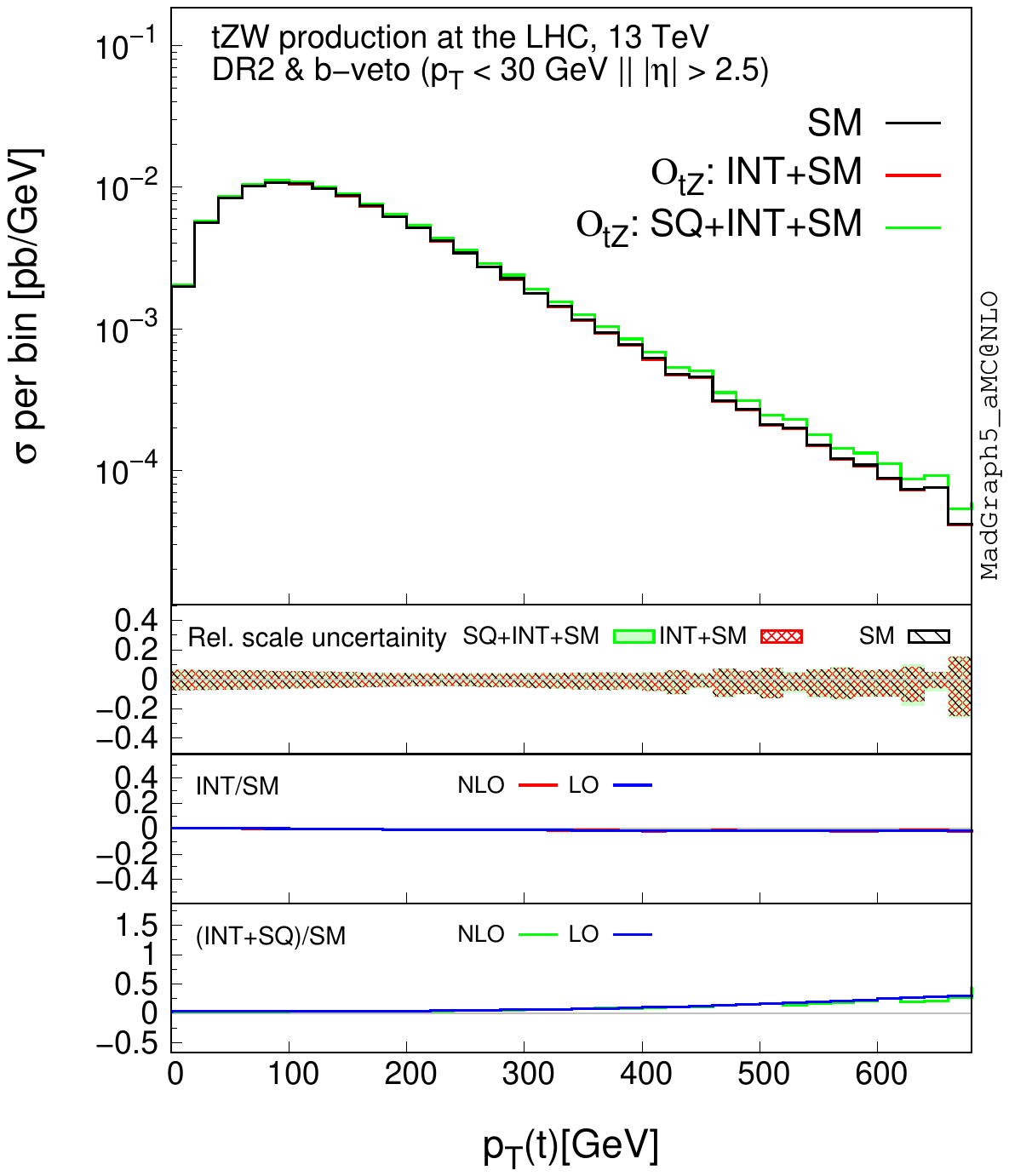}
    \includegraphics[trim=0.0cm 5.0cm 0.0cm 0.0cm, clip,width=.45\textwidth]{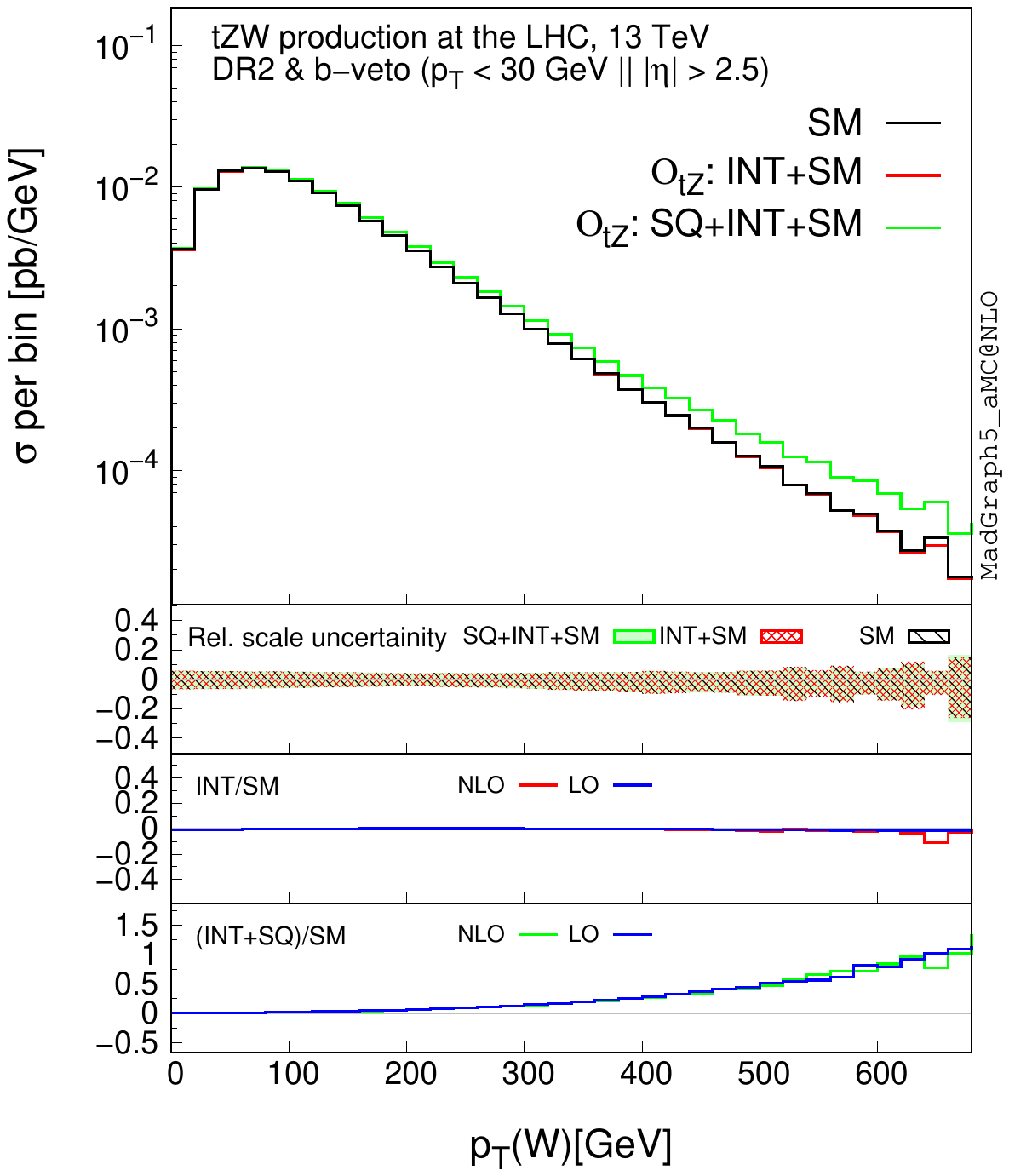}\\
    \includegraphics[trim=0.0cm 5.0cm 0.0cm 0.0cm, clip,width=.45\textwidth]{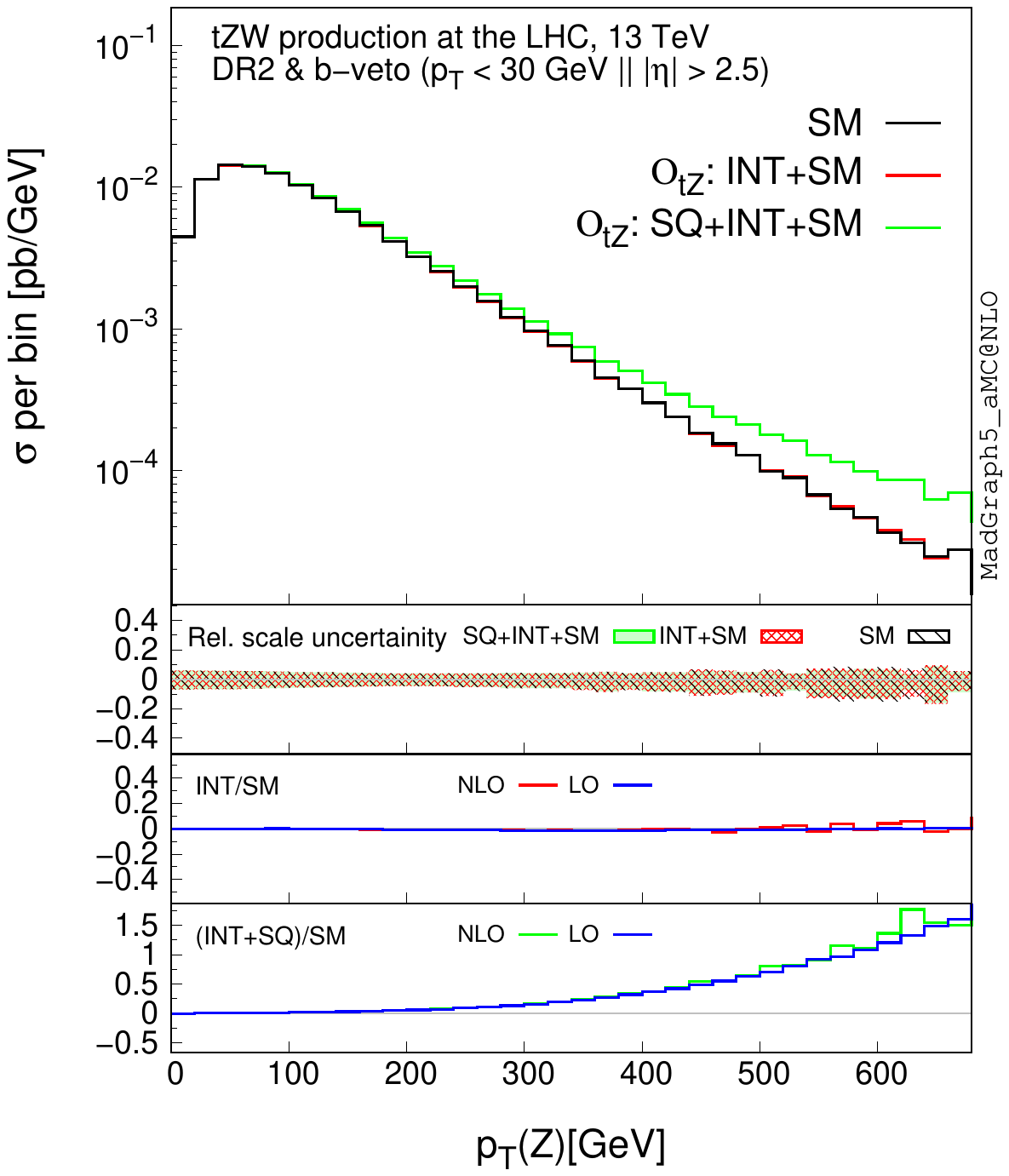}
    \includegraphics[trim=0.0cm 5.0cm 0.0cm 0.0cm, clip,width=.45\textwidth]{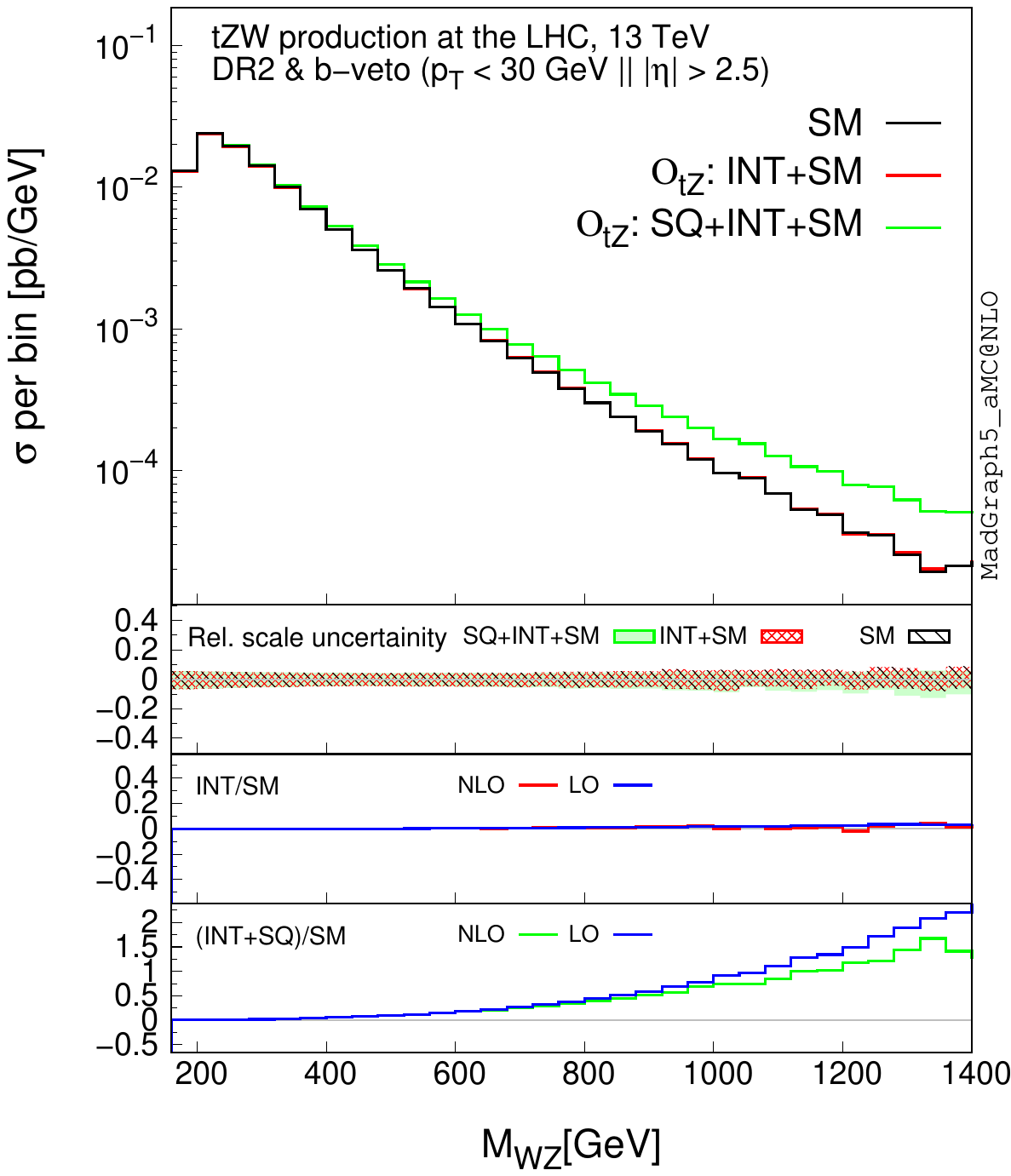}
    \caption{\label{fig:ctz} Same as Fig.\ref{fig:cpq3} but for the $\Op{tZ}$ operator}
\end{figure}

\begin{figure}[h!]
    \centering
    \includegraphics[trim=0.0cm 5.0cm 0.0cm 0.0cm, clip,width=.45\textwidth]{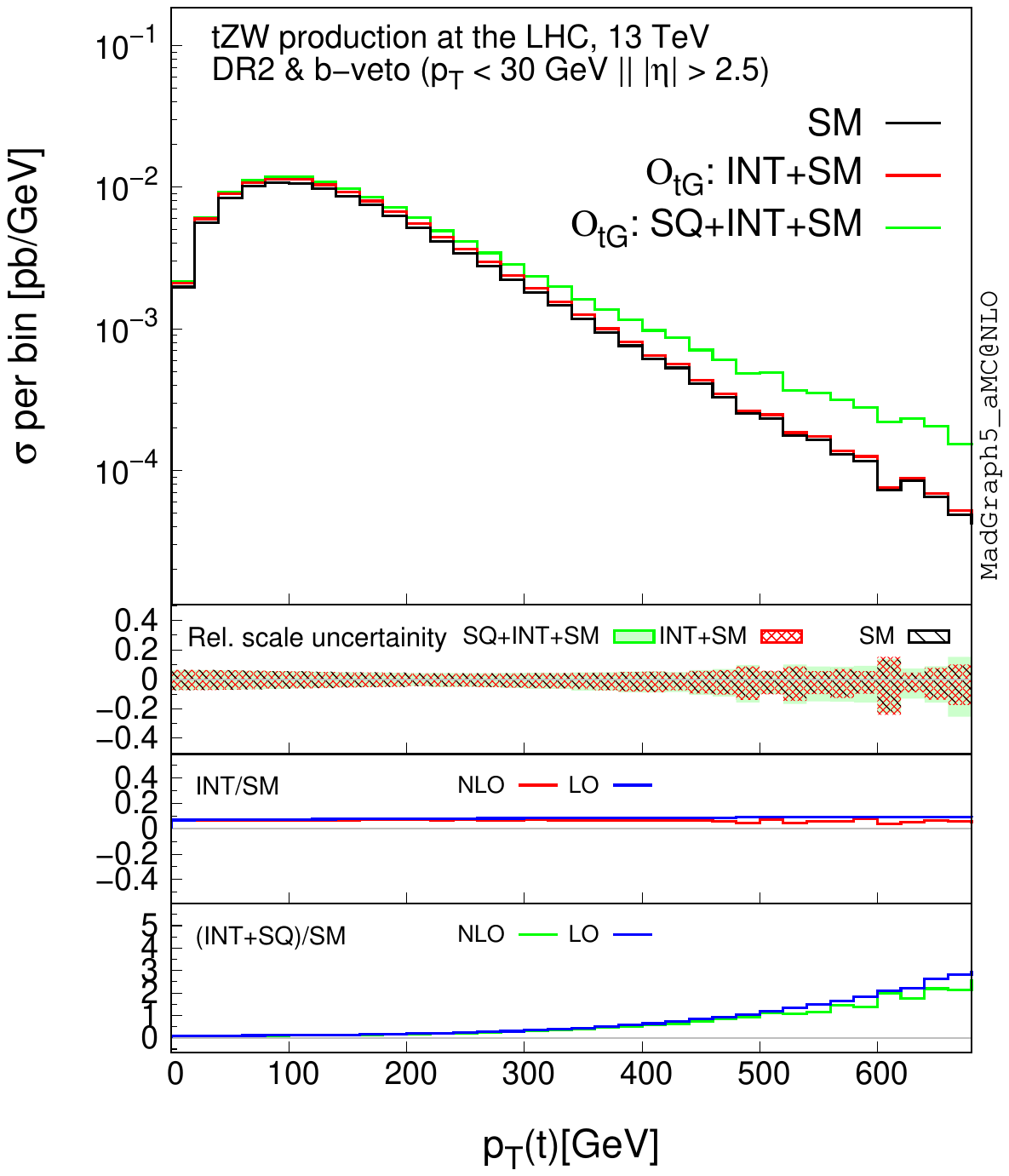}
    \includegraphics[trim=0.0cm 5.0cm 0.0cm 0.0cm, clip,width=.45\textwidth]{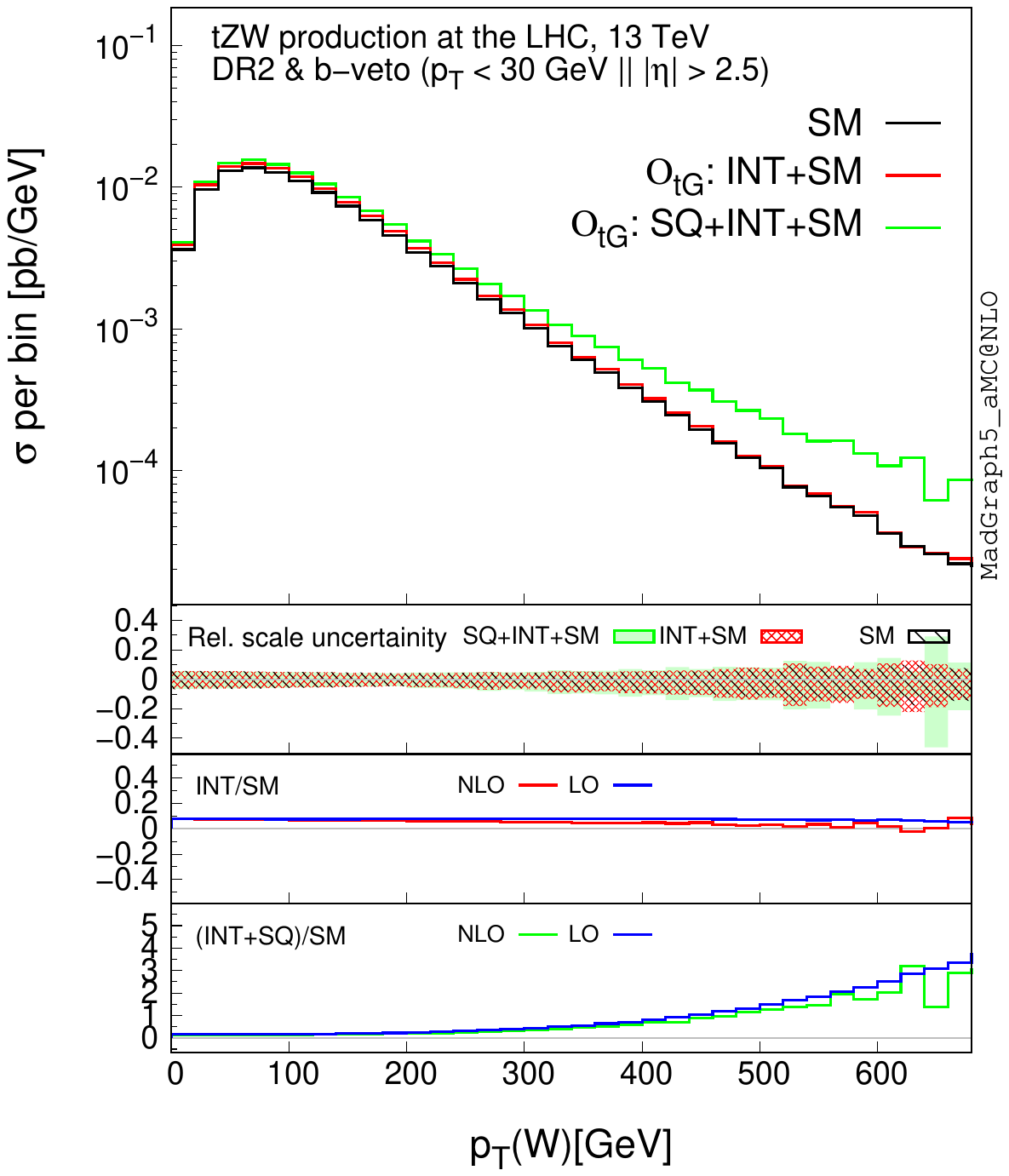}\\
    \includegraphics[trim=0.0cm 5.0cm 0.0cm 0.0cm, clip,width=.45\textwidth]{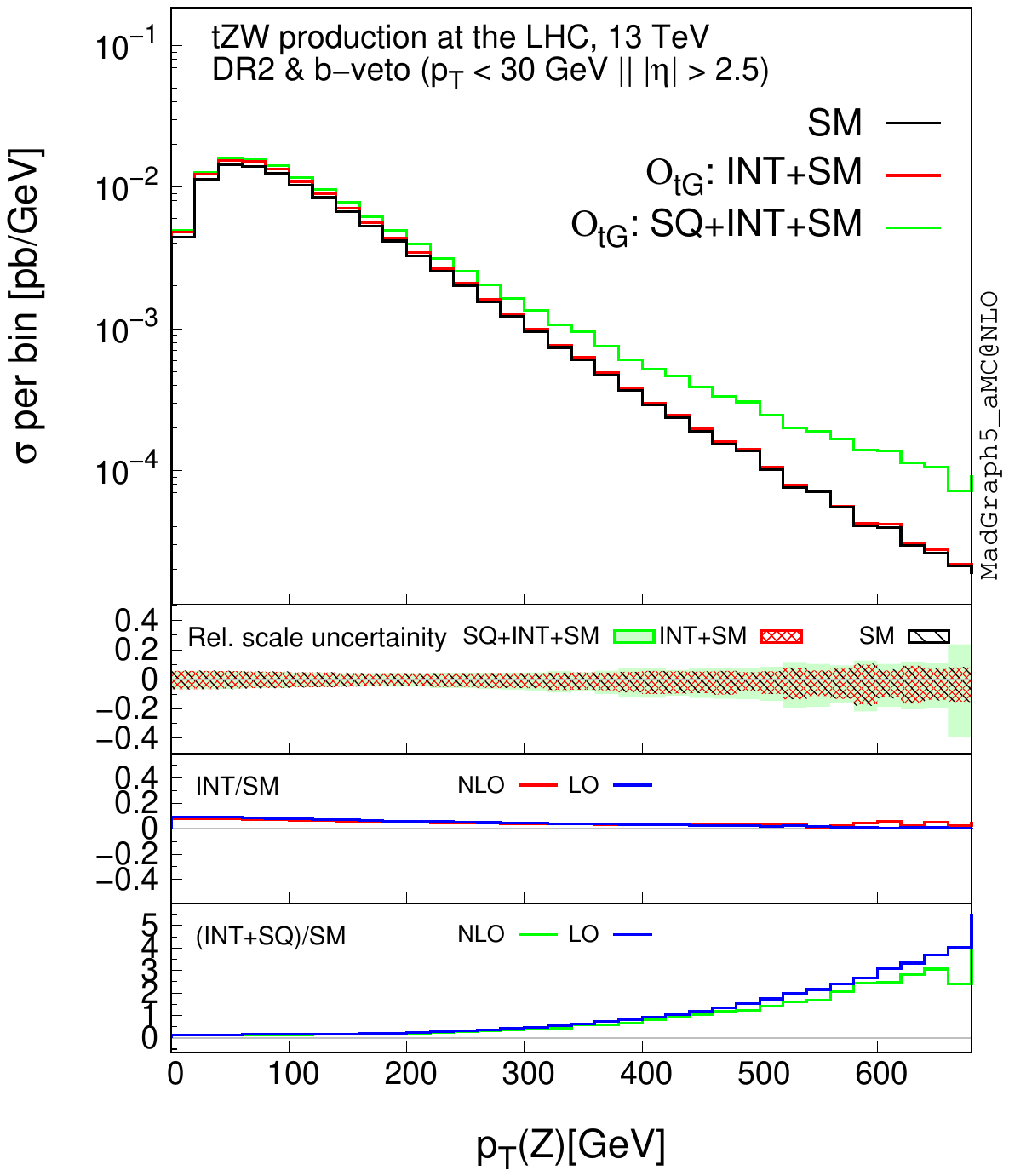}
    \includegraphics[trim=0.0cm 5.0cm 0.0cm 0.0cm, clip,width=.45\textwidth]{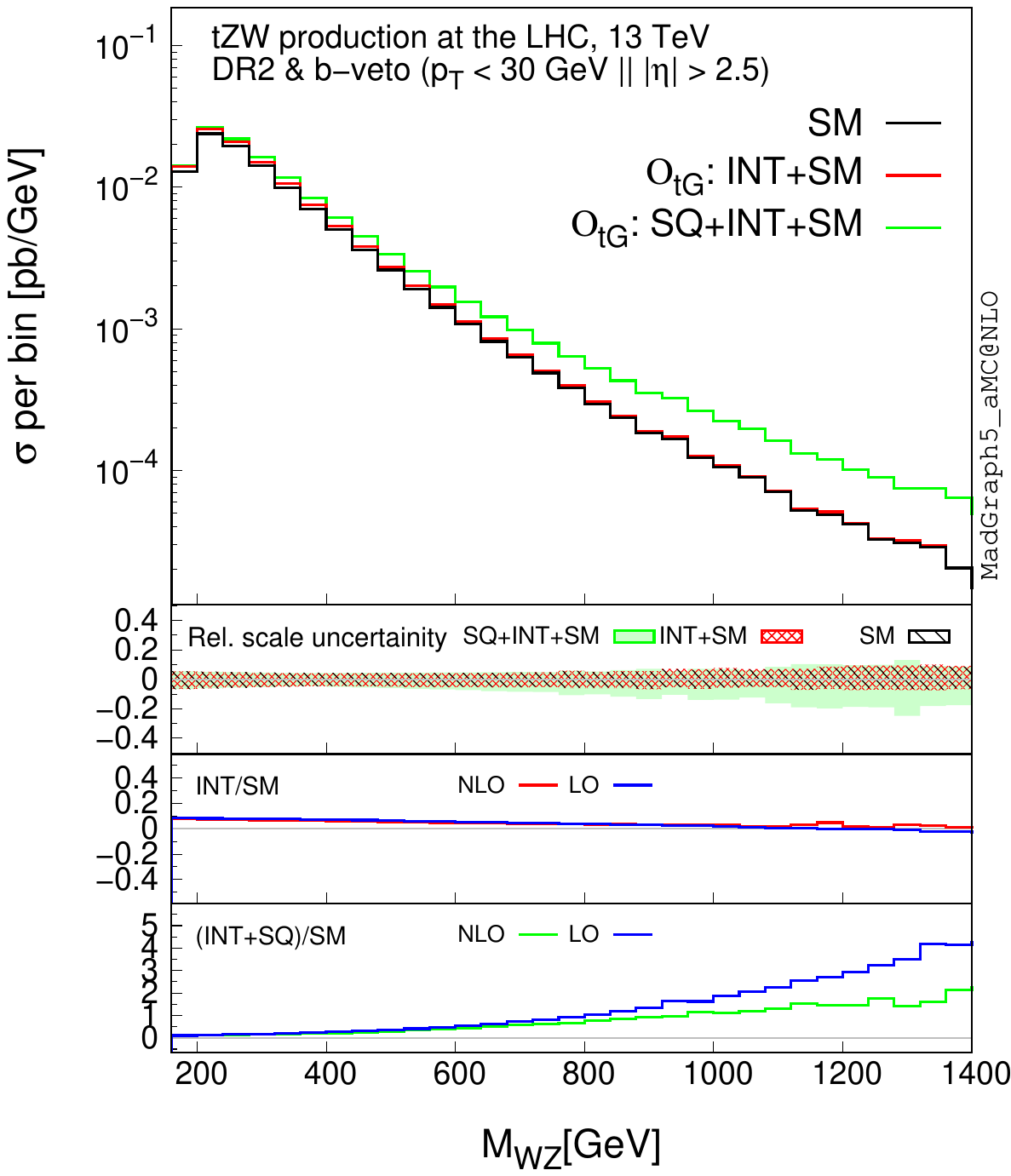}
    \caption{\label{fig:ctg} Same as Fig.\ref{fig:cpq3} but for the $\Op{tG}$ operator}
\end{figure}

Conversely, the quadratic contributions show energy growth across the board, beyond what can be seen from Fig.~\ref{fig:hel_sum}, thanks to the logarithmic scale. As seen in Tab.~\ref{tab:bwtz_hel}, every operator has an energy-growing configuration, whose quadratic contribution should start to dominate at some point. This reflects the fact that the cut chosen for the radar charts likely probes a relatively higher-energy region of phase-space than the tail of Fig.~\ref{fig:hel_sum}. As can be seen from Tab.~\ref{tab:smeft_impact}, the impacts are broadly found to be stable under QCD corrections, with most of the significant deviations from 1 occurring in cases where the overall contribution is relatively small to begin with. Notable exceptions are the $\Opp{\phi Q}{(3)}$ and $\Op{tG}$ impacts in the high-energy region, whose interference(squared) contributions have $K$-factors of 0.82(1.33) and 1.53(0.70), respectively. This radiative stability further confirms that our DR treatment of the $t\bar{t}Z$ and $t\bar{t}$ overlap is correctly identifying the phase-space of the $tWZ$ process in the SMEFT case. 
We now move to presenting differential SMEFT predictions in some key kinematic observables, namely the transverse momenta of top, $W$ or $Z$, and the $WZ$ invariant mass, $M_{WZ}$. Given the relatively insignificant $\Op{\phi t}$ and $\Opp{\phi Q}{(-)}$ impacts, we omit them in what follows. Distributions are presented for the remaining operators, $\Opp{\phi Q}{(3)}$, $\Op{tW}$, $\Op{tZ}$ and $\Op{tG}$ in Figs.~\ref{fig:cpq3},~\ref{fig:ctw},~\ref{fig:ctz} and~\ref{fig:ctg}, respectively. 

As with the inclusive numbers, the predictions are generated for the 13 TeV LHC, subtracting only the $t\bar{t}Z$ overlap in the DR2 scheme and applying the $b$-veto. 
The main figure shows the NLO prediction in the SM in black, alongside predictions for $c_i/\Lambda^2 = 1$ TeV$^{-2}$. The red and green lines correspond to defining the SMEFT prediction taking only interference into account or both interference and squared contributions, respectively. The first inset displays the relative uncertainty obtained from 9-point renormalisation and factorisation scale variation around the central scale of $172$ GeV, with the separate EFT renormalisation scale, \verb|mueft|, kept fixed. The second and third insets show the ratio of the two coloured lines in the main figure to the SM prediction, also including the same ratio at LO, for reference.

The $\Opp{\phi Q}{(3)}$ distributions in Fig.~\ref{fig:cpq3} confirm the expectations we have formed from the helicity amplitude and LO studies presented in Secs.~\ref{sec:bwtz} and~\ref{sec:hel_xs}. The energy growing interference contributions is present in all distributions except for $p_T(t)$, which is generally found to not capture as much of the energy growing effects. This is expected from the fact that the final state top is not really part of the embedded $\bWtZ $ scattering, to which this operator contributes energy growth. The relative impact is most pronounced in $M_{WZ}$, confirming that this is likely the best proxy for the sub-amplitude centre of mass energy. 

Figs.~\ref{fig:ctw} and~\ref{fig:ctz} show the equivalent distributions for the weak dipole operators. The generic behaviours can be identified with our expectations from the helicity amplitudes and previously shown LO calculations. Both operators show the expected growth in the quadratic contributions, that is particularly strong for $\Op{tW}$, which also shows some sub-dominant interference contributions, in contrast to $\Op{tZ}$ whose interference is extremely suppressed. In contrast to the current operator, both weak dipole operators do impact the $p_T(t)$ distributions, albeit in a milder way that the other kinematic variables. 

Finally, Fig.~\ref{fig:ctg} presents the predictions for the top quark chromomagnetic operator, $\Op{tG}$. As previously mentioned, this operator does not contribute to $\bWtZ$ scattering and therefore has slightly different behaviour to the previously three operators. As expected from Fig.~\ref{fig:radar_NLO}, energy growth is evident in all of the quadratic contributions, the main difference with the other operators being that the relative sensitivity of $p_T(t)$ is about the same as the others, highlighting the fact that the final state gauge bosons that participate in the EW top scattering do not have a special importance in this case. Like the other dipoles, the interference contributions are suppressed due to the helicity flipping structure of these types of effective operators. In all of the distributions, the last two insets emphasise the stability of the relative SMEFT impacts under QCD corrections. 

\subsection{NLO matched to Parton Shower (NLOPS)}
We conclude the section on our results by presenting predictions matched to parton shower (PS). We limit ourselves to the case of a single EFT operator, $\Opp{\phi Q}{(3)}$, that best probes unitarity violation in $\bWtZ$. This is done both for the sake of brevity but also since the SMEFT results are found to be stable with respect to this procedure.  We match short-distance events with \texttt{Pythia8}~\cite{Sjostrand:2014zea} using the \texttt{MC@NLO} method~\cite{Frixione:2002ik} as 
automated inside \texttt{MadGraph5\_aMC@NLO}. In our setup, the (anti-)top quark is allowed to decay, while the $W$ and the $Z$ bosons either appearing in the partonic event or originating from the decay are kept stable. Jets are defined using the anti-$k_T$ algorithm~\cite{Cacciari:2008gp} implemented in \texttt{FastJet}~\cite{Cacciari:2011ma}, with cone radius $R = 0.4$, and required to have $p_T(j) > 30$ GeV and $|\eta(j)|< 4.5$. A jet is $b$-tagged if a $b$-hadron appears among its constituents and if $|\eta(j_b)|<2.5$. Our analysis assumes 100\% $b$-tagging efficiency. We select events with exactly one $b$-jet and a central $W$ and $Z$ boson, $|\eta(W)| < 2.5$ \& $|\eta(Z)| < 2.5$, hereafter referred to as the \texttt{1-bjet} scenario. This is the equivalent of imposing the $b$-jet veto previously used at fixed order since, at NLOPS, the top decay will typically give rise to an extra $b$-jet, and most of the time this jet will be the hardest in the event. 

\begin{figure}[h!]
    \centering
    \includegraphics[trim=0.0cm 5.0cm 0.0cm 0.0cm, clip,width=.45\textwidth]{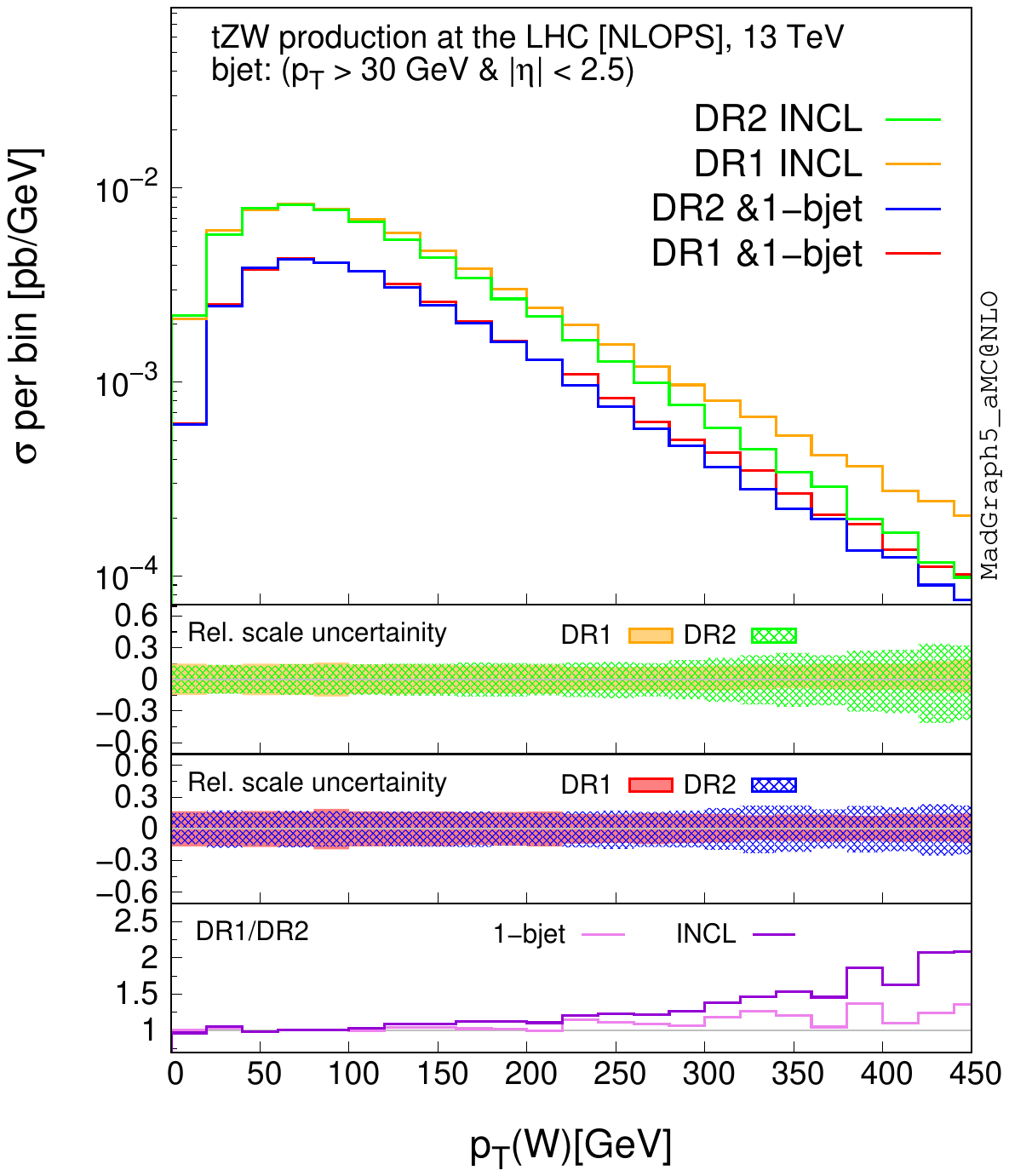}
    \includegraphics[trim=0.0cm 5.0cm 0.0cm 0.0cm, clip,width=.45\textwidth]{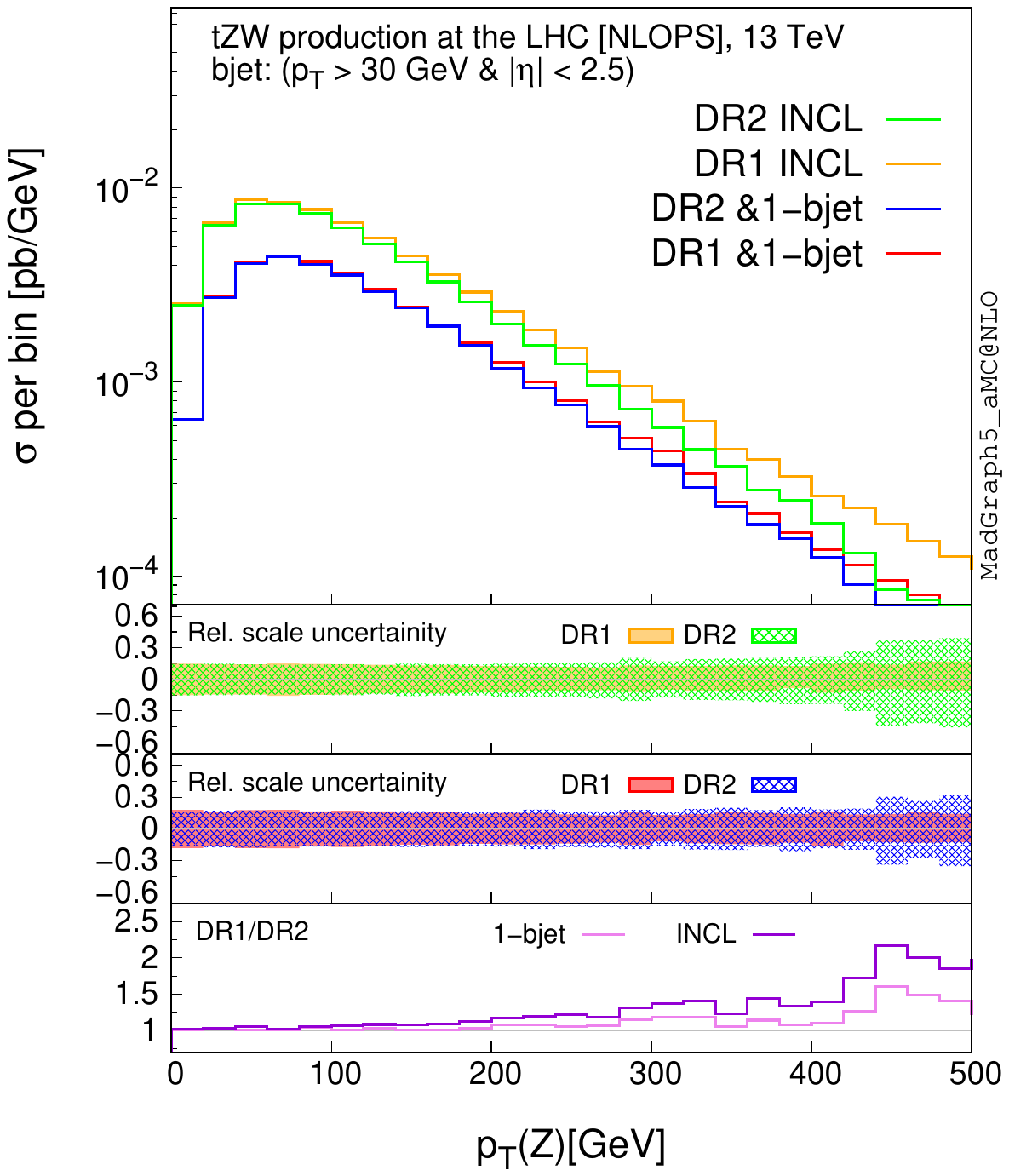}\\
    \includegraphics[trim=0.0cm 5.0cm 0.0cm 0.0cm, clip,width=.45\textwidth]{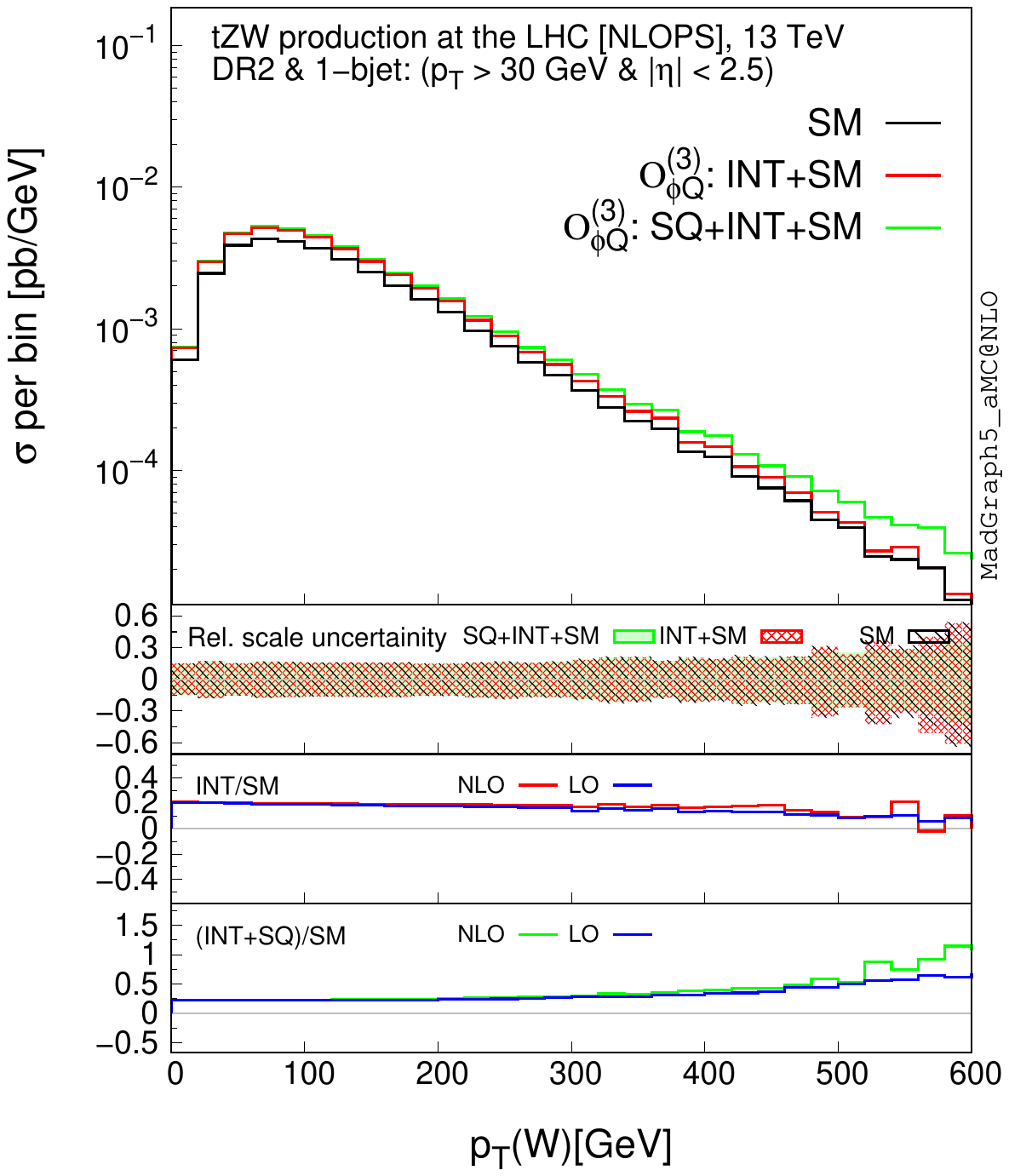}
    \includegraphics[trim=0.0cm 5.0cm 0.0cm 0.0cm, clip,width=.45\textwidth]{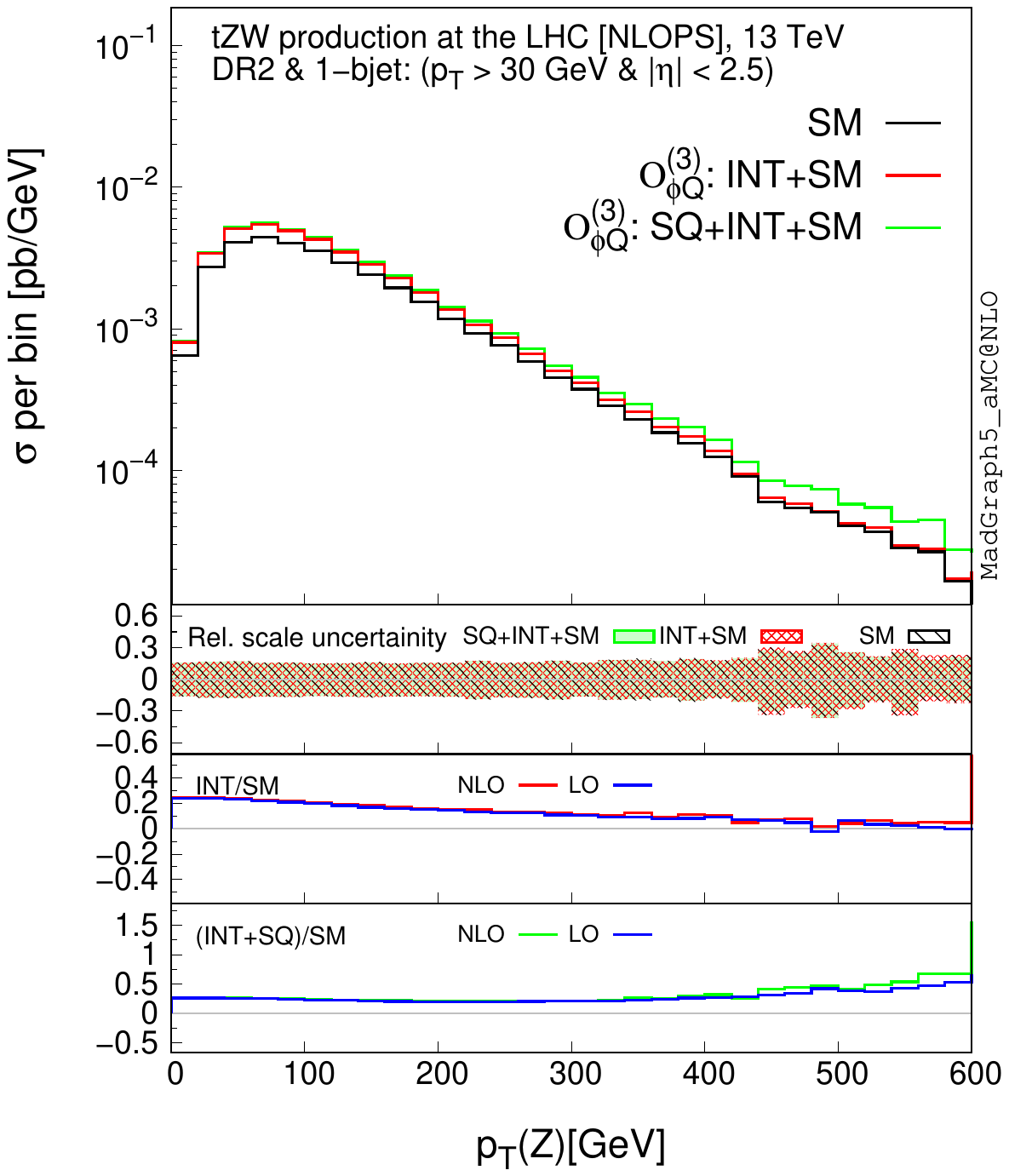}
    \caption{\label{fig:ps_1} The top row shows the NLOPS predictions for the transverse momentum of the $W^{\pm}$ boson, $p_{T}(W)$ (\emph{left}) and of the $Z$ boson, $p_{T}(Z)$ (\emph{right}) in the SM at DR1 and DR2 for both inclusive and \texttt{1-bjet} scenarios. The insets in the top row are the same as in the left panel of Fig.~\ref{fig:sm_main_text}. In all the predictions, only the $t\bar{t}Z$ overlap is removed as the $Z$ boson is kept stable. The bottom row shows the same observables at DR2 for the $\Opp{\phi Q}{(3)}$ SMEFT operator when the selection of one hard $b$-jet is imposed, at NLOPS. The insets in the bottom row are the same as in Fig.~\ref{fig:cpq3}}
\end{figure}

Our NLOPS predictions are presented in Fig.~\ref{fig:ps_1}, where the transverse-momentum distributions of the $W$ boson and of the $Z$ boson are displayed. The former refers to the 
$W$ boson from the hard-scattering event, which is differentiated from the one stemming from the top decay using Monte-Carlo truth. The top row shows the SM distributions in the two diagram removal schemes, showing the efficacy of the single $b$-jet requirement in bringing the two predictions together, thus acting like the parton-level $b$-veto. Figures in the lower row show the effect of $\Opp{\phi Q}{(3)}$. The behaviour of higher-dimension operators closely resembles the corresponding one at fixed-order, shown in Fig.~\ref{fig:cpq3}. We remark that a common feature of the NLOPS results is the increased scale uncertainty relative to the fixed order predictions, as well as larger discrepancies between DR1 and DR2 predictions in the tails of distributions, which were observed also in Ref.~\cite{Demartin:2016axk} for the $tWH$ process. This likely occurs because of the presence of the additional $b$-jets from, \emph{e.g.}, top quark decay, and the fact that at high-$p_T$, one may occasionally select the `wrong' one. Ref.~\cite{Demartin:2016axk} showed that this discrepancy disappears when using MC truth information on the $b$-jet origin.  

All in all we find the $tWZ$ process to be very stable under PS, which is expected from the fact that it is mostly driven by EW interactions, and that our results serve a proof of principle for the realistic generation of precise SMEFT predictions that can be used in future measurements and new physics searches via this process.  

Before concluding, we comment on a feature of \texttt{MadGraph5\_aMC@NLO} that we have exploited for the NLOPS results. Since version 3.1.0 of the code, LHE events store weights for the individual 
coupling combinations of the cross section, in the same format already employed for scale and PDF variations~\cite{Frederix:2011ss}. In our case, this has made it possible to have histograms for the three contributions (SM, EFT-SM interference and EFT squared) from the same event sample. Moreover, as for the scale and PDF uncertainties,
correlations between these contributions help to keep their ratios numerically stable. This makes it possible to simulate scenarios where different values of the EFT coefficients are injected using
results from a single run. 

\section{Summary and conclusions}
\label{sec:conclusions}
We have presented a dedicated study of the $tWZ$ and its sensitivity to indirect new physics effects via the SMEFT. In doing so, we constructed an operative definition of the $tWZ$ process, valid beyond LO in perturbation theory. This was necessary due to the overlap with potentially resonant contributions from processes such as $\ttZ$ and $\ttbar$ that occurs at NLO. Compared to other previously studied processes featuring resonance overlap, such as $tW$, $tWH$ and gluino-squark associated production, $tWZ$ is unique in that it overlaps with two distinct underlying processes, especially when the $\twll$ final state is considered as an example of involving the $Z$ boson decay products. We found that, as with the other top-quark related examples, a veto on an additional hard or central $b$-jets was sufficient to completely suppress the overlap in our fixed order predictions. We also observed that the dominant contribution to the NLO cross section after the veto came from events where the $Z$ decay products lay around the $Z$ mass, such that the effect of $\ttbar$ overlap was subdominant. Such a $Z$ mass window selection is relatively straightforward to implement in the likely leptonic final state in which this process will be searched for. Moreover, the SMEFT contributions favour the high energy phase-space where the $>2$-body top-quark decay is always off-shell. For these reasons, we found it sufficient to continue our study assuming the production of a stable $Z$ boson, alongside the top and $W$ boson that were already taken to be as such, which also justified only subtracting the $\ttZ$ overlap. 

Having defined the phase-space region for $tWZ$ at NLO, we moved on to studying the way in which SMEFT deformations affected the process. The first step was to consider the underlying $\bWtZ$ scattering amplitude and identify potential energy-growing effects that could be exploited at future collider searches. We focused on six dimension-6 SMEFT operators, five of which directly affect $\bWtZ$ and the sixth affecting a different part of the $tWZ$ process. A detailed calculation and study of the $\bWtZ$ helicity amplitudes allowed us to formulate expectations for the sensitivity that might be obtained from $tWZ$, especially in the high-energy regime. We found that each operator contributes to the amplitude in an energy-growing way, showing much potential for new sensitivity in this channel.
Only one operator, $\Opp{\phi Q}{(3)}$, was found to do so in the fully longitudinal configuration, in which the SM energy growth is maximal (constant) such that the overall cross section contribution of the interference between the SM and the SMEFT would also grow with energy. We also discussed the general advantages of the $tWZ$ process in the context of contributing to our global understanding of top EW interactions. Finally, we completed our preliminary investigations with LO simulations of polarised cross sections in $tWZ$, focusing on the gauge boson polarisations, confirming the intuition and expectations developed from the helicity amplitude study and also revealing some additional interference contributions from the weak dipole operator $\Op{tW}$ that appeared to grow with energy. In most cases, this meant that quadratic contributions could potentially be important, depending on the value of the Wilson coefficient. This confirmed our expectations that two of the 6 operators, $\Op{\phi t}$ and $\Opp{\phi Q}{(-)}$, do not actually lead to significant high-energy behaviour and therefore $tWZ$ is not likely to be the best place to probe them.

Our main results constitute a series of NLO predictions for $tWZ$ in the SM and for each operator contribution at linear and quadratic order in the dimension-6 couplings $c_i/\Lambda^2$. All results were obtained using the recently released \verb|SMEFTatNLO| model. At inclusive level, we quoted results in both DR schemes, which showed excellent agreement in the fiducial phase-space region. We also considered a very high-energy phase-space region where $p_T^{W,Z} > 500$ GeV as a simple, first look at how each operator contribution evolves with energy. The relative impact of each contribution to the corresponding SM prediction indicated the presence of the expected high-energy behaviour determined in the previous LO studies. Since the operative NLO prediction depends on the details of the $b$-jet veto implementation, we avoided quoting traditional $K$-factors, but did note that the relative impact of each operator was generally stable with respect to the NLO corrections, with some mild departures of at most 30\%, mainly in the high-energy region. We also found our results to be consistent with our previous, LO study of this channel. 

Next, we presented differential predictions for the four operators for which we did expect interesting energy growth. Our results confirmed that the stability of the relative impact of each operator under QCD corrections persists at the differential level. We showed differential predictions in the three final state $p_T$s, as well as in $\mwz$, which was found to be the best observable to probe energy growth, again confirming our understanding from $\bWtZ$. The top quark $p_T$ was often found to be less sensitive to the energy growth, since the final state top does not participate directly in the underlying sub-amplitude. Finally, both for the SM and $\Opp{\phi Q}{(3)}$, we go beyond fixed order and show differential NLOPS predictions, including top quark decay, as a proof of principle that these numbers can be relatively simply obtained. The positive impact of the $b$-veto was observed, albeit with a slightly reduced efficacy, due to the small chance that the wrong $b$-jet is vetoed at high $p_T$. The relative effect of $\Opp{\phi Q}{(3)}$ was practically unchanged compared to the fixed order case, suggesting that our existing tools are able to generate precise predictions for this process to be used as inputs to future analyses and therefore eventually improve our understanding of top quark EW interactions.

\acknowledgments
M.Z. thanks Olga Bessidskaia Bylund and Lidia dell’Asta for discussions on $tWZ$.
This work has received funding from the European Union’s Horizon 2020 research and innovation program as part of the Marie Skłodowska-Curie Innovative Training Network MCnetITN3 (grant agreement no. 722104) and by the F.R.S.-FNRS under the ``Excellence of Science'' EOS be.h project no. 30820817. K.M. is supported by the UK STFC via grant ST/T000759/1. M.Z.~is supported by the ``Programma per Giovani Ricercatori Rita Levi Montalcini'' granted by the Italian Ministero dell'Universit\`a e della Ricerca (MUR). 

\clearpage
\appendix
\section{$\bWtZ$ helicity amplitudes and differential cross sections\label{app:bwtz}}
\begin{table}[H]
    \centering
    \input{tab_bwtz_hel}
    \caption{\label{tab:bwtz_hel}
    Top quark operator contributions to the individual helicity amplitudes of $\bWtZ$ scattering in the high-energy limit, $s,-t\gg v$. $\lambda_i$ denotes the helicity/polarisation of the external leg $i$ and the contribution of each operator $\mathcal{O}_j$ omits an overall factor of $c_{i}/\Lambda^2$. ``$-$'' entries denote SMEFT contributions that decrease with energy. The energy dependence of the corresponding SM helicity amplitude is given in a schematic form such that the high-energy behaviour of each interference term can be inferred. As indicated by the last row, all amplitudes with a right-handed $b$-quark are not generated since $m_b=0$ is enforced by our flavor symmetry assumption.
    }
\end{table}

\begin{figure}[h!]
    \centering
    \includegraphics[width=.45\textwidth]{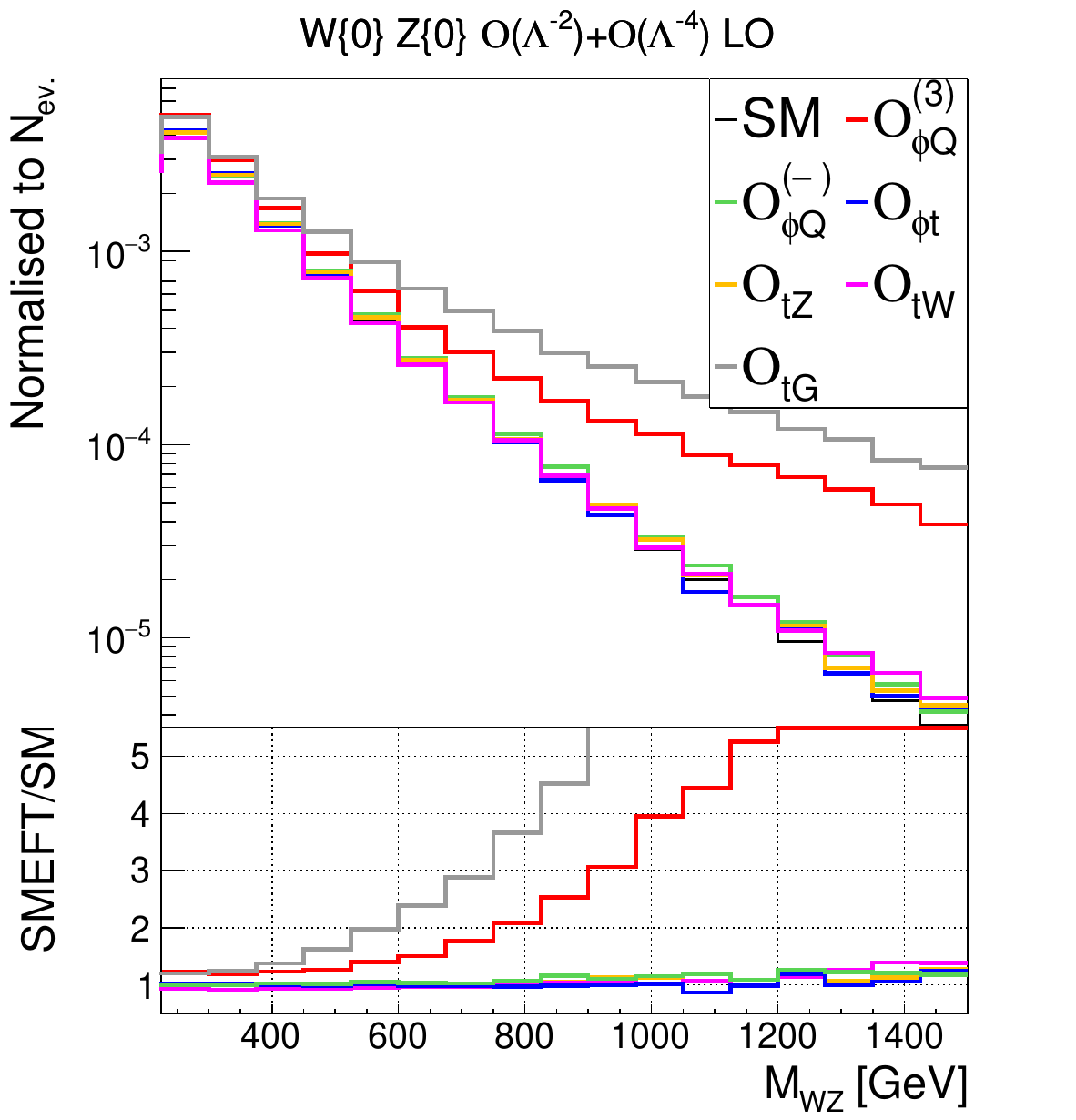}
    \includegraphics[width=.45\textwidth]{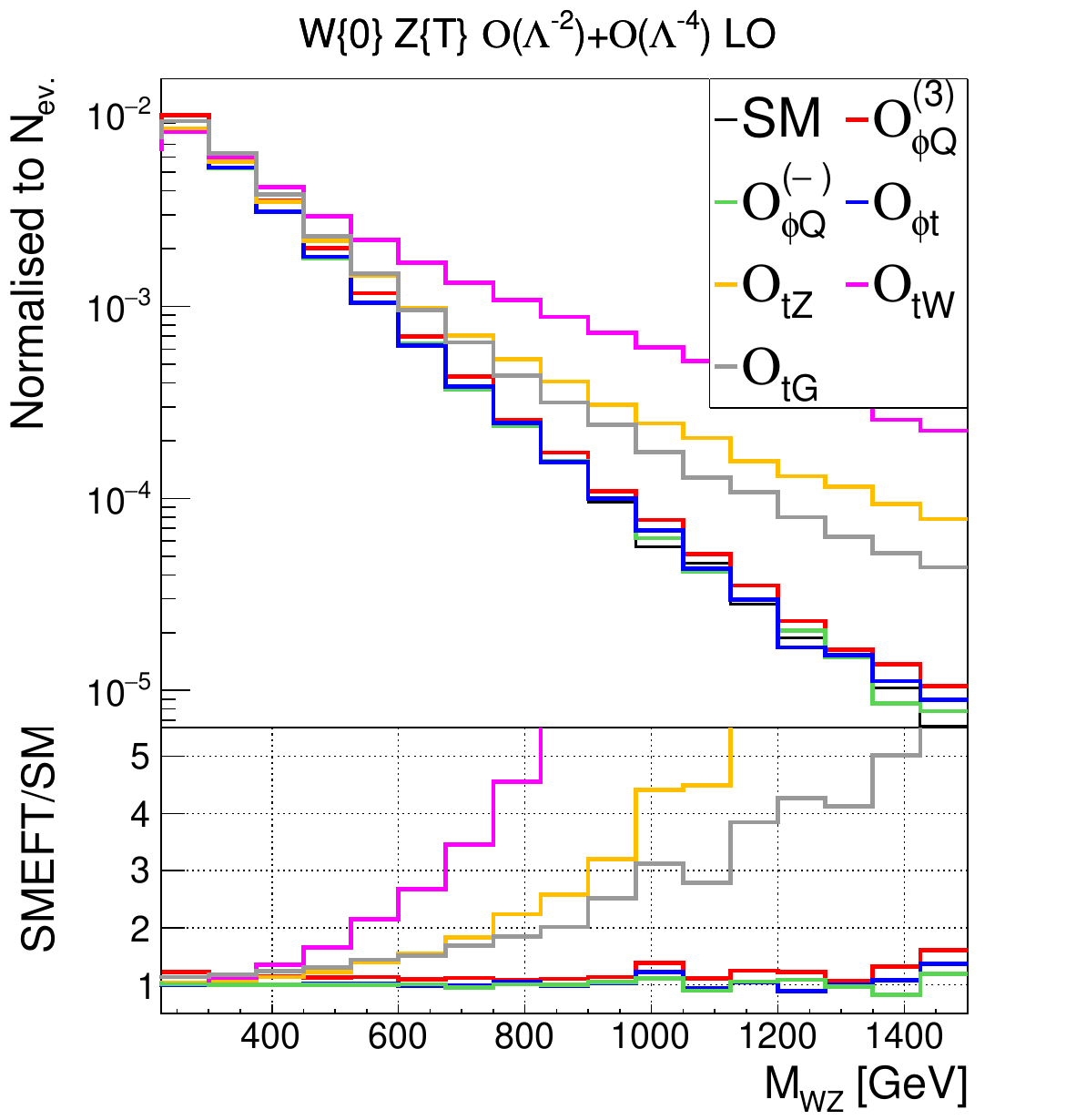}\\
    \includegraphics[width=.45\textwidth]{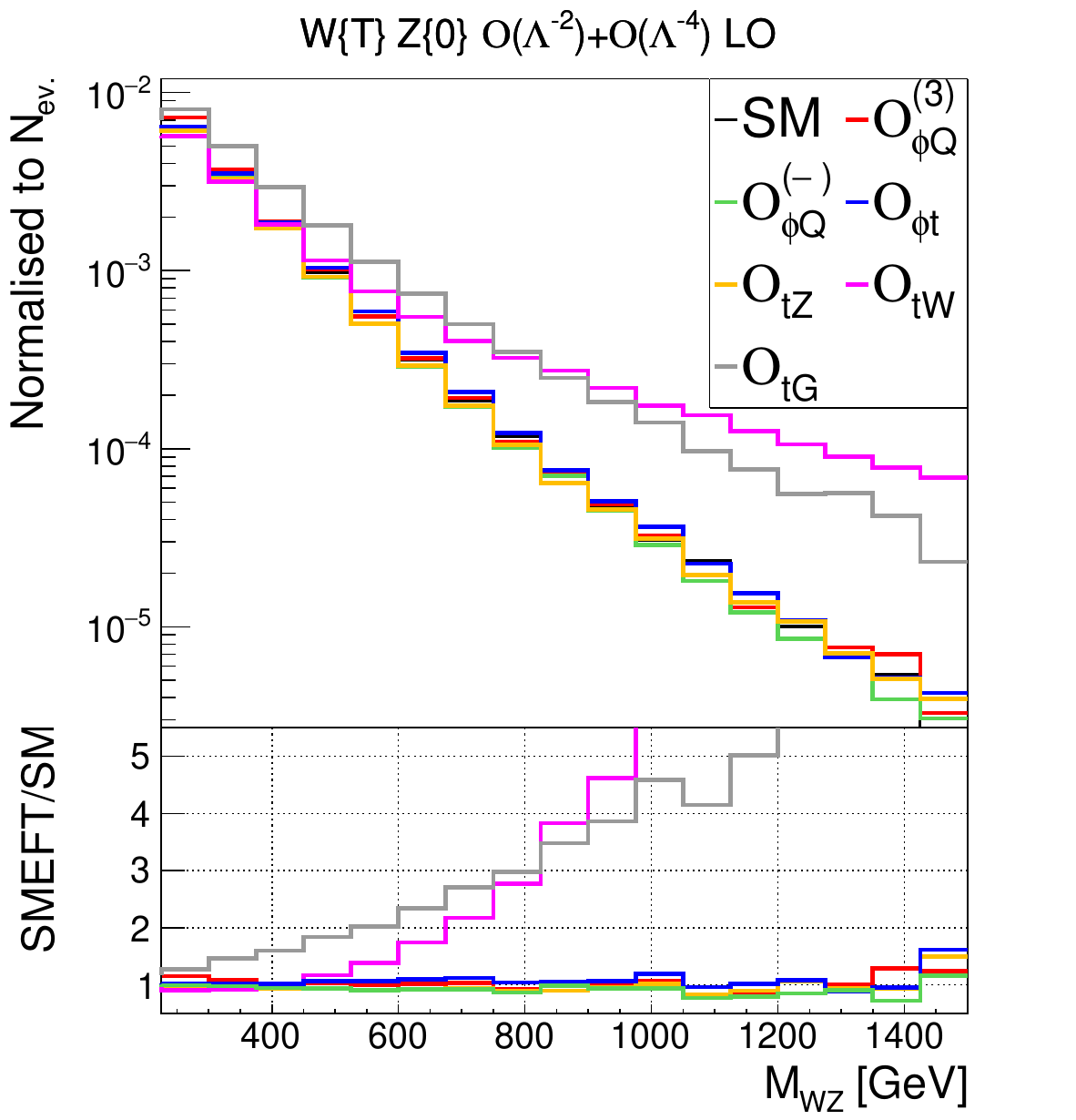}
    \includegraphics[width=.45\textwidth]{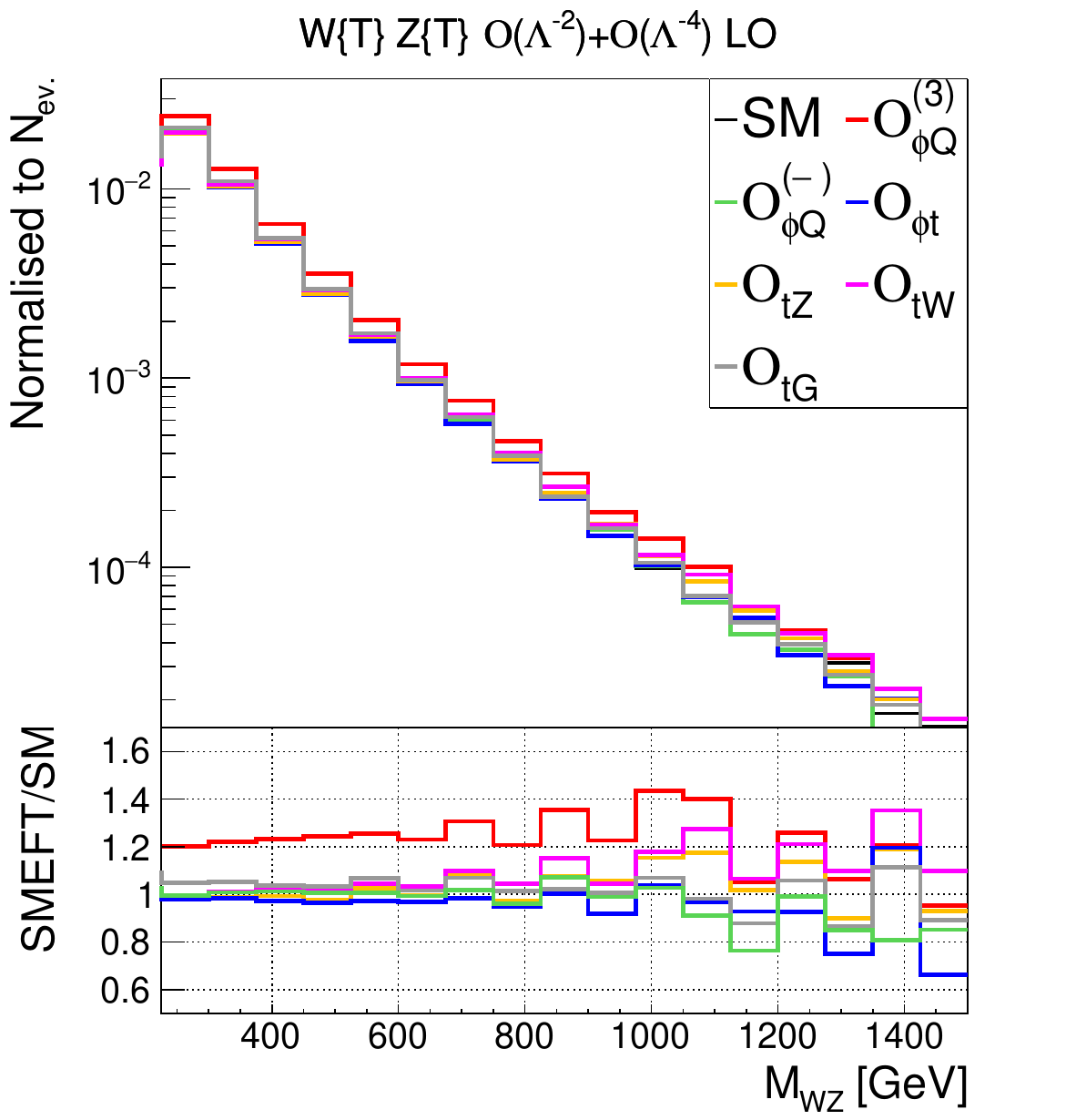}
    \caption{\label{fig:hel_full} Same as Fig.~\ref{fig:hel_int} but for the SMEFT predictions truncated at the quadratic-level ($\mathcal{O}(\Lambda^{-4})$)}
\end{figure}
\clearpage
\bibliographystyle{JHEP}
\bibliography{bibliography}

\end{document}

%% file: tab_operators.tex
\renewcommand{\arraystretch}{1.3}
\resizebox{\textwidth}{!}{
\setlength{\tabcolsep}{3pt}
{\small
\begin{tabular}{|l|c|c|c|c|c|c|c|}
\hline
     \multirow{3}{*}{Op.}& \multirow{3}{*}{Definition} & \multicolumn{6}{c|}{Allowed range $\left(\frac{\Lambda}{1\text{TeV}}\right)^2$}
     \tabularnewline \cline{3-8}
    & &\multicolumn{3}{c|}{$\mathcal{O}(\Lambda^{-2})$}&\multicolumn{3}{c|}{$\mathcal{O}(\Lambda^{-4})$} 
     \tabularnewline \cline{3-8}
    &&Ind.&Marg.& Ref.&Ind.&Marg.& Ref.\tabularnewline\hline
     $\Op{\phi D}$       
    &  $(\phi^\dagger D^\mu\phi)^\dagger(\phi^\dagger D_\mu\phi)$
    & [-2.3,0.27]$\sst\times 10^{-2}$ & [-1.6,0.81]&\cite{Ellis:2020unq}& $-$& $-$ &$-$ \tabularnewline
     $\Op{\phi WB}$       
    & $(\phi^\dagger \tau_{\sss I}\phi)\,B^{\mu\nu}W_{\mu\nu}^{\sss I}\,$ 
    & [-4.3,2.6]$\sst\times 10^{-3}$& [-0.36,0.73]&\cite{Ellis:2020unq}&  $-$& $-$ &$-$  \tabularnewline
     $\Op{W}$            
    & $\varepsilon_{\sss IJK}\,W^{\sss I}_{\mu\nu}\,{W^{{\sss J},}}^{\nu\rho}\,{W^{{\sss K},}}^{\mu}_{\rho}$
    & [-0.071,0.42] & [-0.11,0.4] & \cite{Ellis:2020unq} &  [-0.21,0.24] & [-0.18,0.22] & \cite{Ethier:2021bye}  \tabularnewline
     $\Op{tG}$           
    & $ig{\sss S}\,\big(\bar{Q}\tau^{\mu\nu}\,T_{\sss A}\,t\big)\,\tilde{\phi}\,G^A_{\mu\nu}  + \text{h.c.}$
    & [0.007,0.11] & [-0.13,0.40]& \cite{Ethier:2021bye} & [0.006,0.11] & [0.062,0.24] & \cite{Ethier:2021bye}  \tabularnewline
     $\Qp{tW}$           
    & $i\big(\bar{Q}\tau^{\mu\nu}\,\tau_{\sss I}\,t\big)\,\tilde{\phi}\,W^I_{\mu\nu} + \text{h.c.}$
    & [-0.12,0.51] & [-0.088,0.55]& \cite{Ellis:2020unq} &   [-1.3,1.8] & [-4,3.4] & \cite{Buckley:2015lku} \tabularnewline
     $\Qp{tB}$           
    &  $i\big(\bar{Q}\tau^{\mu\nu}\,t\big)\,\tilde{\phi}\,B_{\mu\nu} + \text{h.c.}$
    & [-4.5,1.2] & [-5.2,2.5] & \cite{Ellis:2020unq} & [7.1,4.7] & $-$ & \cite{Buckley:2015lku} \tabularnewline
     $\Op{tW}$           
    & $\Qp{tW} + (\cw/\sw)\Qp{tB}$
    & [-9.3,2.6]$\sst\times 10^{-2}$ & [-0.31,0.12] & \cite{Ethier:2021bye} & [-8.4,2.9]$\sst\times 10^{-2}$ & [-0.24,0.086] & \cite{Ethier:2021bye}  \tabularnewline
     $\Op{tZ}$           
    &  $-\Qp{tB}/\sw$
    & [-0.039,0.10]& [-16,5.6] & \cite{Ethier:2021bye} & [-4.4,9.4]$\sst\times 10^{-2}$ & [-1.1,0.86] & \cite{Ethier:2021bye} \tabularnewline
     $\Qpp{\phi Q}{(1)}$ 
    & $i\big(\phi^\dagger\lra{D}_\mu\,\phi\big)\big(\bar{Q}\,\gamma^\mu\,Q\big)$
    & [-3.1,4.9]$\sst\times 10^{-2}$  & [-0.59,0.58]& \cite{Ellis:2020unq} & $-$ & $-$ & $-$  \tabularnewline
     $\Qpp{\phi Q}{(3)}$ 
    &  $i\big(\phi^\dagger\lra{D}_\mu\,\tau_{\sss I}\phi\big)\big(\bar{Q}\,\gamma^\mu\,\tau^{\sss I}Q\big)$
    & [-3.2,4.8]$\sst\times 10^{-2}$  & [-0.67,0.46]& \cite{Ellis:2020unq} & $-$ & $-$ & $-$  \tabularnewline
     $\Opp{\phi Q}{(-)}$ 
    & $\Qpp{\phi Q}{(1)}$ 
    &  [-1.0,1.4] & [-1.7,12] & \cite{Ethier:2021bye} & [-1.2,1.6] & [-2.3,2.9]& \cite{Ethier:2021bye}  \tabularnewline
     $\Opp{\phi Q}{(3)}$ 
    &  $\Qpp{\phi Q}{(3)}+\Qpp{\phi Q}{(1)}$ 
    &  [-0.35,0.35] & [-1.2,0.74] & \cite{Ethier:2021bye} & [-0.38,0.34] & [-0.62,0.48]& \cite{Ethier:2021bye}  \tabularnewline
     $\Op{\phi t}$       
    & $i\big(\phi^\dagger\,\lra{D}_\mu\,\,\phi\big)\big(\bar{t}\,\gamma^\mu\,t\big)$
    &  [-1.2,2.9] & [2,11] & \cite{Ellis:2020unq} & [-3.0,2.2] & [-13,4.0] & \cite{Ethier:2021bye}  \tabularnewline
\hline
\end{tabular}
}}
\renewcommand{\arraystretch}{1.}

%% file: sm_results_tab.tex
\renewcommand{\arraystretch}{1.5}
\resizebox{\textwidth}{!}{
\centering
\begin{tabular}{|c|c|c|c|c|c|c|}
\hline
\multirow{2}{*}{} & \multicolumn{3}{c|}{\textbf{Inclusive}} & \multicolumn{3}{c|}{\textbf{High-Energy}} \\ \cline{2-7} 
            & LO    & NLO DR1    & NLO DR2   & LO       & NLO DR1       & NLO DR2      \\ \hline
SM{}  
&$103.36(4)^{+12.76\%}_{-12.82\%}$       
&$106.70(15)^{+4.97\%}_{-6.28\%}$            
&$106.80(9)^{+5.04\%}_{-5.62\%}$
&$0.073(0)^{+15.92\%}_{-14.23\%}$          
&$0.048(0)^{+10.86\%}_{-18.35\%}$               
&$0.036(0)^{+26.82\%}_{-45.63\%}$              \\ \hline
\end{tabular}}

%% file: merged_tab_smeft.tex
\resizebox{\textwidth}{!}{
\renewcommand{\arraystretch}{1.5}
\centering
\begin{tabular}{|c| c| c| c| c| c| c|}
\hline
\multirow{2}{*}{\textbf{$c_i$}} & \multicolumn{3}{|c|}{\textbf{$\mathcal{O}(\Lambda^{-2}$)}} & \multicolumn{3}{|c|}{\textbf{$\mathcal{O}(\Lambda^{-4}$)}} \\ \cline{2-7} 
    & \textbf{LO}   & \textbf{NLO DR1}  & \textbf{NLO DR2}   & \textbf{LO} &\textbf{NLO DR1} &\textbf{NLO DR2}              
    \\ \hline 
$c_{\phi Q}^{(3)}$        
&19.78$(1) ^{+12.98\%}_{-13.02\%}$           
&21.20$(2) ^{+5.66\%}_{-6.13\%}$   
&21.37$(3) ^{+5.90\%}_{-6.27\%}$   
&4.94$(1) ^{+10.53\%}_{-10.62\%}$ 
&4.71$(2) ^{+4.88\%}_{-6.37\%}$
&4.72$(2) ^{+4.90\%}_{-6.50\%}$       \\ 
$c_{\phi Q}^{(-)}$        
&2.19$(0) ^{+12.65\%}_{-12.72\%}$           
&2.69$(1) ^{+8.92\%}_{-8.18\%}$                    
&2.74$(1) ^{+8.52\%}_{-8.17\%}$ 
&0.44$(0) ^{+12.18\%}_{-12.29\%}$ 
&0.48$(0) ^{+5.89\%}_{-5.89\%}$ 
&0.48$(0) ^{+5.86\%}_{-6.03\%}$        \\ 
$c_{\phi t}$         
&1.77$(0) ^{+13.11\%}_{-13.13\%}$           
&1.81$(0) ^{+4.81\%}_{-5.53\%}$                    
&1.84$(0) ^{+5.08\%}_{-5.74\%}$  
&0.19$(0) ^{+11.44\%}_{-11.61\%}$
&0.18$(0) ^{+4.45\%}_{-7.08\%}$
&0.17$(0) ^{+5.09\%}_{-7.54\%}$       \\ 
$c_{tW}$         
&-11.34$(1) ^{+12.27\%}_{-12.15\%}$           
&-11.49$(2) ^{+5.84\%}_{-5.57\%}$                   
&-11.68$(1) ^{+5.72\%}_{-5.41\%}$
&24.06$(3) ^{+10.53\%}_{-9.90\%}$
&23.38$(5) ^{+4.15\%}_{-5.18\%}$ 
&22.79$(3) ^{+4.59\%}_{-6.41\%}$       \\ 
$c_{tZ}$         
&-0.26$(0) ^{+11.03\%}_{-11.01\%}$           
&-0.35$(2) ^{+4.99\%}_{-6.66\%}$   
&-0.34$(1) ^{+5.47\%}_{-7.06\%}$
&5.23$(1) ^{+10.53\%}_{-10.05\%}$
&4.90$(2) ^{+4.59\%}_{-6.69\%}$           
&4.94$(1) ^{+4.64\%}_{-6.57\%}$      \\ 
$c_{tG}$         
&7.95$(0) ^{+13.00\%}_{-13.04\%}$           
&7.36$(1) ^{+4.00\%}_{-5.01\%}$                    
&7.26$(1) ^{+4.65\%}_{-6.29\%}$ 
&15.04$(3) ^{+11.61\%}_{-11.22\%}$
&12.19$(8) ^{+6.97\%}_{-11.93\%}$
&12.17$(7) ^{+6.95\%}_{-11.96\%}$       \\  \hline
\end{tabular}
}

%% file: SMEFT_impacts.tex
\renewcommand{\arraystretch}{1.4}
\resizebox{\textwidth}{!}{
\centering
\begin{tabular}{|c|c|c|c|c|c|c|c|c|c|c|c|c|}
\hline
\multirow{3}{*}{$c_{i}$} & \multicolumn{6}{c|}{\textbf{Inclusive}}  & \multicolumn{6}{c|}{\textbf{High-Energy}} \\ \cline{2-13}
    & \multicolumn{3}{c|}{\textbf{$\mathcal{O}(\Lambda^{-2})$}}     & \multicolumn{3}{c|}{\textbf{$\mathcal{O}(\Lambda^{-4})$}}       & \multicolumn{3}{c|}{\textbf{$\mathcal{O}(\Lambda^{-2})$}}       & \multicolumn{3}{c|}{\textbf{$\mathcal{O}(\Lambda^{-4})$}}  \\ \cline{2-13}  
    &LO  &NLO  &$K$     &LO &NLO &$K$   &LO &NLO &$K$   &LO &NLO &$K$  \\ \hline
    $c_{\phi Q}^{(3)}$      
    &0.191      &0.200      &1.05     
    &0.048      &0.044      &0.92      
    &-0.870     &-0.715     &0.82     
    &5.626      &7.476      &1.33       \\ \hline
    $c_{\phi Q}^{(-)}$           
    &0.021      &0.026      &1.24     
    &0.004      &0.004      &1.00      
    &0.028      &0.056      &2.00     
    &0.058      &0.057      &0.98       \\ \hline
    $c_{\phi t}$                 
    &0.017      &0.017      &1.00     
    &0.002      &0.002      &1.00      
    &0.017      &0.023      &1.35     
    &0.056      &0.054      &0.96       \\ \hline
    $c_{tW}$                     
    &-0.110     &-0.109     &0.99     
    &0.233      &0.213      &0.91      
    &-0.528     &-0.524     &0.99     
    &30.905     &40.695     &1.32        \\ \hline
    $c_{tZ}$                     
    &-0.003     &-0.003     &1.00     
    &0.051      &0.046      &0.90      
    &0.098      &0.076      &0.78     
    &7.739      &10.482     &1.35        \\ \hline
    $c_{tG}$                     
    &0.077      &0.068      &0.88     
    &0.145      &0.114      &0.79      
    &-0.232     &-0.354     &1.53     
    &7.488      &5.242      &0.70       \\ \hline
\end{tabular}
}

%% file: tab_bwtz_hel.tex
\renewcommand{\arraystretch}{1.3}
\setlength{\tabcolsep}{3pt}
\begin{tabular}{cccc|c|ccccc}    
$\lambda_{\sss b}\,$,& $\lambda_{\sss W}\,$,& $\lambda_{\sss t}\,$,& $\lambda_{\sss Z}\,$ & SM & $ \Op{tW}$ & $ \Op{tZ}$ & $ \Op{\phi Q}^{(-)}$ & $ \Op{\phi Q}^{(3)}$ & $ \Op{\phi t}$\tabularnewline
\hline
$-\,$,& $0\,$,& $-\,$,& $0\,$& $\scriptstyle s^{0\phantom{-,}}$ & $\scriptstyle -s^0$ & $\scriptstyle -$ & $\scriptstyle -$ & $\scriptstyle 2 \sqrt{2} \sqrt{s (s+t)}$ & $\scriptstyle -$\tabularnewline
$-\,$,& $0\,$,& $+\,$,& $0\,$& $\scriptstyle s^{\sss -\frac{1}{2}}$ & $\scriptstyle 4 \mw\left(\frac{\scriptstyle s}{ \scriptstyle \sqrt{-t}} - 2 \sqrt{-t}\right)$ & $\scriptstyle 2 \mz \sqrt{-t}$ & $\scriptstyle \sqrt{2}\mt \sqrt{-t} $ & $\scriptstyle 2 \sqrt{2} \mt \sqrt{-t} $ & $\scriptstyle -\sqrt{2}  \mt \sqrt{-t}$\tabularnewline
$-\,$,& $-\,$,& $-\,$,& $0\,$& $\scriptstyle s^{\sss -\frac{1}{2}}$ & $\scriptstyle -$ & $\scriptstyle -$ & $\scriptstyle -$ & $\scriptstyle 4 \mw\sqrt{-t} $ & $\scriptstyle -$\tabularnewline
$-\,$,& $-\,$,& $+\,$,& $0\,$& $\scriptstyle s^{\sss-1\phantom{,}}$ & $\scriptstyle 2 \sqrt{2} \sqrt{s (s+t)}$ & $\scriptstyle s^0$ & $\scriptstyle s^0$ & $\scriptstyle s^0$ & $\scriptstyle s^0$\tabularnewline
$-\,$,& $+\,$,& $-\,$,& $0\,$& $\scriptstyle s^{\sss -\frac{1}{2}}$ & $\scriptstyle -$ & $\scriptstyle -$ & $\scriptstyle -$ & $\scriptstyle -$ & $\scriptstyle -$\tabularnewline
$-\,$,& $+\,$,& $+\,$,& $0\,$& $\scriptstyle s^{0\phantom{-,}}$ & $\scriptstyle s^0$ & $\scriptstyle -$ & $\scriptstyle -$ & $\scriptstyle s^0$ & $\scriptstyle -$\tabularnewline
$-\,$,& $0\,$,& $-\,$,& $-\,$& $\scriptstyle s^{\sss -\frac{1}{2}}$ & $\scriptstyle -$ & $\scriptstyle -2 \sqrt{2} \mt\sqrt{-t} $ & $\scriptstyle -$ & $\scriptstyle -4 \mz \sqrt{-t}$ & $\scriptstyle -$\tabularnewline
$-\,$,& $0\,$,& $-\,$,& $+\,$& $\scriptstyle s^{\sss -\frac{1}{2}}$ & $\scriptstyle -$ & $\scriptstyle -$ & $\scriptstyle -$ & $\scriptstyle -$ & $\scriptstyle -$\tabularnewline
$-\,$,& $0\,$,& $+\,$,& $-\,$& $\scriptstyle s^{0\phantom{-,}}$ & $\scriptstyle s^0$ & $\scriptstyle -$ & $\scriptstyle s^0$ & $\scriptstyle s^0$ & $\scriptstyle -$\tabularnewline
$-\,$,& $0\,$,& $+\,$,& $+\,$& $\scriptstyle s^{\sss-1\phantom{,}}$ & $\scriptstyle -4 \sqrt{2} \cw \sqrt{s (s+t)}$ & $\scriptstyle 2 \sqrt{2} \sqrt{s (s+t)}$ & $\scriptstyle s^0$ & $\scriptstyle -$ & $\scriptstyle s^0$\tabularnewline
$-\,$,& $-\,$,& $-\,$,& $-\,$& $\scriptstyle s^{0\phantom{-,}}$ & $\scriptstyle s^0$ & $\scriptstyle s^0$ & $\scriptstyle s^0$ & $\scriptstyle s^0$ & $\scriptstyle -$\tabularnewline
$-\,$,& $-\,$,& $-\,$,& $+\,$& $\scriptstyle s^{\sss-1\phantom{,}}$ & $\scriptstyle -$ & $\scriptstyle -$ & $\scriptstyle -$ & $\scriptstyle -$ & $\scriptstyle -$\tabularnewline
$-\,$,& $-\,$,& $+\,$,& $-\,$& $\scriptstyle s^{\sss -\frac{1}{2}}$ & $\scriptstyle  8 \mw\cw \frac{\scriptstyle s}{\scriptstyle \sqrt{-t}} - 4    \mz\left(1-\frac{4}{3} \sw^2\right) \sqrt{-t}$ & $\scriptstyle -$ & $\scriptstyle -$ & $\scriptstyle -$ & $\scriptstyle -$\tabularnewline
$-\,$,& $-\,$,& $+\,$,& $+\,$& $\scriptstyle -$ & $\scriptstyle 8  \mz \left(\frac{1}{3}\sw^2-1\right)\sqrt{-t}$ & $\scriptstyle 4  \mw\sqrt{-t}$ & $\scriptstyle -$ & $\scriptstyle -$ & $\scriptstyle -$\tabularnewline
$-\,$,& $+\,$,& $-\,$,& $-\,$& $\scriptstyle s^{\sss-1\phantom{,}}$ & $\scriptstyle -$ & $\scriptstyle -$ & $\scriptstyle -$ & $\scriptstyle -$ & $\scriptstyle -$\tabularnewline
$-\,$,& $+\,$,& $-\,$,& $+\,$& $\scriptstyle s^{0\phantom{-,}}$ & $\scriptstyle -$ & $\scriptstyle -$ & $\scriptstyle s^0$ & $\scriptstyle s^0$ & $\scriptstyle -$\tabularnewline
$-\,$,& $+\,$,& $+\,$,& $-\,$& $\scriptstyle s^{\sss -\frac{1}{2}}$ & $\scriptstyle -$ & $\scriptstyle -$ & $\scriptstyle -$ & $\scriptstyle -$ & $\scriptstyle -$\tabularnewline
$-\,$,& $+\,$,& $+\,$,& $+\,$& $\scriptstyle s^{\sss -\frac{1}{2}}$ & $\scriptstyle 8\mw  \cw \left(\frac{\scriptstyle s}{\scriptstyle\sqrt{-t}}-\sqrt{-t}\right)$ & $\scriptstyle -$ & $\scriptstyle -$ & $\scriptstyle -$ & $\scriptstyle -$\tabularnewline
$+\,$,& $\mkern2mu\,\cdot\,\,\mkern1mu$,& $\mkern2mu\,\cdot\,\,\mkern1mu$,& $\mkern2mu\,\cdot\,\,\mkern1mu$ & $\scriptstyle -$ & $\scriptstyle -$ & $\scriptstyle -$ & $\scriptstyle -$ & $\scriptstyle -$ & $\scriptstyle -$\tabularnewline
\end{tabular}
\renewcommand{\arraystretch}{1.}